\PassOptionsToPackage{pagebackref=true}{hyperref}

\documentclass[runningheads]{llncs}

\usepackage{eccv}

\newcommand{\status}{1}

\input{tools/package}
\definecolor{INCOMPLETECOLOR}{RGB}{178,34,34}
\definecolor{UNDERREVISIONCOLOR}{RGB}{210,121,121}
\definecolor{FEEDBACKNEEDEDCOLOR}{RGB}{230,170,50}
\definecolor{FEEDBACKGIVENCOLOR}{RGB}{121,210,121}
\definecolor{COMPLETECOLOR}{RGB}{121,124,210}
\definecolor{LOCKEDCOLOR}{RGB}{153,102,255}

\definecolor{TODOCOLOR}{RGB}{255,0,0}
\definecolor{BINGXUANCOLOR}{RGB}{0,0,255}
\definecolor{QICOLOR}{RGB}{118,185,0}
\definecolor{JIAHAOCOLOR}{RGB}{127,127,0}
\definecolor{KENNYCOLOR}{RGB}{127,0,127}
\definecolor{COLINCOLOR}{RGB}{227,0,207}
\definecolor{GUESTCOLOR}{RGB}{0,127,127}

\definecolor{WHITE}{RGB}{255,255,255}

\definecolor{rOne}{RGB}{12, 116, 207}   
\definecolor{rTwo}{RGB}{226, 119, 51}    
\definecolor{rThree}{RGB}{76,155,80}   

\newcommand{\colorTableGreen}{green!10}
\newcommand{\colorTableYellow}{yellow!20}
\newcommand{\colorTableGray}{gray!20}
\newcommand{\colorTableLightGray}{gray!8}

\newcommand{\colorTableFirst}{orange!32}
\newcommand{\colorTableSecond}{orange!16}
\newcommand{\colorTableThird}{orange!8}

\newcommand{\tableGreen}{\cellcolor{\colorTableGreen}}
\newcommand{\tableYellow}{\cellcolor{\colorTableYellow}}
\newcommand{\tableGray}{\cellcolor{\colorTableGray}}
\newcommand{\tableLightGray}{\cellcolor{\colorTableLightGray}}

\newcommand{\rankFirst}{\cellcolor{\colorTableFirst}}
\newcommand{\rankSecond}{\cellcolor{\colorTableSecond}}
\newcommand{\rankThird}{\cellcolor{\colorTableThird}}

\newcommand{\triangcellWH}[3]{%
\begingroup
\setlength{\tabcolsep}{0pt}%
\begin{tikzpicture}[baseline=(A.base)]
  \node[
    inner sep=1pt,          
    outer sep=0pt,
    minimum width=#1,
    minimum height=#2
  ] (A) {#3};               
  \begin{scope}[on background layer]
    \path[fill=\colorTableGreen]
      (A.north west) -- (A.north east) -- (A.south west) -- cycle;
    \path[fill=\colorTableYellow]
      (A.south east) -- (A.north east) -- (A.south west) -- cycle;
  \end{scope}
\end{tikzpicture}%
\endgroup
}

\newcommand{\nothing}[1]{}
\newcommand{\isolated}[1]{\hfill\break#1\xspace}
\newcommand{\Caption}[2]{\caption[#1]{{\em #1} #2}}

\newcommand{\tinyspace}{\hspace{0.25em}}

\ifthenelse{\equal{\status}{0}} {                     

    \newcommand{\todo}[1]{%
        \addcontentsline{toc}{subsection}{
            \protect\numberline{}
            \textcolor{TODOCOLOR}{[TODO] #1}}
            \textcolor{TODOCOLOR}{[TODO] \emph{#1}}}
    \newcommand{\warning}[1]{\todo{#1}}

    \newcommandx{\bingxuan}[2][1=]
        {\setulcolor{BINGXUANCOLOR}{\ul{#1}}
         \isolated{\textcolor{BINGXUANCOLOR}{\textbf{Bingxuan:} #2}}}
    \newcommandx{\qisun}[2][1=]
        {\setulcolor{QICOLOR}{\ul{#1}}
         \isolated{\textcolor{QICOLOR}{\textbf{Qi:} #2}}}
    \newcommandx{\jiahao}[2][1=]
        {\setulcolor{JIAHAOCOLOR}{\ul{#1}}
         \isolated{\textcolor{JIAHAOCOLOR}{\textbf{Jiahao:} #2}}}
    \newcommandx{\kenny}[2][1=]
        {\setulcolor{KENNYCOLOR}{\ul{#1}}
         \isolated{\textcolor{KENNYCOLOR}{\textbf{Kenny:} #2}}}
    \newcommandx{\cg}[2][1=]
        {\setulcolor{COLINCOLOR}{\ul{#1}}
         \isolated{\textcolor{COLINCOLOR}{\textbf{Colin:} #2}}}

    \newcommandx{\guest}[3][1=]
        {\setulcolor{LOCKEDCOLOR}{\ul{#1}} \textcolor{LOCKEDCOLOR}
        {[\textbf{#2:} #3]}}
}{                                                    
    \newcommand{\todo}[1]{}
    \newcommand{\warning}[1]{}

    \newcommandx{\bingxuan}[2][1=]{#1}
    \newcommandx{\qisun}[2][1=]{#1}
    \newcommandx{\jiahao}[2][1=]{#1}
    \newcommandx{\kenny}[2][1=]{#1}
    \newcommandx{\cg}[2][1=]{#1}
    \newcommandx{\guest}[3][1=]{#1}
}

\ifthenelse{\equal{\status}{0}} {          
    
}{}
\ifthenelse{\equal{\status}{1}} {          
    
}{}
\ifthenelse{\equal{\status}{2}} {          
    
}{}
\ifthenelse{\equal{\status}{3}} {          
    
}{}
\ifthenelse{\equal{\status}{4}} {          
    
}{}

\ifthenelse{\equal{\status}{0}} {
    \newcommand{\badge}[2]{\colorbox{#1}{\small\textcolor{WHITE}{\texttt{#2}}}}
    \newcommand{\headerBadge}[2]{\hspace*{\fill}\badge{#1}{#2}}
    
    \newcommand{\complete}{\headerBadge{COMPLETECOLOR}{complete}}

}{
    \newcommand{\badge}[2]{}{}
    \newcommand{\headerBadge}[2]{}{}
    
    \newcommand{\complete}{}

}

\newcommand{\matName}{TiO\textsubscript{2}\xspace}

\newcommand{\image}{I}
\newcommand{\polarDirection}{k}
\newcommand{\polarImageX}{\image_x}
\newcommand{\polarImageY}{\image_y}
\newcommand{\simulatorDepthMask}{M}
\newcommand{\sceneIrradiance}{S}

\newcommand{\physicalPoint}{\mathbf{X}}

\newcommand{\depth}{z}

\newcommand{\inE}{E_{\mathrm{in}}}
\newcommand{\outE}{E_{\mathrm{out}}}
\newcommand{\metaPhase}{\psi_m}
\newcommand{\metaPhaseX}{\psi_{m,x}}
\newcommand{\metaPhaseY}{\psi_{m,y}}

\newcommand{\metaVector}{\vec{r}_m}

\newcommand{\metaVectorZero}{\vec{r}_{m0}}

\newcommand{\amplitudePSF}{U}
\newcommand{\imageVector}{\vec{r}_{i}}
\newcommand{\imageDistance}{\Delta\hat{z}}

\newcommand{\phaseLens}{\psi_f}

\newcommand{\sepy}{\Delta y}
\newcommand{\focus}{f}
\newcommand{\phaseFresnel}{\psi_r} 
\newcommand{\pupilR}{r_m} 
\newcommand{\pupilAngle}{\phi_m} 
\newcommand{\pupilx}{x_m}
\newcommand{\pupily}{y_m}
\newcommand{\RingNumber}{N} 
\newcommand{\RingIndex}{n} 
\newcommand{\infocus}{z_f} 
\newcommand{\radius}{R} 
\newcommand{\wavelength}{\lambda}
\newcommand{\defocus}{\zeta} 
\newcommand{\normimageR}{\tilde{r}_i} 
\newcommand{\imageAngle}{\phi_i}
\newcommand{\imageR}{r_i}
\newcommand{\imageX}{x_i}
\newcommand{\imageY}{y_i}

\newcommand{\psf}{\mathcal{P}}
\newcommand{\psfX}{\psf_x}
\newcommand{\psfY}{\psf_y}
\newcommand{\psfboth}{\psf_k}

\newcommand{\point}{\mathbf{p}}
\newcommand{\pointX}{x(\point)}
\newcommand{\pointY}{y(\point)}
\newcommand{\pointZ}{z(\point)}

\newcommand{\inEx}{E_{\mathrm{in},x}}
\newcommand{\inEy}{E_{\mathrm{in},y}}
\newcommand{\outEx}{E_{\mathrm{out},x}}
\newcommand{\outEy}{E_{\mathrm{out},y}}
\newcommand{\metaPhaseboth}{\psi_{k}}
\newcommand{\Fresnelphaseboth}{\psi_{r,k}}
\newcommand{\phaseLensboth}{\psi_{f,k}}
\newcommand{\Fresnelphasex}{\psi_{r,x}}
\newcommand{\Fresnelphasey}{\psi_{r,y}}
\newcommand{\atomPhaseX}{\psi_x}
\newcommand{\atomPhaseY}{\psi_y}
\newcommand{\phaseSpace}{\mathcal{PS}}
\newcommand{\pitch}{a}
\newcommand{\height}{h}
\newcommand{\slx}{L_x}
\newcommand{\sly}{L_y}
\newcommand{\clx}{L_{x1}}
\newcommand{\cly}{L_{y1}}
\newcommand{\cclx}{L_{x2}}
\newcommand{\ccly}{L_{y2}}
\newcommand{\fabCon}{\delta_f}

\newcommand{\trans}{t}

\DeclareMathOperator{\dd}{d}

\newcommand{\Fourier}{\mathcal{F}}

\begin{document}
\title{Physically Grounded Monocular Depth via Nanophotonic Wavefront Encoding}
\titlerunning{Physically Grounded Monocular Depth}

\author{
Bingxuan Li\textsuperscript{*,1} \enspace
Jiahao Wu\textsuperscript{*,2} \enspace
Yuan Xu\textsuperscript{*,2} \enspace 
Zezheng Zhu\textsuperscript{2} \enspace 
Yunxiang Zhang\textsuperscript{1} \enspace 
Kenneth Chen\textsuperscript{1} \enspace 
Yanqi Liang\textsuperscript{2} \enspace 
Nanfang Yu\textsuperscript{$\dagger$,2} \enspace 
Qi Sun\textsuperscript{$\dagger$,1} \enspace 
}

\authorrunning{\hspace{-12em}B.~Li et al.}
\institute{\textsuperscript{1} New York University \quad \textsuperscript{2} Columbia University}



\maketitle
\begingroup
\renewcommand{\thefootnote}{}
\footnotetext{* Equal contribution.}
\footnotetext{$\dagger$ Corresponding authors.}
\endgroup

\begin{abstract}
Depth foundation models (DFMs) offer strong learned priors for 3D perception from single RGB images but lack physical depth cues, leading to ambiguities in metric scale. We introduce metalenses, an emerging class of ultrathin planar optical elements, as a solution to physically encode missing metric depth cues via nanophotonics. In this paper, we bridge the gap between metalens and DFMs to achieve accurate metric monocular depth sensing. In a single monocular shot, our metalens embeds depth-dependent positional shifts into two polarized optical wavefronts. With an input adaptation strategty, we enable direct fine-tuning that aligns a pretrained DFM with the optical signals. To scale the training data, we further develop a comprehensive simulation pipeline that synthesizes metalens responses from RGB-D datasets, incorporating physical factors to minimize the sim-to-real gap. Experiments demonstrate that this approach outperforms both monocular metric depth estimation and depth-from-defocus baselines, showing an effective pathway for accurate monocular metric depth sensing.
\keywords{Depth foundation models \and Metasurface}
\end{abstract}

\begin{figure}
\centering
\includegraphics[width=.9\linewidth]{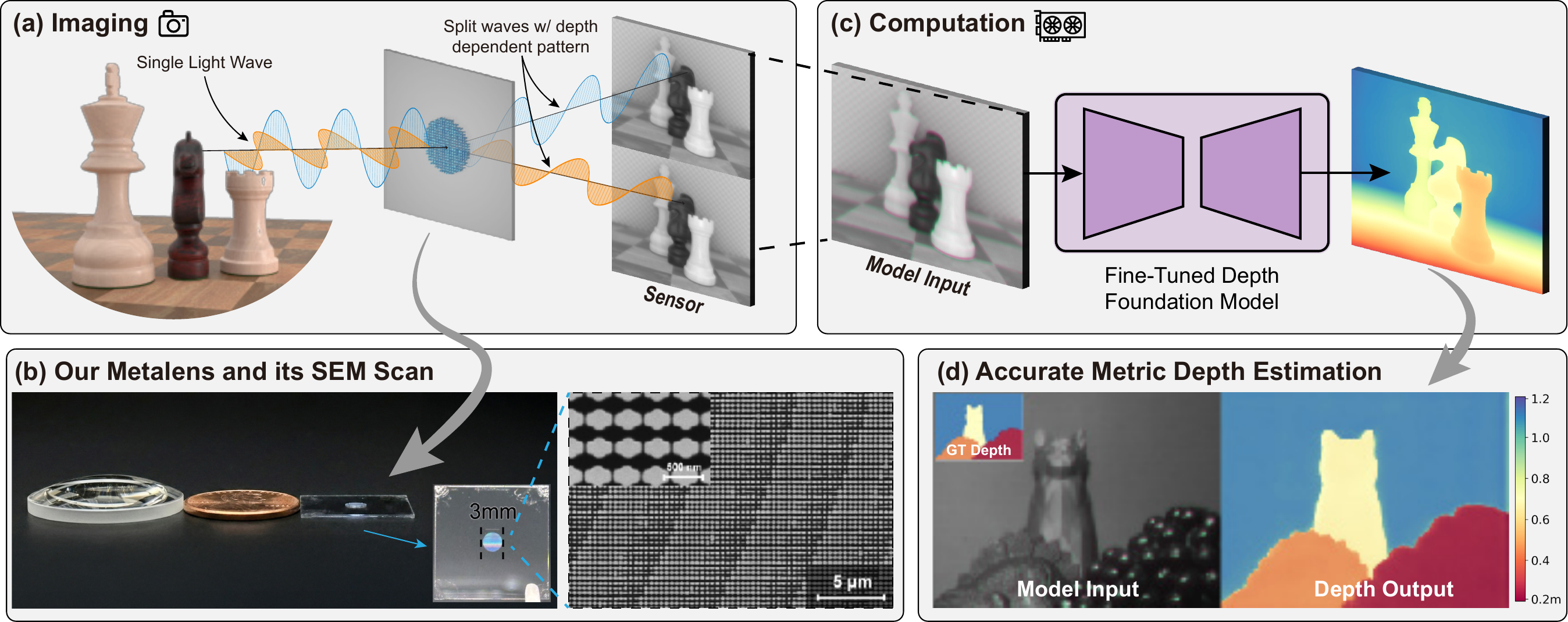}
\caption{\textit{Overview of our system and method.} (a) Our birefringent metalens converts a 3D scene into two polarized images, encoding depth information in pixel-wise shifts between the images (see \cref{fig:lib:psf}). 
(b) The compact 3-mm-diameter metalens (right) consists of a two-dimensional array of 700-nm-tall TiO$_2$ nanopillars with anisotropic cross-sections, engineered to provide independent phase control for x- and y-polarized light. For scale, it is shown alongside a 1-inch plano-convex lens (left) and a U.S. 1-cent coin (middle).
(c) These depth-dependent optical signals are converted into model inputs and processed by a fine-tuned depth foundation model. 
(d) Our method recovers metrically accurate depth by combining physical depth cues with learned image priors, enabling high-quality physically grounded monocular depth estimation.
}
\label{fig:teaser}
\vspace{-1.5em}
\end{figure}


\vspace{-6pt}
\section{Introduction \complete}
\label{sec:intro}

Depth foundation models (DFMs) \cite{depth_anything_v2, ke2023repurposing} have recently achieved remarkable progress in monocular depth estimation by learning rich geometric priors from large-scale data, showing strong capabilities from \textit{relative} to \textit{metric} depth estimation \cite{ZoeDepth, DepthPro, Guizilini2023depth, piccinelli2024unidepth, yin2023metric}. However, the lack of physical depth cues from a monocular capture makes \textit{metric} depth estimation inherently ill-posed, resulting in ambiguity and inaccuracy in applications requiring precise metric depth.

To enable physically grounded monocular depth estimation, providing DFMs with diverse modalities has emerged as a promising direction. Recent works leverage auxiliary sensors such as LiDAR \cite{lin2025prompting,liang2025distilling, park2024depth} to provide accurate metric supervision. Yet, such systems depend on active, energy-consuming hardware, and the inclusion of additional sensors increases form factor and system complexity, deviating from a strict monocular setting. This raises a natural question: \textit{can we ground DFMs in physics solely through passive optics within a compact monocular device, without relying on active sensing and additional sensors}?

To answer this question, we introduce a new framework that physically grounds DFMs through passive light wave encoding in a single monocular capture.
This is enabled by our custom-designed and fabricated birefringent metalens --- an ultra-thin, planar element composed of nanophotonic structures for modulating optical wavefronts with subwavelength resolution (see \cref{sec:related:metalens} for background).
As illustrated in \cref{fig:teaser}, our metalens decomposes incoming light into two orthogonal polarization channels, each formed by a distinct depth-dependent point spread function (PSF). 
These two channels are formed along the same optical path and are projected onto the sensor in a single exposure, where the positional shift between the conjugate PSFs encodes metric depth. Both images originate from one viewpoint without multi-view parallax, making our approach fundamentally distinct from stereo.

Subsequently, through an input adaptation strategy, we transform the two polarization channels into a three-channel representation that embeds physical depth cues while retaining scene semantics. This enables a pretrained DFM to leverage its robust learned priors while simultaneously recovering metric scale from the optical signals without necessitating any architectural modifications.
Specifically, we choose the Depth Anything V2~\cite{depth_anything_v2} as our model backbone. To solve the challenge in collecting large-scale training data, we develop a simulation pipeline that synthesizes the polarization channels from RGB-D datasets by physically modeling the birefringent metalens. While the simulation-to-real gap can degrade performance, we analyze its sources and introduce a novel disocclusion-aware simulator that more accurately models the optical formation of asymmetric PSFs, supplemented by polarization-aware data augmentation.

We evaluate our approach in both simulated and physical experiments, demonstrating consistent improvement over state-of-the-art metric monocular depth estimators. 
Notably, we achieve performance comparable to PromptDA~\cite{lin2025prompting}, which relies on LiDAR as an auxiliary sensor. 
Our method also outperform depth-from-defocus baselines~\cite{ikoma2021depth,splitaperturecameras} in simulation, with ablation study verifying that both the optical frontend and the pretrained model backend drive the performance gains.
These results underscore the potential of metalens in depth perception and its applicability to VR/AR, miniature robotics, medical endoscopy, and other embedded 3D vision systems.
In summary, we make the following contributions: 

\begin{itemize}
\item We introduce a new approach for physically grounded monocular depth estimation with birefringent metalens, featuring an input adaptation strategy that enables direct fine-tuning of a DFM.
\item We present a disocclusion-aware optical forward model that accurately captures the image formation of asymmetric PSFs, paired with polarization-aware augmentation for improved simulation-to-real transfer.
\item We demonstrate an integrated hardware-software depth sensing system, achieving highly accurate metric depth through the synergy of physical optical grounding and learned depth priors.
\end{itemize}

\section{Background \& Related Work \complete}
\label{sec:related}

\subsection{Metasurface and Metalens}
\label{sec:related:metasurface}
\label{sec:related:metalens}

A metasurface is a planar nanophotonic device composed of a 2D array of subwavelength dielectric pixels with different sizes and shapes chosen to locally control the optical phase delay, so that the array of pixels collectively mold the optical wavefront into a desired shape with subwavelength resolution ~\cite{yu2011light, ni2012broadband}. The pixels can also be designed to control the amplitude and polarization state of the scattered light wave so that the metasurface can impart designer polarization and amplitude profiles over the wavefront ~\cite{balthasarmueller2017metasurface, huang2023int, cao2024aberration}.

Metasurfaces have enabled ultra-compact optics for displays \cite{nam2023depolarized,gopakumar2024full,zheng2023close}, optical computation \cite{wei2024spatially}, and color imaging \cite{chakravarthula2023thin,tseng2021neural}. 
They also show promise for depth sensing, with prior work on active metasurfaces for structured-light projection \cite{Li2018cloud,Ni2020fov,Kim2022fullspace}, LiDAR beam steering \cite{Kim2021rev,Park2021slm}, and compact high-speed or high-accuracy systems \cite{Chen2022semi,Martins2022metalidar,Yan2024nano}. These approaches rely on external illumination or electro-optic control. In contrast, passive metasurfaces can encode depth information in the optical response of metasurface-based lenses --- known as \textbf{metalenses} --- for example, in defocus \cite{Guo2019spidereye} and chromatic aberration \cite{Tan2021dispersion} of individual metalenses, and light fields of metalens arrays \cite{Chen2023intdepth}. 
One promising route for depth sensing use a helical PSF to encode depth information \cite{Berlich2016single,Jin2019dielectric,Jin2019dhpsf,Colburn2020paired,shen2023monocular}. 

However, lacking powerful computational backends with large-scale training, these prior methods are largely confined to single-object depth estimation or sparse feature matching. Consequently, they fail to reconstruct accurate, dense depth maps for full complex scenes. We overcome this limitation by combining metalens-encoded physical depth cues with the rich depth priors embedded in DFMs. Our approach leverages these priors to enable high-resolution, metric depth estimation across the entire scene within a passive, single-sensor system.


%

\subsection{Monocular Depth Estimation}
\label{sec:related:depth}
\bingxuan{add more citations}

Recovering 3D geometry from 2D images has long been a fundamental problem. Recent progress in monocular depth estimation has advanced 3D perception using. Models trained on large-scale datasets, including diffusion-based and vision transformer–based approaches, have evolved into DFMs \cite{ZoeDepth,depth_anything_v1,depth_anything_v2,bochkovskii2024depth,fu2024geowizard,yin2023metric}, demonstrating strong generalization across a wide range of scenes \cite{ke2023repurposing,piccinelli2025unidepthv2,DepthPro,he2024lotus,hu2024metric3d}. However, because single-view intensity lacks absolute depth cues, these models remain fundamentally scale-ambiguous.
To resolve this, prompting DFMs with additional sensors such as LiDAR has been explored \cite{lin2025prompting,park2024depth}. Recent work also explored inference-time optimization strategy that uses defocus blur cues to resolve the scale ambiguity of Marigold \cite{ke2023repurposing,talegaonkar2025repurposingmarigoldzeroshotmetric}.
However, LiDAR-based prompting requires a multi-sensor setup with active illumination\cite{sun}, while inference-time optimization is prohibitively time-consuming (taking approximately five minutes on a modern GPU). In contrast, our approach employs a purely passive optical modulator, leveraging polarization-dependent PSF shifts to encode depth without the need for active sensors.

A parallel line of work lies in computational optics. Conventional depth-from-defocus (DfD) estimates depth from blur \cite{gur2019single, tang2017depth, alexander2016focal}, but destroys high-frequency details and suffers from low sensitivity. To improve sensitivity, researchers design specialized masks \cite{DiffuserCam, Phasecam3d, ikoma2021depth, LenslessCameras, splitaperturecameras, Wijayasingha_2024_WACV} to engineer distinct PSFs. Other approaches use dual-pixel sensors \cite{splitaperturecameras, ghanekar2024passive, DualPixels, DualPixels2} or conventional birefringent materials \cite{bire1, bire2, ghanekar2022ps}. Despite these advancements, most systems train task-specific networks from scratch, without leveraging the depth priors of DFMs, which are crucial for reconstructing fine details and estimating depth in texture-less regions. Furthermore, engineered DfD introduces blur, while dual-pixel and polarization methods rely on specific sensors or extra components. Overcoming this, our metalens integrates PSF engineering, light focusing, and polarization multiplexing into one element which avoids severe blur and extra components.


\section{Method \complete}
\label{sec:method}
%

%
As illustrated in \cref{fig:teaser}, our system integrates three components: a birefringent metalens that converts depth into and polarization channels (\cref{sec:method:metalens}), a depth foundation model backbone and a encoding mechanism to inject physical cues (\cref{sec:method:prompting}), and a disocclusion-aware optical forward model with alpha compositing and data augmentation that reduce simulation-to-real gaps (\cref{sec:method:sim2real}). 

\begin{figure}[t]
\centering
\includegraphics[width=0.9\linewidth]{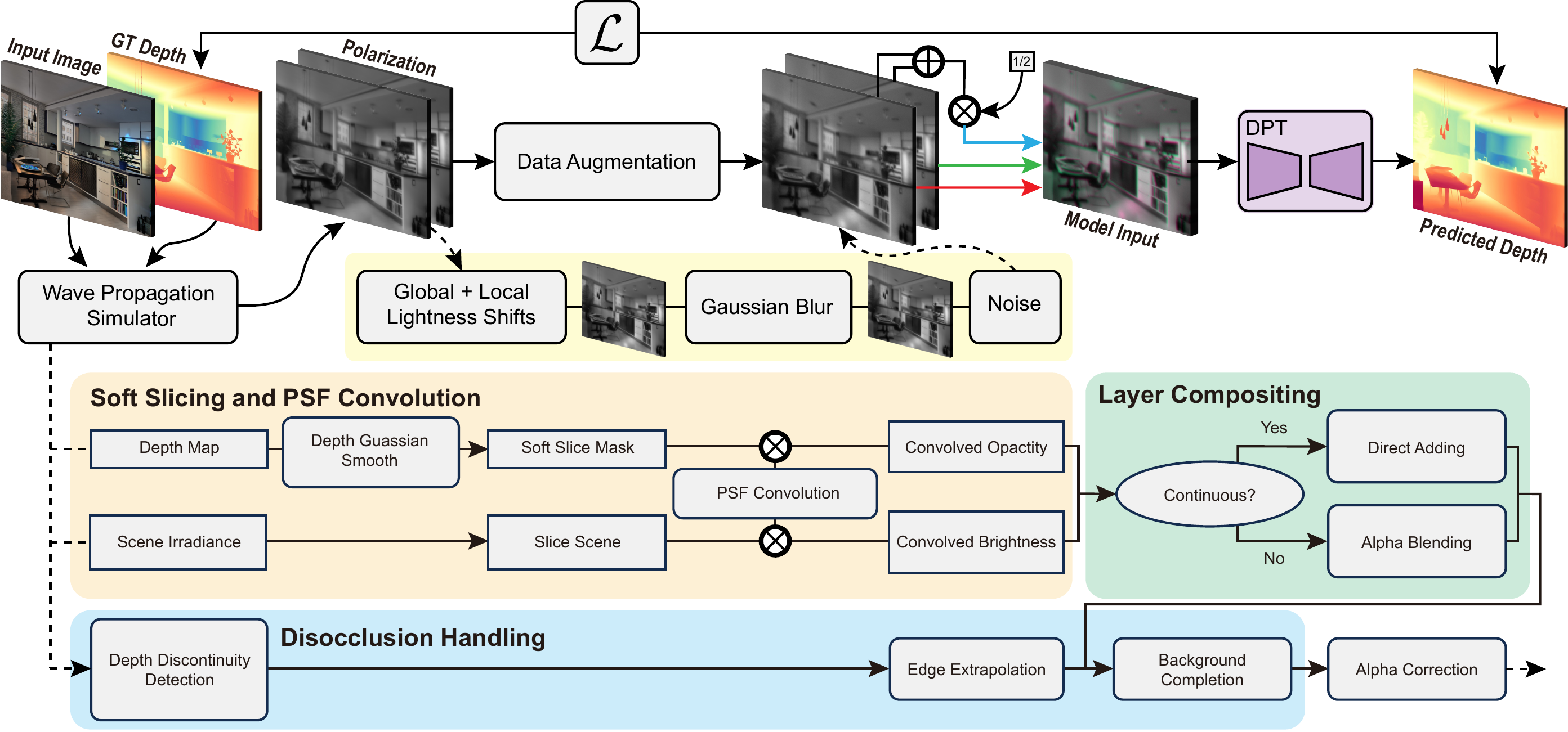}
\Caption{Illustration of our learning pipeline.}
{
The top half illustrates our learning pipeline: synthetic RGB-D data are processed by our simulator to generate polarization image pairs, which are then augmented and transformed to model input. 
We adopt the DPT architecture of DepthAnything v2 \cite{depth_anything_v2} with pretrained weights for fine-tuning.
The bottom half shows the workflow of our optical forward model, which integrates soft slicing, PSF convolution, disocclusion handling, and blending to eliminate simulation artifacts, narrowing the sim-to-real gap.
}
\vspace{-10pt}
\label{fig:pipeline}
\end{figure}

\subsection{Birefringent Metalens for Polarization-Based Depth Encoding}
\label{sec:method:metalens}

To optimally leverage DFMs, we adopt polarization-multiplexed single-helix PSFs \cite{Prasad2013psf,shen2023monocular}. Rotating PSFs provide noise-robust cues superior to standard defocus \cite{rotationdepth,rotationdepth2}, while their sharp profiles preserve high-spatial-frequency details essential for DFMs. Isolating single lobes via polarization eliminates ghosting to maximize the DFM's accuracy \cite{ghanekar2022ps}. Additionally, replacing prior near-infrared designs \cite{shen2023monocular} with our visible-light $\text{TiO}_2$ metasurface aligns input features with DFM priors while boosting depth sensitivity. Ablations (\cref{tab:ablation_compare_with_dfd}) verify that this configuration provides strong physical grounding for DFMs without joint training overhead.

\vspace{-5pt}

\begin{figure}[h]
\centering
\subfloat[birefringent metalens]{
    \begin{minipage}[c][3.2cm][c]{0.32\textwidth}
        \centering
        \includegraphics[height=2.5cm, trim={8cm 1cm 6cm 1cm},clip]{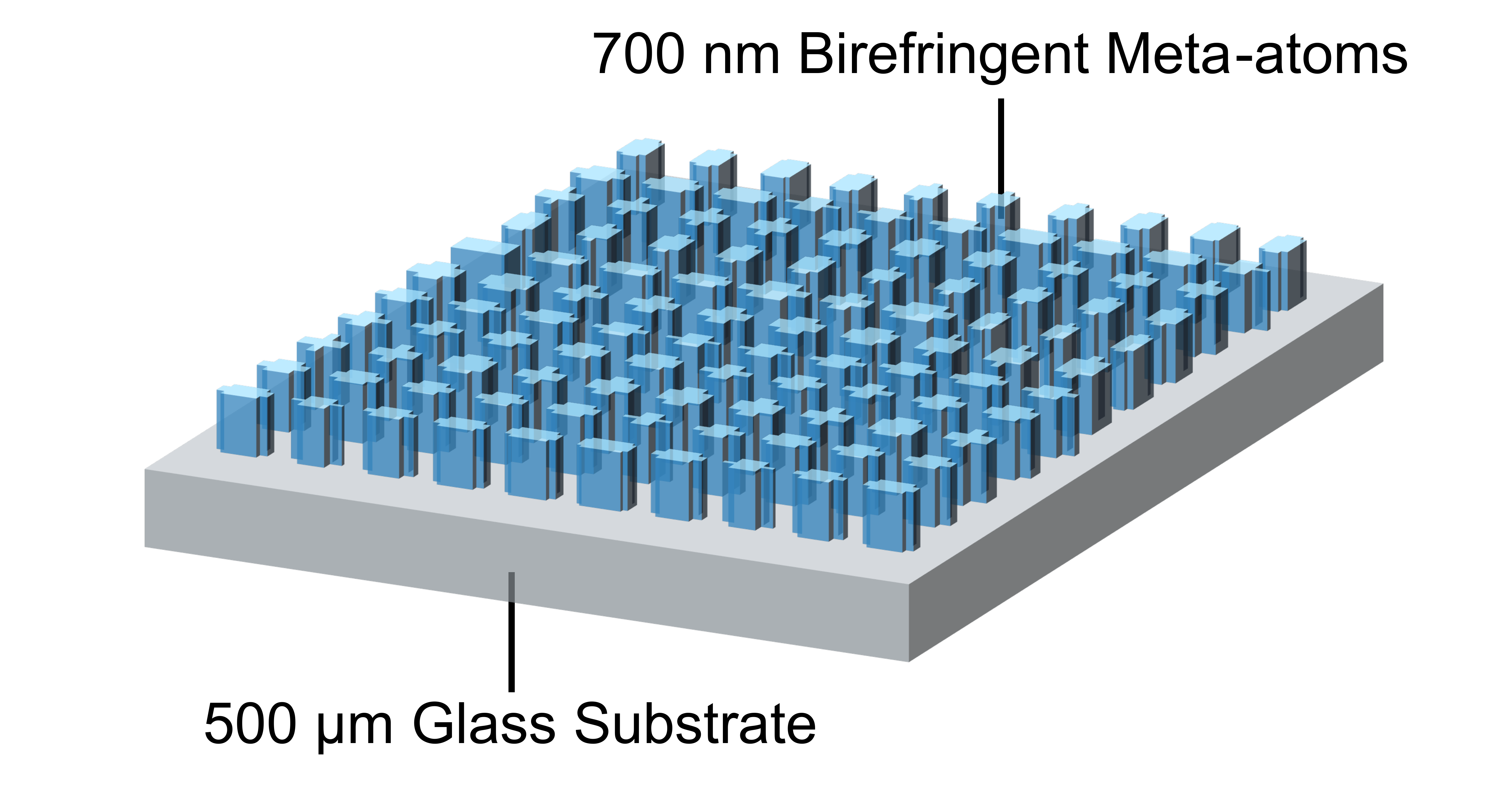}
        \label{fig:lib:bimeta}
    \end{minipage}
}
\subfloat[X/Y phases]{
    \begin{minipage}[c][3.2cm][c]{0.22\textwidth}
        \centering
        \includegraphics[height=3.5 cm]{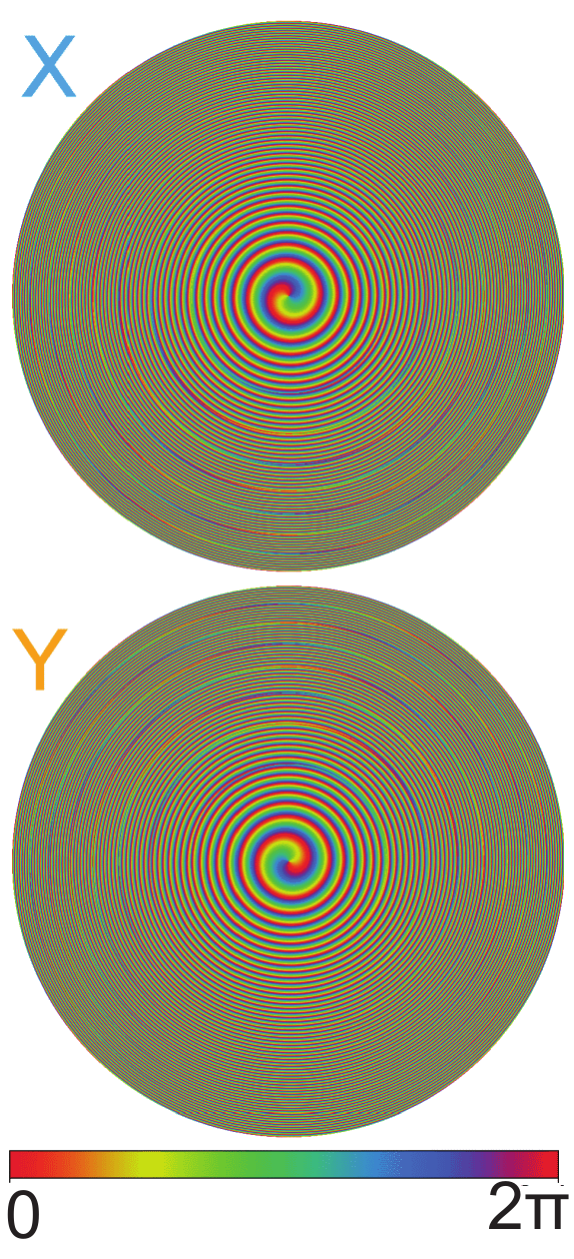}
        \label{fig:lib:phase}
    \end{minipage}
}
\subfloat[rotating PSFs]{
    \begin{minipage}[c][3.2cm][c]{0.3\textwidth}
        \centering
        \includegraphics[height=3.5cm]{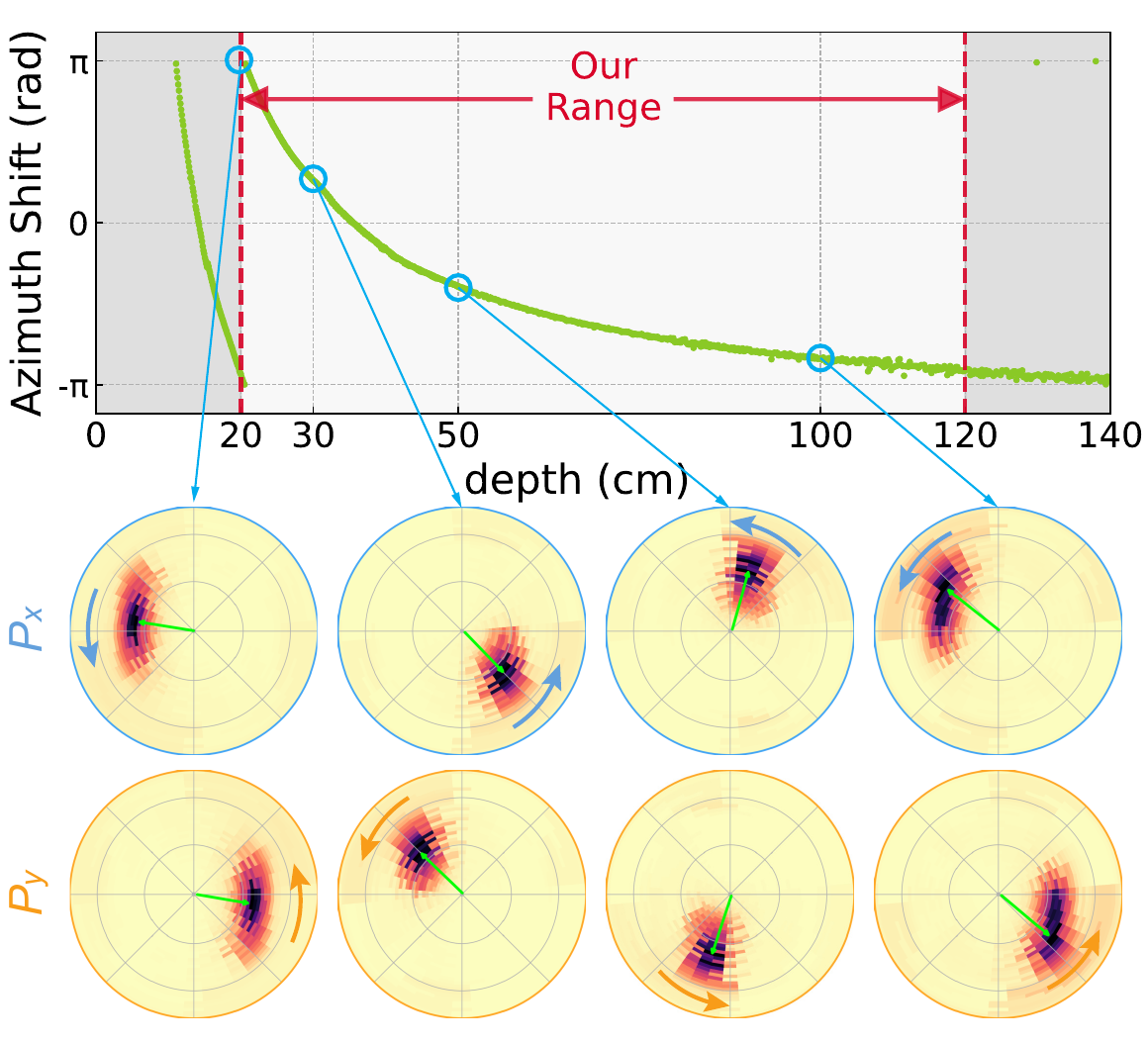}
        \label{fig:lib:psf}
    \end{minipage}
}
\Caption{Visualization of our metalens and PSFs at different depths.}{(a) Schematic of the metalens. (b) Phase profiles for X- and Y-polarized light. (c) Monotonic relation between depth and PSF rotation angle in our designed depth range.
}
\label{fig:metasurface-psf}
\vspace{-10pt}
\end{figure}

\paragraph{Birefringent Metalens.}
We employ a birefringent metalens to \textit{independently} modulate the phase $\metaPhaseboth$ ($k\in\{x,y\}$) for $x$- and $y$-polarized light (\cref{fig:lib:bimeta}). 
For each polarization $k$, we decompose its phase profile as $\metaPhaseboth=\phaseLensboth+\Fresnelphaseboth$. The $\phaseLensboth$ term provides the focusing power. The $\Fresnelphaseboth$ component is engineered to create a \textit{depth-dependent point spread function} (PSF, the blur on the sensor formed by a point light source), $\psfboth(z)$. This PSF's shape varies with source depth $\depth$, an effect arising from the interplay between our engineered phase $\Fresnelphaseboth$ and the natural defocus that occurs as $z$ deviates from the in-focus plane.
\jiahao{
$\defocus(z)$, which occurs as depth $z$ deviates from the in-focus plane. The resulting PSF is computed via the Fourier transform $\Fourier$:
\begin{equation}
\psfboth(z)=|\Fourier\{\mathrm{exp}[i(\Fresnelphaseboth-\defocus(z))]\}|^2.
\label{eq:method:psf}
\end{equation}
\qisun{remember adding punctuations by treating equations as nouns. }
}

\paragraph{Depth Encoding with Rotating PSFs.}
Following \cite{Prasad2013psf,shen2023monocular}, we design the phase $\Fresnelphaseboth$ to encode depth $\depth$ as a PSF rotation. In the imaging plane's polar coordinates $(\imageR,\imageAngle)$, the engineered PSF for both polarizations, $\psfboth$, rotates by the same depth-dependent angle $\Delta\imageAngle(\depth)$: 
\begin{equation}
\psfboth(\imageR,\imageAngle;\depth) \approx \psfboth(\imageR,\imageAngle-\Delta\imageAngle(\depth);\infocus),
\end{equation}
where $\infocus$ is the in-focus depth. We set the two polarized patterns $180^\circ$ apart, so their relative disparity vector's angle directly tracks their co-rotation $\Delta\imageAngle(\depth)$, enabling robust depth estimation \cite{shen2023monocular}.

To realize the PSF rotation, we partition the metalens at the pupil (radius $\radius$) into $\RingNumber=8$ concentric rings, each with a topological charge of $\RingIndex$ ($\RingIndex=1,\dots,\RingNumber$) \cite{Prasad2013psf}. In the pupil polar coordinates $(\pupilR,\pupilAngle)$, the x-polarized phase profile is\nothing{ (see \cref{fig:lib:phase})}:
\begin{align}
\Fresnelphasex\left(\pupilR,\pupilAngle\right) 
= \left\{\RingIndex\,\pupilAngle \mid \sqrt{\frac{\RingIndex-1}{\RingNumber}} \le \frac{\pupilR}{\radius} < \sqrt{\frac{\RingIndex}{\RingNumber}}\right\}.
\label{eq:method:rotationphase}
\end{align}
The y-polarized phase profile $\Fresnelphasey$ is this pattern rotated by $180^\circ$: $\Fresnelphasey(\pupilR,\pupilAngle)=\Fresnelphasex(\pupilR,\pupilAngle-\pi)$. This design yields a PSF rotation angle $\Delta\imageAngle(z)$ given by: 
\begin{align}
\Delta\imageAngle (\depth)=\frac{\pi\radius^2}{\RingNumber\lambda}(\frac{1}{\depth}-\frac{1}{\infocus}),
\end{align}
where $\lambda$ is the wavelength of light. The rotating PSF is illustrated in \cref{fig:lib:psf}. Further details are provided in the supplementary material.

\paragraph{Polarization-Multiplexing Depth Encoding.}
The 2D image $\image_\polarDirection$ is formed by integrating the depth-wise convolutions between scene slices $\sceneIrradiance(\depth)$ and their corresponding depth-dependent PSFs $\psfboth(\depth)$. 
The rotating PSF induces slight, depth-dependent shifts in 2D images (\cref{fig:sim:compare}). Since $\psfX$ and $\psfY$ are 180° apart, these shifts occur in opposite directions for polarized image pairs, creating a monotonic disparity vector that serves as a geometric depth cue. To capture both polarized images  simultaneously, we engineer the focusing phase $\phaseLensboth$ to introduce opposite vertical deflections, spatially separating them onto the sensor halves: 
\begin{align}
\phaseLensboth = -\frac{2\pi}{\wavelength}
\begin{cases}
      \sqrt{\pupilx^2+(\pupily-\sepy)^2+\focus^2}, & \polarDirection=x \\
 \sqrt{\pupilx^2+(\pupily+\sepy)^2+\focus^2}, & \polarDirection=y, 
\end{cases}
\label{eq:method: focusing}
\end{align}
where $(\pupilx,\pupily)$ are the coordinates on the metalens.

\subsection{Physically Grounded Monocular Depth}
\paragraph{Monocular Backbone.} 
Recent depth foundation models \cite{depth_anything_v1,depth_anything_v2} largely follow the architecture of Dense Prediction Transformer(DPT) \cite{ranftl2021vision}. Given an input RGB image $\image \in \mathbb{R}^{C \times H \times W}$, a Vision Transformer (ViT) \cite{dosovitskiy2020image} encoder processes it into a hierarchy of token features ${T_i}$, where each stage $S_i$ produces tokens $T_i \in \mathbb{R}^{C_i \times (\frac{H}{p} \times \frac{W}{p} + 1)}$ with feature dimension $C_i$ and patch stride $p$. The DPT decoder then reconstructs spatial feature maps $F_i \in \mathbb{R}^{C_i \times \frac{H}{p} \times \frac{W}{p}}$ from tokens and progressively fuses multi-level representations through a series of convolutional layers, culminating in a dense depth prediction $D \in \mathbb{R}^{H \times W}$. While diffusion-based monocular depth approaches \cite{ke2023repurposing, he2024lotus} have also emerged, their computational demands make them less suitable for real-time deployment. As such, we only adopt DPT-based architectures as our base model in this work.

\paragraph{Adapting Polarization Measurements for Monocular Models.}
\label{sec:method:prompting}
Our camera produces two polarization observations, $(\polarImageX, \polarImageY)$, whereas monocular depth models are pretrained to take a three-channel RGB image as input. Therefore, a central question is how to make these pretrained models compatible with our wavefront-encoded measurements, while preserving their learned visual priors.

\begin{wrapfigure}{r}{0.45\textwidth}
\vspace{-22pt}
\centering
\includegraphics[width=0.99\linewidth]{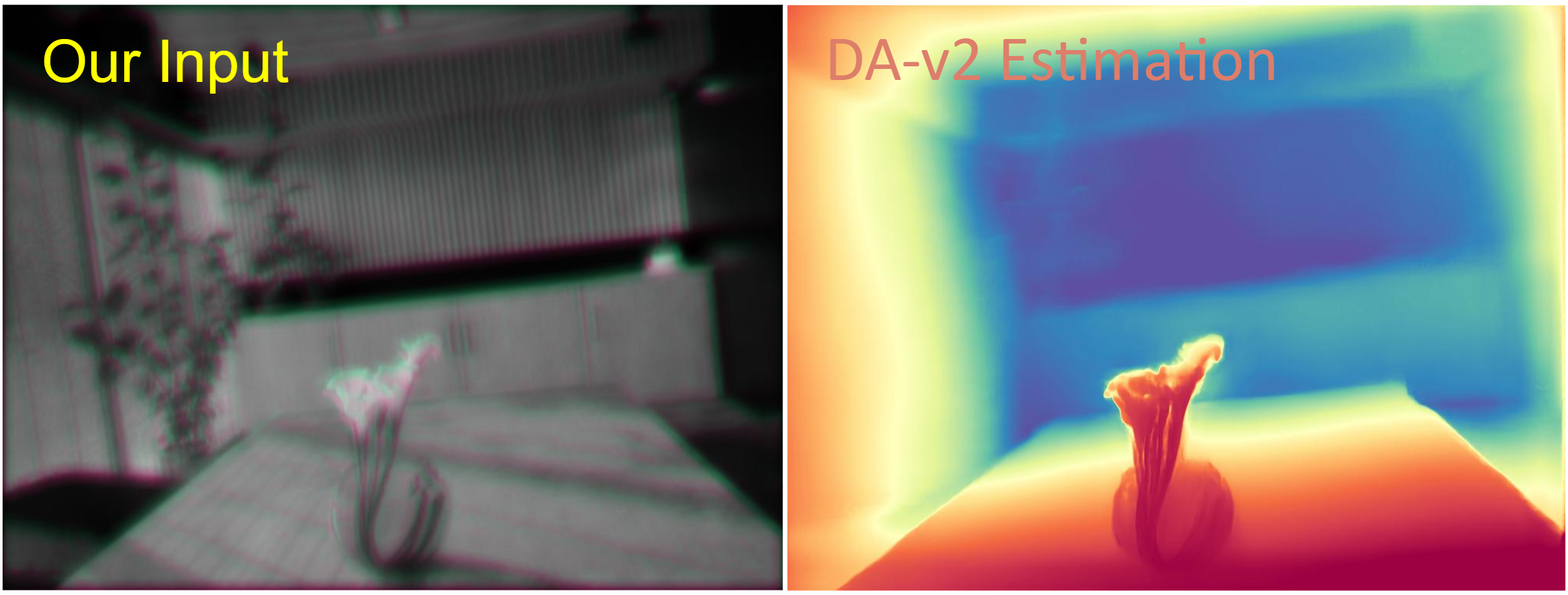}
\vspace{-15pt}
\Caption{Qualitative validation of input compatibility.} {DepthAnything V2 generates high-quality depth maps from our adapted input without fine-tuning. This demonstrates that our encoding preserves essential scene structure and remains aligned with the model's pretrained natural image priors.}
\label{fig:justify_input}
\vspace{-26pt}
\end{wrapfigure}

To this end, we propose a simple input adaptation strategy that converts the two polarization channels into a three-channel input:
\begin{align*}
    \left(\polarImageX, \polarImageY\right) 
    \Rightarrow 
    \left(\polarImageX, \polarImageY, \left(\polarImageX+\polarImageY\right)/2\right).
\end{align*}
This input matches the expected input format of monocular depth models without modifying any network layers or adding auxiliary branches. More importantly, it preserves scene structure in a form that remains compatible with the model's learned priors while injecting physically encoded metric depth cues. As illustrated in \cref{fig:justify_input}, a pretrained Depth Anything V2 model can still estimate high-quality depth from this input, suggesting that it retains sufficient natural image details for effective zero-shot transfer.

We also explored alternative designs. In particular, inspired from PromptDA\cite{lin2025prompting}, we attached a decoder-side fusion module following their design, while adapting the fusion input from one channel to two channels to accommodate $(\polarImageX, \polarImageY)$. However, as shown in our ablation study (\cref{tab:ablation_design_choice}), this additional fusion module brings no meaningful advantage. In practice, the input adaptation is already sufficient to align pretrained monocular models with our polarization measurements while avoiding any extra parameters or architectural overhead.



\begin{figure}
\vspace{-5pt}
\centering
\includegraphics[width=0.85\linewidth]{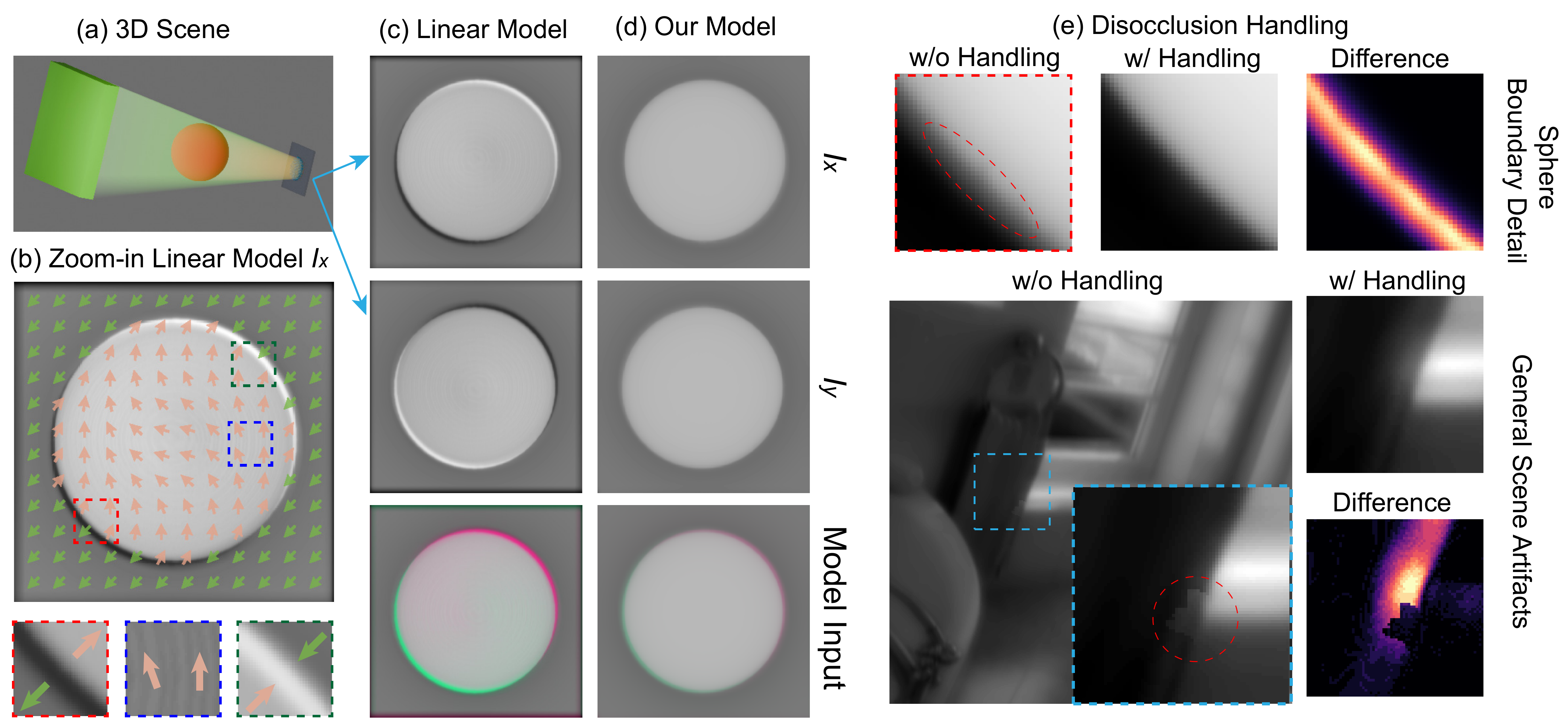}
\Caption{Reducing the simulation-to-real gap with a disocclusion-aware optical forward model.}{(a) A 3D scene comprising a foreground sphere and a background. (b) Zoom-in of the linear model's $\polarImageX$ channel, with arrows indicating PSF position shifts. Insets highlight inherent artifacts: bright occlusion edges, dark disocclusion gaps, and aliasing fringes on steep surfaces. (c) Polarization pairs and model input generated by the standard linear convolution model. (d) Corresponding outputs from our disocclusion-aware model, which significantly mitigates these artifacts. (e) Qualitative ablation of the disocclusion handling module. Omitting this step introduces erroneous sphere boundary details (top) and severe artifacts in general scenes (bottom).
}
\label{fig:simulator-artifact}
\label{fig:sim:compare}
\vspace{-10pt}
\end{figure}
\vspace{-2pt}
\paragraph{Simulator for Training Data.}
\label{sec:method:data}
Training our model requires large-scale paired polarization and depth data. Since collecting this in the real world with pixel-wise accuracy is infeasible, we synthesize our training dataset by converting RGB-D images into depth-encoded polarization pairs using an optical forward model.
We model the 3D scene capture as a discrete sum of layer-wise 2D convolutions between the per-plane scene irradiance and its corresponding PSF $\psfboth(z)$:
\begin{align}
\image_\polarDirection
=
\sum_{n=1}^{N}
\left(\sceneIrradiance \odot \simulatorDepthMask_n\right) * \psfboth(z_n),
\label{eq:depth_binned_conv}
\end{align}
where $\image_\polarDirection$ is the image intensity for polarization channel $k \in \{x, y\}$, $\sceneIrradiance$  is the scene irradiance, and $M_n$ is the binary mask isolating the $n$-th depth bin $z_n$.  The PSFs $\mathcal{P}_k(z)$ are simulated using a Fast Fourier Transform (FFT) implementation of the Kirchhoff diffraction integral \cite{born2013principles}. 

Although our simulated PSFs align closely with the measured ones (see supplementary), a pronounced simulation-to-real gap remains (\cref{fig:simulator-artifact}). To address this and improve real-world generalization, we introduce a disocclusion-aware model with alpha-compositing and polarization-aware augmentation. These improvements are comprehensively validated through our ablation studies ( \cref{tab:ablation}).

\subsection{Bridging Sim-to-Real Gap}
\label{sec:method:sim2real}
\paragraph{Disocclusion-Aware Optical Forward Model}
Pronounced simulation-to-real discrepancies primarily arise in regions with rapid depth changes (\cref{fig:sim:compare}b,c). The standard linear model (\cref{eq:depth_binned_conv}) \cite{Phasecam3d,chang2019deep,ghanekar2022ps} fails here because it ignores occlusion geometry. While prior alpha-compositing approaches \cite{ikoma2021depth} mitigate this, they assume symmetric blur and thus only handle occlusions. For general asymmetric optics (e.g., rotating PSFs), boundaries present a dual challenge: overlapping PSF shifts create bright occlusion edges, whereas diverging shifts leave dark \textit{disocclusion} gaps. Additionally, steep depth gradients severely undersample the rapid PSF rotation, producing aliasing-like fringes.

To accurately render these complex dynamics, our pipeline (\cref{fig:pipeline}) introduces a dedicated disocclusion handling module. We first generate layer-wise opacity and brightness maps by convolving Gaussian-smoothed slice masks and scene irradiance with the PSF. While our baseline employs hybrid compositing (alpha blending for occlusions and direct summation for continuous regions), it inherently fails at disocclusion gaps. We tackle this by explicitly detecting depth discontinuities, applying edge extrapolation and background completion to reconstruct the missing background. Finally, an alpha correction step normalizes the output to ensure full opacity and suppress undersampling fringes. Our updated simulator significantly reduces simulation-to-real discrepancies (\cref{fig:sim:compare}d); conversely, omitting this module produces inaccurate boundaries and severe artifacts (\cref{fig:sim:compare}e), the consequence of which is illustrated in our quantitative ablation (\cref{tab:ablation}).
\vspace{-1.em}
\paragraph{Polarization-Aware Augmentation.}
Beyond simulator issues, several factors contribute to the sim-to-real gap, including (1) polarization imbalance from illumination and surface properties, (2) sensor and environmental noise, and (3) fabrication imperfections.
To improve robustness, we introduce polarization-aware data augmentations: (i) global scaling for illumination changes, (ii) local brightness perturbations via a Gaussian mask for spatial polarization imbalance, (iii) Poisson and Gaussian noise for sensor and environmental effects, and (iv) Gaussian blur for fabrication-induced aberrations. As shown in \cref{fig:pipeline}, these augmentations regularize the model and improve tolerance to physical imperfections.

\vspace{-1.em}
\paragraph{Few-Shot Real Adaptation.}
With the refined model and augmentation, most physics-induced gaps are mitigated. We address the remaining domain shift between simulated and real scenes by mixing a few real shots into the training set. Because dense depth is difficult to obtain, we manually segment objects and assign approximate planar depths (\cref{fig:eval:label}). 

\subsection{Implementation Details}

\paragraph{Metalens Fabrication.} 
We design and fabricate a 3-mm-diameter metalens operating at $\wavelength$=590 nm. The metalens consists of 700-nm-tall cross-shaped birefringent $\text{TiO}_2$ nanopillars patterned on a 500-$\mu$m-thick glass substrate; the nanopillars are arranged in a square lattice with a subwavelength pitch of 400 nm (\cref{fig:lib:bimeta}). The fabrication (detailed in supplementary material) involves three steps: (1) Electron-beam lithography patterning of a resist template, (2) atomic layer deposition of $\text{TiO}_2$ into the template, and (3) dry etching and plasma ashing to remove the resist and excess $\text{TiO}_2$, leaving the free-standing $\text{TiO}_2$ nanopillars.

\vspace{-10pt}
\paragraph{Imaging Setup.}
We build a compact monocular depth imager (\cref{fig:eval:hw_setting}), which consists of the metasurface mounted at a distance of 37.6 mm in front of a 20-MP, 1-inch monochrome CMOS sensor equipped with a 590-nm bandpass filter. The imager's in-focus depth is set to 35 cm, and the depth-sensing range is from 20 cm to 120 cm (\cref{fig:metasurface-psf}). As a research prototype, the chosen hardware parameters aim to prove feasibility rather than maximize performance, which remains an important future optimization direction.

\vspace{-10pt}
\paragraph{Training.}
We use DepthAnything v2 \cite{depth_anything_v2} as backbone and evaluate all three variants---ViT-Small, ViT-Base, and ViT-Large (denoted as Small, Base and Large). Starting from the metric-pretrained weights, we fine-tune the model on Hypersim \cite{hypersim} dataset. The depth range is linearly mapped to 0.2--1.2\,m, followed by our data-preparation pipeline in \cref{fig:pipeline}. We use an $L_1$ and gradient loss $L_{\mathrm{grad}}$ \cite{bochkovskii2024depth} as $L = L_1 + 0.5\,L_{\mathrm{grad}}$. We additionally mix in 5 manually annotated real samples with probability 0.05. The model is trained for 80k steps with a learning rate of $4\times10^{-6}$ and batch sizes of 2 (Large) or 8 (Small/Base). Additional details \nothing{and real training data }are provided in the supplementary material.
\label{sec:method:details}

\section{Experiments \complete}
\label{sec:evaluation}

We evaluate our method through both simulation and real captures. We first compare against monocular metric depth estimation (MMDE) baselines in both simulated and physical experiments. We then compare with representative depth-from-defocus (DfD) baselines in simulation. Finally, we present ablations to isolate the contributions of our design choices.

\subsection{Comparison with MMDE Baselines}

\paragraph{Baselines and metrics.}
We compare against recent metric depth estimators, including Depth Anything v2/v3\cite{depth_anything_v2,depthanything3} (DepthAny.\ v2/v3),  DepthPro \cite{DepthPro}, Lotus \cite{he2024lotus}, Marigold \cite{ke2023repurposing}, Metric3D v2 \cite{hu2024metric3d}, MoGe v2 \cite{wang2025moge2accuratemonoculargeometry}, UniDepth v2 \cite{piccinelli2025unidepthv2}, and ZoeDepth \cite{ZoeDepth}. For each method, we use its largest available model variant. We report standard depth metrics, including $\text{MAE}$, $\text{RMSE}$, $\text{AbsRel}$, and $\delta_{0.5}$. For our method, we evaluate with all three DepthAny. v2 backbones. Since different models may operate over different metric depth ranges, we adopt a linear alignment $\{s,t\}$ to map predictions to the ground truth, following prior work \cite{splitaperturecameras, lin2025prompting}. For a challenging and fair comparison, baseline results are optimized per image by the normalization that best aligns its predictions $\hat{D}$ with the ground-truth $D$:
$
(s^{*}, t^{*}) = \arg\min_{s, t} \|\, s \hat{D} + t - D \,\|_2^2.
$
This alignment removes global scale/shift ambiguity and can make baseline performance appear \textit{higher} than their raw metric-depth accuracy.
We further compare our \textit{single-sensor} method with PromptDA \cite{lin2025prompting}, a recent \textit{dual-sensor} RGB+LiDAR method. To emulate LiDAR on synthetic data, we downsample ground-truth depth by $10\times$ to match the typical image-to-LiDAR ratio, adding uniform 1–2 cm noise to approximate iPhone LiDAR accuracy \cite{abdel2024indoor}. Finally, we fine-tune DepthAny. v2 without depth encoding on our dataset to isolate the benefit of the physical depth cues.

\paragraph{Simulation.}

\begin{figure*}[t]
\centering
\includegraphics[width=1.0\linewidth]{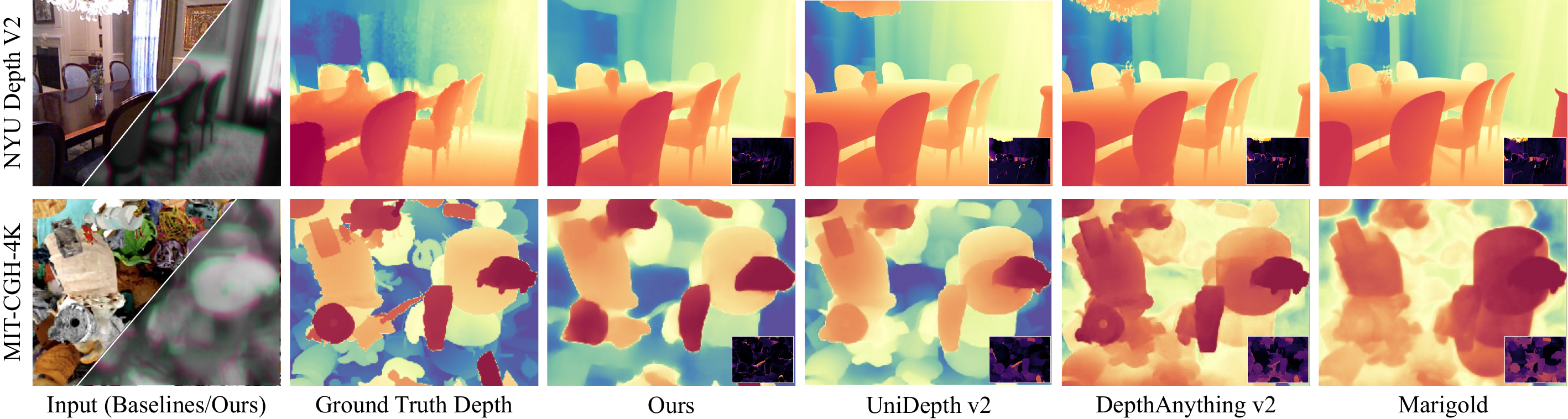}
\Caption{Qualitative comparison with monocular depth estimation baselines.}{Bottom-right insets show the error map where dark colors indicate low error.}
\label{fig:qualitative-sim}
\vspace{-12pt}
\end{figure*}

We first evaluate our approach in simulation, with quantitative and qualitative results presented in \cref{tab:sim-1} and \cref{fig:qualitative-sim}, respectively. We experiment with two datasets: NYU Depth V2 \cite{nyu_depth_v2}, a standard indoor benchmark featuring dense LiDAR ground truth; MIT-CGH-4k \cite{shi2021towards,shi2022end}, a synthetic dataset containing randomly placed 3D objects, serving as a zero-shot benchmark to assess generalization and our utilization of physical depth cues. For both benchmarks, our refined simulator is used to generate the necessary polarization images. 

\begin{table*}[b!]
\vspace{-15pt}
\centering
\Caption{Quantitative comparisons on simulated experiment.}{
\colorbox{\colorTableGreen}{Train}: fine-tuned on our dataset; \colorbox{\colorTableYellow}{Post.}: post-aligned with GT using least-square fitting; \colorbox{\colorTableGray}{w/ LiDAR}: with additional simulated LiDAR input. Method with * is finetuned on our dataset. We highlight the top three results among LiDAR-free methods. Note that post-alignment removes global scale/shift ambiguity which can substantially improve the results..
}
\vspace{-8pt}
\scriptsize

{
\setlength{\tabcolsep}{0pt}          
\renewcommand{\arraystretch}{1.15}      

\begin{tabular}{c|l|cccc|c|cccc}
\toprule
\multirow{2}{*}{\tinyspace \makecell{Zero\\Shot}\tinyspace} & 
\multirow{2}{*}{\makecell{\tinyspace \tiny \colorbox{\colorTableGreen}{Train}/ \colorbox{\colorTableYellow}{Post.}/ \\ \tiny \colorbox{\colorTableGray}{w/ LiDAR}}}&
\multicolumn{4}{c|}{\textbf{NYU Depth v2}} & \multirow{2}{*}{\tinyspace \makecell{Zero\\Shot}\tinyspace } & 
\multicolumn{4}{c}{\textbf{MIT-CGH-4k}} \\

& & {\tinyspace MAE$\downarrow$\tinyspace } & {\tinyspace RMSE$\downarrow$\tinyspace } & {\tinyspace AbsRel$\downarrow$\tinyspace } & {\tinyspace $\delta_{0.5}\!\uparrow$\tinyspace } & 
  & {\tinyspace MAE$\downarrow$\tinyspace } & {\tinyspace RMSE$\downarrow$\tinyspace } & {\tinyspace AbsRel$\downarrow$\tinyspace } & {\tinyspace $\delta_{0.5}\!\uparrow$\tinyspace } \\
\hline
\multirow{13}{*}{Yes}  
 & \tableGreen{\tinyspace \textbf{Ours-Large}} & \rankSecond{ {0.023} } & \rankSecond{ {0.040} } & \rankSecond{ {0.039} } & \rankSecond{ {0.951} } & \multirow{14}{*}{Yes}  & \rankFirst\textbf{0.067} & \rankFirst\textbf{0.126} & \rankSecond{0.105} & \rankSecond{0.764} \\
 & \tableGreen{\tinyspace \textbf{Ours-Base}} & \rankFirst{\textbf{0.022}} & \rankFirst{\textbf{0.039}} & \rankFirst{\textbf{0.036}} & \rankFirst{\textbf{0.957}} &  & \rankSecond{0.068} & \rankSecond{0.129} & \rankFirst\textbf{0.102} & \rankFirst\textbf{0.772} \\
 & \tableGreen{\tinyspace \textbf{Ours-Small}} & \rankThird{0.025} & \rankThird{0.043} & \rankThird{0.043} & \rankThird{0.936} &  & \rankThird{0.076} & \rankThird{0.137} & \rankThird{0.125} & \rankThird{0.724} \\ 
 & \tableGreen{\tinyspace DepthAny. v2$^\ast$} & 0.128 & 0.148 & 0.267 & 0.341 &   & 0.301 & 0.371 & 0.410 & 0.100 \\ 
 & \triangcellWH{5em}{1.15em}{\hspace{0.12em}DepthAny. v2$^\ast$\hspace{1.5em} } & 0.054 & 0.079 & 0.098 & 0.731 &   & 0.180 & 0.220 & 0.371 & 0.241 \\
 & \tableYellow{\tinyspace DepthAny. v2 \cite{depth_anything_v2}\tinyspace } & 0.043 & 0.067 & 0.079 & 0.805 &  & 0.151 & 0.190 & 0.308 & 0.300 \\ 
 & \tableYellow{\tinyspace DepthAny. v3 \cite{depthanything3}\tinyspace } & {0.037} & {0.062} & {0.069} & {0.846} & & {0.134} & {0.172} & {0.276} & {0.349}\\ 
 & \tableYellow{\tinyspace Depth Pro} \cite{DepthPro} & 0.038 & 0.061 & 0.071 & 0.841 &  & 0.144 & 0.181 & 0.292 & 0.309 \\ 
 & \tableYellow{\tinyspace Lotus} \cite{he2024lotus} & 0.069 & 0.093 & 0.127 & 0.575 &  & 0.162 & 0.201 & 0.330 & 0.268 \\ 
 & \tableYellow{\tinyspace Marigold} \cite{ke2023repurposing} & 0.045 & 0.070 & 0.085 & 0.785 & & 0.174 & 0.213 & 0.355 & 0.243 \\ 
 & \tableYellow{\tinyspace Metric3D v2} \cite{hu2024metric3d} & 0.056 & 0.082 & 0.110 & 0.766 & & 0.212 & 0.252 & 0.442 & 0.183 \\ 
 & \tableYellow{\tinyspace MoGe v2} \cite{wang2025moge2accuratemonoculargeometry} & {0.034} & {0.058} & {0.063} & {0.867} & & {0.133} & {0.169} & {0.272} & {0.346} \\ 
 & \tableYellow{\tinyspace UniDepth} v2 \cite{piccinelli2025unidepthv2} & 0.034 & 0.059 & 0.063 & 0.865 & & 0.127 & 0.164 & 0.259 & 0.360 \\
\cline{0-1}
\multirow{1}{*}{No} & \tableYellow{\tinyspace ZoeDepth} \cite{ZoeDepth} & 0.041 & 0.062 & 0.076 & 0.805 & & 0.205 & 0.247 & 0.428 & 0.204 \\ 
\hline
 & \tableGray{\tinyspace PromptDA} \cite{lin2025prompting} & \tableLightGray{0.021} & \tableLightGray{0.042} & \tableLightGray{0.036} & \tableLightGray{0.955} &  & \tableLightGray{0.058} & \tableLightGray{0.113} & \tableLightGray{0.099} & \tableLightGray{0.802} \\
\bottomrule
\end{tabular}
}
\label{tab:sim-1}
\vspace{-6pt}
\end{table*}

As shown in \cref{tab:sim-1}, our method achieves the best performance among all LiDAR-free baselines. On NYU Depth V2, our method demonstrate a clear advantage and is highly competitive with the LiDAR-assisted PromptDA, achieving lower RMSE, AbsRel, and $\delta_{0.5}$ errors alongside a comparable L1 error. On MIT-CGH-4k, where a lack of semantics severely degrades most baselines even after scale/shift alignment, our method retains strong accuracy. This confirms our model's ability to reliably decode metric depth from polarization wavefronts. Fine-tuning DepthAny. v2 does not improve performance on either benchmark, indicating that the gains of our method do not arise from dataset-specific fine-tuning, but from the physically encoded depth information.

\begin{figure*}[t]
\centering
    \subfloat[our physical setup]{
        \includegraphics[height=1.55cm,trim={14cm 2cm 2cm 0},clip]{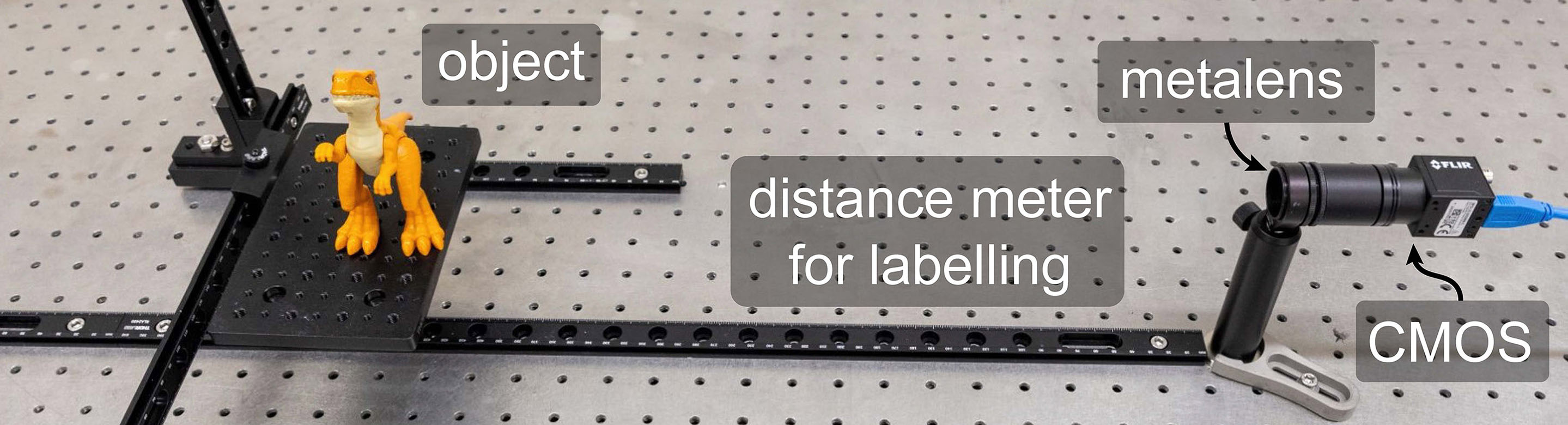}
        \label{fig:eval:hw_setting}
    } 
    \subfloat[model input, our depth prediction, and label]{
        \includegraphics[height=1.55cm,trim={0cm 0.1cm 0cm 0.01cm},clip]{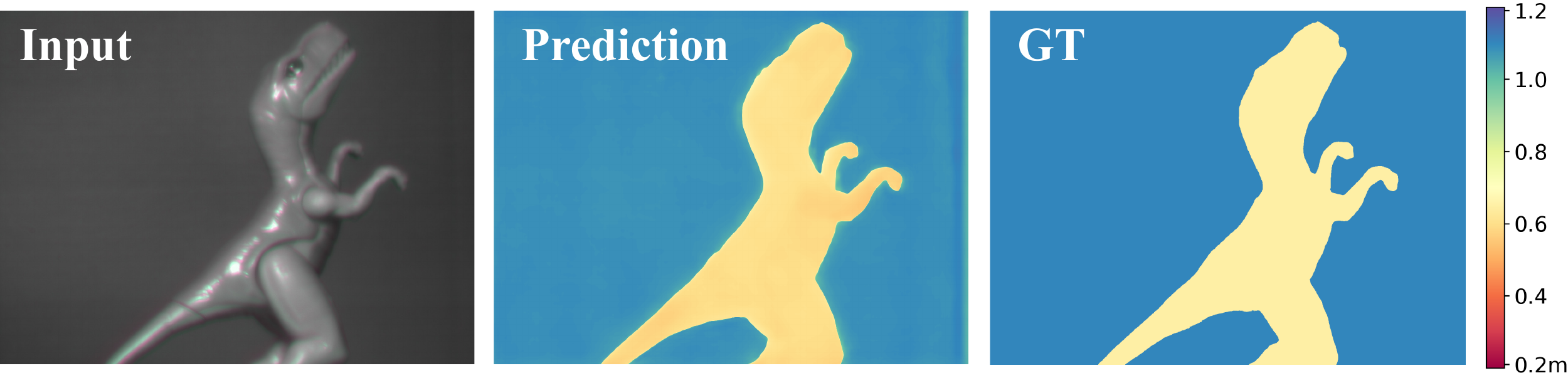}
        \label{fig:eval:label}
    } 
    
    \subfloat[our results on diverse backgrounds and objects; GT depth is shown on the top-left corner]{
        \includegraphics[width=0.98\linewidth]{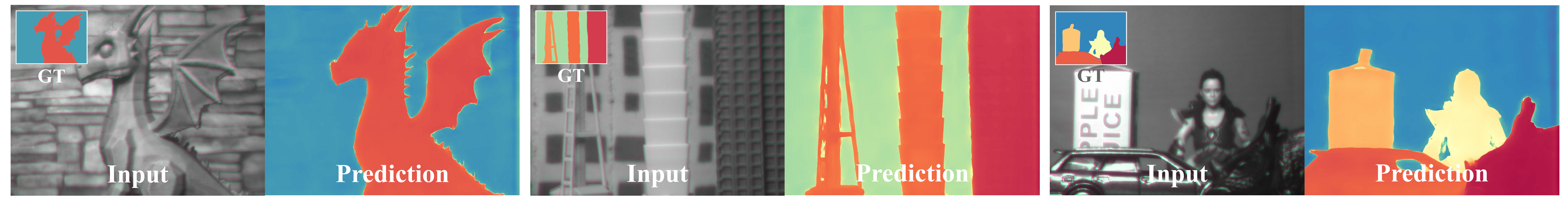}
        \label{fig:eval:hw:results1}
    }
    \vspace{-5pt}
    \Caption{Physical experiment setup and qualitative results.}{We encourage readers to see our supplementary material for additional results.
    }
    \label{fig:qualitative-real}
    \vspace{-10pt}
\end{figure*}

\paragraph{Physical experiments.}
To acquire physical measurements, we mounted our metalens-based depth camera prototype and target objects on an optical table with precise distance control (\cref{fig:qualitative-real}). We captured $42$ scenes featuring $25$ distinct objects, including single- and multi-object setups placed at various depths; $20$ objects are unseen in our five-shot training set. Our model uses both polarization channels as input (\cref{sec:method:prompting}), whereas baselines are provided with a grayscale image from one polarization channel. Obtaining dense LiDAR or stereo ground truth is challenging due to field-of-view mismatch and sparsity \cite{splitaperturecameras}. Following established practices \cite{ikoma2021depth, shen2023monocular}, we assign reference depths to nearly planar objects using masks and known mounting distances, with averaged label uncertainty (< 1 cm) well below our performance margins.
\jiahao{We can cite some references here}

\begin{wraptable}{r}{0.49\textwidth}
\vspace{-40pt}
\centering
\scriptsize
\Caption{Quantitative comparisons on physical experiment.}{
Notations are consistent with the previous table.
}
\setlength{\tabcolsep}{2.0pt} 
\renewcommand{\arraystretch}{1.15}
\begin{tabular}{l|cccc}
\toprule
\multirow{2}{*}{\makecell{\tiny \colorbox{\colorTableGreen}{Train}/ \colorbox{\colorTableYellow}{Post.}/ \\ \tiny \colorbox{\colorTableGray}{w/ LiDAR}}} & \multirow{2}{*}{\makecell{MAE$\downarrow$}} & \multirow{2}{*}{\makecell{RMSE$\downarrow$}} & \multirow{2}{*}{\makecell{AbsRel$\downarrow$}} & \multirow{2}{*}{\makecell{$\delta_{0.5}\uparrow$}}  \\
& & & & \\ 
\hline
 \tableGreen{\textbf{Ours-Large}}     & \rankSecond{0.036} & \rankFirst{\textbf{0.075}} & \rankSecond{0.060} & \rankSecond{{0.888}} \\ 
 \tableGreen{\textbf{Ours-Base}}     & \rankFirst{\textbf{0.032}} & \rankSecond{0.089} & \rankFirst{\textbf{0.055}} & \rankFirst{\textbf{0.895}} \\ 
 \tableGreen{\textbf{Ours-Small}}     & \rankThird{0.048} & \rankThird{0.098} & \rankThird{0.081} & \rankThird{0.818} \\ 
 \tableGreen{DepthAny. v2*}     & 0.061 & 0.100 & 0.128 & 0.678   \\ 
 \tableYellow{DepthAny. v2}    & 0.135 & 0.169 & 0.234 & 0.483 \\ 
 \tableYellow{DepthAny. v3}    & 0.111 & 0.146 & 0.198 & 0.561 \\
 \tableYellow{Depth Pro}       & 0.089 & 0.122 & 0.159 & 0.661 \\ 
 \tableYellow{Lotus}           & 0.140 & 0.181 & 0.261 & 0.479 \\ 
 \tableYellow{Marigold}        & 0.062 & 0.101 & 0.117 & 0.744 \\ 
 \tableYellow{Metric3D v2}     & 0.159 & 0.193 & 0.276 & 0.424 \\ 
 \tableYellow{MoGe v2}         & 0.063 & 0.095 & 0.119 & 0.709 \\ 
 \tableYellow{UniDepth v2}     & 0.107 & 0.145 & 0.196 & 0.592 \\ 
 \tableYellow{ZoeDepth}        & 0.109 & 0.146 & 0.175 & 0.561 \\ 
 \hline
 \tableGray{PromptDA}       & \tableLightGray{0.030} & \tableLightGray{0.086} & \tableLightGray{0.054} & \tableLightGray{0.951} \\
\bottomrule
\end{tabular}
\vspace{-5pt}
\label{tab:physical}
\vspace{-20pt}
\end{wraptable}




As shown in \cref{tab:physical}, our method consistently outperforms all LiDAR-free baselines and achieves performance close to the LiDAR-assisted PromptDA. Even after applying optimal scale/shift alignment to baseline predictions, our approach maintains a clear margin, highlighting its effective use of physical depth cues. Qualitative results in \cref{fig:qualitative-real} further show that our predictions are metrically accurate while preserving sharp object boundaries. Overall, these results demonstrate that our physically prompted model transfers robustly from simulation to real captures.

\vspace{-10 pt}

\subsection{Comparison with DfD}

\begin{wraptable}{r}{0.49\textwidth}
\vspace{-10pt}
\centering
\scriptsize
\Caption{Comparison on FlyingThings3D.}{}
\setlength{\tabcolsep}{2.0pt} 
\renewcommand{\arraystretch}{1.15}
\begin{tabular}{l c c c c}
\toprule
\textbf{Method} & MAE$\downarrow$ & RMSE$\downarrow$ & log$_{10}$$\downarrow$ & $\delta_1\uparrow$ \\
\midrule
\textbf{Ours-Small} & \textbf{0.022} & \textbf{0.132} & \textbf{0.005} & \textbf{0.995} \\
DeepDfD \cite{ikoma2021depth} & 0.089 & 0.191 & 0.034 & 0.941 \\
Split-Aperture \cite{splitaperturecameras} & 0.086 & 0.147 & 0.011 & 0.993 \\
\bottomrule
\end{tabular}
\vspace{-0pt}
\label{tab:compare_with_dfd}
\vspace{-20pt}
\end{wraptable}
We consider two representative DfD baselines: DeepDfD \cite{ikoma2021depth} and Split-Aperture \cite{splitaperturecameras}. For fair evaluation, we follow their original metrics and dataset, FlyingThings3D \cite{MIFDB16}. To match their $1$--$5$\,m range, we retain the phase design in \cref{eq:method:rotationphase} and scale our metalens to a 5 mm diameter with a 50 mm focal length, comparable to baseline optics. We report results using our small backbone to match the baseline model capacity. As shown in \cref{tab:compare_with_dfd}, our method consistently outperforms the DfD baselines across all metrics. To attribute these gains, we next conduct ablations on both the optical design and the backbone model (\cref{sec:ablation}).

\subsection{Ablations and Analysis}
\label{sec:ablation}

\paragraph{Deconstructing the performance gains over DfD baselines.}
To isolate the source of our improvements over DfD baselines, we conduct controlled ablations (\cref{tab:ablation_compare_with_dfd}) across three factors: metalens design (\textbf{Meta}), backbone architecture (\textbf{ViT}), and learned pretraining prior (\textbf{Prior}). For baselines, we use DeepDfD’s optical design \cite{ikoma2021depth} which yields a compatible three-channel observation, and a U-Net of comparable capacity to the ViT. 

\begin{wraptable}{r}{0.55\textwidth}
\vspace{-40pt}
\centering
\Caption{Ablation study on performance gains over DFD baselines.}{}
\scriptsize
\setlength{\tabcolsep}{3pt}
\renewcommand{\arraystretch}{1.15}
\begin{tabular}{c c c c c c c c }
\toprule
\textbf{Meta} & \textbf{ViT} & \textbf{Prior} & MAE$\downarrow$ & RMSE$\downarrow$ & log$_{10}$$\downarrow$ & $\delta_1\uparrow$ \\
\midrule
\checkmark & \checkmark & \checkmark  & \textbf{0.022} & \textbf{0.132} & \textbf{0.005} & \textbf{0.995}\\
\texttimes & \checkmark & \checkmark  & 0.035 & 0.133 & 0.009 & 0.994 \\
\checkmark & \checkmark & \texttimes  & {0.061} & {0.268} & {0.014} & {0.984} \\
\checkmark & \texttimes & \texttimes  & {0.063} & {0.254} & {0.014} & {0.983} \\
\texttimes & \texttimes & \texttimes  & 0.089 & 0.191 & 0.034 & 0.941 \\
\bottomrule
\end{tabular}
\label{tab:ablation_compare_with_dfd}
\vspace{-15pt}
\end{wraptable}



\noindent Our results reveal that the pretrained prior is the primary driver of performance; omitting it causes a steep drop in accuracy (row~1 vs.\ row~4), whereas fine-tuning a pretrained ViT with DeepDfD optics yields substantial gains (row~2 vs.\ row~5). Furthermore, our metalens design provides superior physical grounding for depth estimation, outperforming DeepDfD optics when using the same pretrained backbone (row~1 vs.\ row~2, further discussed in supplementary material). Finally, while the ViT marginally outperforms a similarly sized U-Net, the architectural difference alone is not significant (row~3 vs.\ row~4). 

\paragraph{Preservation of pretrained priors.}
We further analyze why the pretrained depth foundation model remains effective under our pseudo-RGB input adaptation. Although the chromatic channels are remapped to polarization views, the input preserves the spatial statistics most relevant to dense prediction, including edges, object boundaries, and texture gradients. To quantify the resulting feature shift, we feed the pretrained DA-V2 ViT-L the same Hypersim test scenes represented as grayscale, a single polarization view tiled to three channels, and our pseudo-RGB input, and then measure layer-wise CKA similarity to features extracted from the original RGB input. 

\begin{wrapfigure}{r}{0.58\textwidth}
\vspace{-0pt}
  \centering
  \includegraphics[width=0.55\textwidth,
    trim=10pt 10pt 10pt 10pt,
    clip]{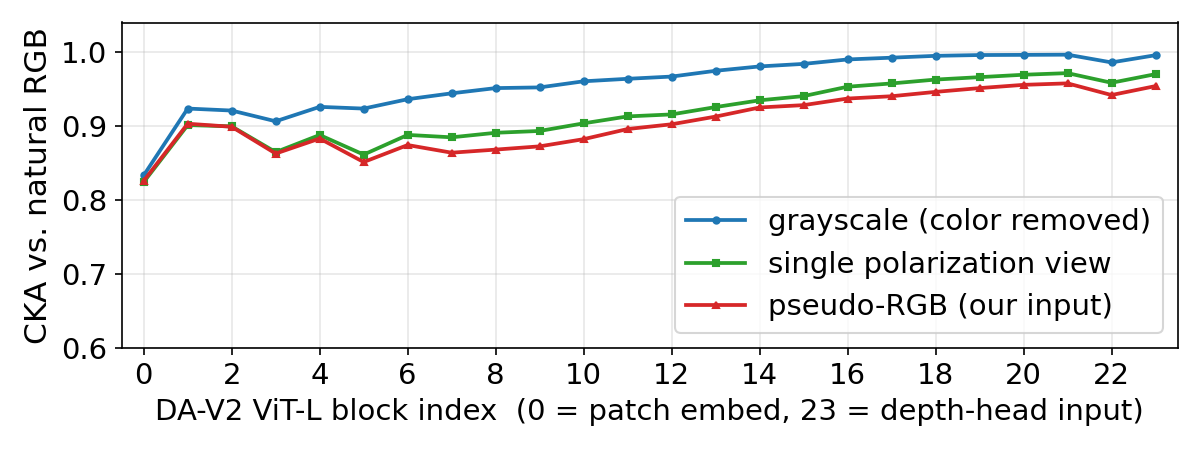}
  \vspace{-10pt}
  \label{fig:cka_layerwise}
\end{wrapfigure}
\noindent As shown in \cref{fig:cka_layerwise}, all variants maintain high similarity to RGB features, with CKA scores above $0.83$ across layers and above $0.95$ in the final three blocks. Moreover, pseudo-RGB remains within approximately $0.04$ CKA of the grayscale baseline. This indicates that the chromatic-to-polarization remapping induces no larger representational shift than removing color alone, supporting the transferability of the pretrained priors.

\paragraph{Alternative designs.}
To validate our input adaptation strategy (\textbf{Adapt.}) in \cref{sec:method:prompting}, we compare it against a decoder-side fusion approach (\textbf{Fusion}) in physical experiments. We adapt the fusion module in PromptDA \cite{lin2025prompting} to our setting by modifying its input layers.

\begin{wraptable}{r}{0.55\textwidth}
\vspace{-30pt}
\centering
\scriptsize
\Caption{Ablation study comparing our input adaptation strategy against decoder-side fusion.}{}
\setlength{\tabcolsep}{3pt}
\renewcommand{\arraystretch}{1.15}
\begin{tabular}{c c c c c c c}
\toprule
\textbf{Adapt.} & \textbf{Fusion} & MAE$\downarrow$ & RMSE$\downarrow$ & AbsRel$\downarrow$ & $\delta_{0.5}\uparrow$ \\
\midrule
\checkmark & \texttimes  & {\textbf{0.032}} & {0.089} & {\textbf{0.055}} & {\textbf{0.895}} \\
\checkmark & \checkmark  & \textbf{0.032} & \textbf{0.087} & 0.058 & 0.875 \\
\texttimes & \checkmark  & {0.063} & {0.113} & {0.109} & {0.727} \\
\bottomrule
\end{tabular}
\vspace{-15pt}

\label{tab:ablation_design_choice}
\end{wraptable}

\noindent As shown in \cref{tab:ablation_design_choice}, our adaptation strategy (row 1) achieves the best overall performance. Adding the fusion module (row 2) yields only a marginal RMSE improvement while degrading other metrics, and relying solely on fusion (row 3) drastically reduces performance. This confirms that our input adaptation strategy is a simpler, more effective way to inject polarization wavefronts.

\paragraph{Sim-to-real transfer.}
To decouple the benefits of physically accurate modeling from the inherent advantages of real-world fine-tuning, we ablate our pipeline (\cref{tab:ablation}) under two training regimes: synthetic-only and with real data included. Removing disocclusion modeling (\cref{tab:ablation}(a)) consistently degrades performance which confirms that real-world data cannot fully compensate for inaccurate optical modeling. Removing all augmentations (\cref{tab:ablation}(c)) causes a drastic performance collapse in the synthetic-only regime, proving their necessity for zero-shot generalization. (\cref{tab:ablation}(d–f)) reveals that modeling light imbalance is the most critical factor. Since our approach relies on polarization, we hypothesize it prevents the network from overfitting to absolute light intensities and forces it to focus on robust positional shifts.


\begin{table}[b]
\centering
\scriptsize
\Caption{Ablation study on sim-to-real transfer.} {We disable simulator/augmentation components from full pipeline and evaluate with and without including real data.}
\setlength{\tabcolsep}{2.5pt} 
\renewcommand{\arraystretch}{1.15}
\begin{tabular}{l|cccc|cccc}
\toprule
\multirow{2}{*}{\makecell{Component}} & \multicolumn{4}{c|}{{Synthetic only}} & \multicolumn{4}{c}{{Real data included}}\\
& {MAE$\downarrow$} & {\makecell{RMSE$\downarrow$}} & {\makecell{AbsRel$\downarrow$}} & {\makecell{$\delta_{0.5}\uparrow$}} & {\makecell{MAE$\downarrow$}} & {\makecell{RMSE$\downarrow$}} & {\makecell{AbsRel$\downarrow$}} & {\makecell{$\delta_{0.5}\uparrow$}} \\
\hline
(a) {Full} & \textbf{0.107} &\textbf{ 0.149} & 0.148 & \textbf{0.546} 
& \textbf{0.032} &\textbf{0.088} & \textbf{0.055} & 0.895 \\
(b) w/o disocclusion            & 0.121 & 0.173 & 0.159 & 0.479
& 0.046 & 0.096 & 0.074 & 0.854\\
(c) w/o augmentation                 & 0.198 & 0.260 & 0.226 & 0.240
& 0.038 & 0.097 & 0.066 & 0.865 \\
(d) w/o light imbalance              & 0.116 & 0.159 & 0.157 & 0.451 
& 0.034 & {0.089} & 0.058 & 0.868 \\
(e) w/o blur                         & 0.112 & 0.151 & 0.153 & 0.496
& 0.034 & 0.091 & 0.059 & 0.876 \\
(f) w/o noise                        & 0.111 & 0.162 & \textbf{0.143} & 0.534
& 0.033 & 0.092 & 0.056 & \textbf{0.907} \\
\bottomrule
\end{tabular}
\vspace{6pt}
\label{tab:ablation}

\vspace{-15pt}
\end{table}

\section{Limitations and Future Work}
\label{sec:limitation}
%
%

\paragraph{Limitations.}
Our current system is intended as a proof of concept for physically grounding DFMs with nanophotonic wavefront cues, rather than a deployment-ready depth camera. The prototype is constrained by a system-level photon budget. Its 3-mm, $f/11.3$ aperture and 10-nm bandpass filter substantially reduce aperture--spectral throughput compared with mature $f/6$ RGB or DOE-based systems. As a result, the current setup is better suited to well-lit, near-range scenes with relatively longer exposures.

Polarization multiplexing does not directly discard the total collected signal, but it maps the two polarization views to separate sensor regions, reducing the effective field of view or sampling density. It can also increase sensitivity to background and read noise in non-shot-noise-limited regimes. In addition, the prototype currently has a limited field of view, increased tube length, and a narrower operating range than mature multi-lens or RGB+LiDAR systems.

Real-world polarization imbalance is another practical limitation. Our low-pass filtered mask and polarization-aware augmentation mitigate low-frequency lighting discrepancies and encourage the network to rely on positional shifts rather than intensity. However, high-frequency imbalance from specular reflections or actively polarized illumination remains challenging.

\paragraph{Future Work.}
Future work will explore end-to-end metalens--DFM co-design to jointly optimize optical encoding and depth inference. Larger-aperture metasurfaces and polarization-resolved sensors could improve photon throughput, compactness, field of view, and depth range, while reducing the need for narrow spectral filtering. We will also investigate training and calibration strategies that improve robustness under complex environmental polarization, especially for specular and partially polarized scenes.

\section{Conclusion}
\label{sec:conclusion}
We present a metalens-based depth imaging system that physically grounds depth foundation models using polarization-encoded nanophotonic wavefronts. Combining passive optical modulator with pretrained depth priors, we mitigate monocular scale ambiguity and enable accurate metric depth, substantially outperforming monocular depth estimation and prior depth-from-defocus methods that did not leverage depth priors. We hope this work will help bridge emerging foundation models and nanophotonics materials, enabling compact depth sensing for VR/AR, miniature robotics, medical endoscopy, and beyond.






\label{sec:discussion}

\bibliographystyle{splncs04}
\bibliography{main}

\begin{thebibliography}{10}
\providecommand{\url}[1]{\texttt{#1}}
\providecommand{\urlprefix}{URL }
\providecommand{\doi}[1]{https://doi.org/#1}

\bibitem{abdel2024indoor}
Abdel-Majeed, H.M., Shaker, I.F., Abdel-Wahab, A., Awad, A.A.D.I.: Indoor
  mapping accuracy comparison between the apple devices’ lidar sensor and
  terrestrial laser scanner. HBRC Journal  \textbf{20}(1),  915--931 (2024)

\bibitem{alexander2016focal}
Alexander, E., Guo, Q., Koppal, S., Gortler, S., Zickler, T.: Focal flow:
  Measuring distance and velocity with defocus and differential motion. In:
  European conference on computer vision. pp. 667--682. Springer (2016)

\bibitem{DiffuserCam}
Antipa, N., Kuo, G., Heckel, R., Mildenhall, B., Bostan, E., Ng, R., Waller,
  L.: Diffusercam: lensless single-exposure 3d imaging. Optica  \textbf{5}(1),
  ~1--9 (Jan 2018). \doi{10.1364/OPTICA.5.000001},
  \url{https://opg.optica.org/optica/abstract.cfm?URI=optica-5-1-1}

\bibitem{bire2}
Baek, S.H., Gutierrez, D., Kim, M.H.: Birefractive stereo imaging for
  single-shot depth acquisition. ACM Transactions on Graphics (TOG)
  \textbf{35}(6),  1--11 (2016)

\bibitem{balthasarmueller2017metasurface}
Balthasar~Mueller, J.P., Rubin, N.A., Devlin, R.C., Groever, B., Capasso, F.:
  {Metasurface polarization optics: independent phase control of arbitrary
  orthogonal states of polarization}. Phys.~Rev.~Lett.  \textbf{118}(11),
  113901 (Mar 2017), \url{http://dx.doi.org/10.1103/PhysRevLett.118.113901}

\bibitem{Mueller2017metapol}
Balthasar~Mueller, J.P., Rubin, N.A., Devlin, R.C., Groever, B., Capasso, F.:
  Metasurface polarization optics: Independent phase control of arbitrary
  orthogonal states of polarization. Physical Review Letters  \textbf{118}(11),
   113901 (2017). \doi{10.1103/PhysRevLett.118.113901},
  \url{https://link.aps.org/doi/10.1103/PhysRevLett.118.113901}, pRL

\bibitem{Berlich2016single}
Berlich, R., Bräuer, A., Stallinga, S.: Single shot three-dimensional imaging
  using an engineered point spread function. Optics Express  \textbf{24}(6),
  5946--5960 (2016). \doi{10.1364/OE.24.005946},
  \url{https://opg.optica.org/oe/abstract.cfm?URI=oe-24-6-5946}

\bibitem{ZoeDepth}
Bhat, S.F., Birkl, R., Wofk, D., Wonka, P., Müller, M.: Zoedepth: Zero-shot
  transfer by combining relative and metric depth (2023).
  \doi{10.48550/ARXIV.2302.12288}, \url{https://arxiv.org/abs/2302.12288}

\bibitem{bochkovskii2024depth}
Bochkovskii, A., Delaunoy, A., Germain, H., Santos, M., Zhou, Y., Richter,
  S.R., Koltun, V.: Depth pro: Sharp monocular metric depth in less than a
  second. arXiv preprint arXiv:2410.02073  (2024)

\bibitem{DepthPro}
Bochkovskii, A., Delaunoy, A., Germain, H., Santos, M., Zhou, Y., Richter,
  S.R., Koltun, V.: Depth pro: Sharp monocular metric depth in less than a
  second. arXiv preprint arXiv:2410.02073  (2024)

\bibitem{born2013principles}
Born, M., Wolf, E.: Principles of optics: electromagnetic theory of
  propagation, interference and diffraction of light. Elsevier (2013)

\bibitem{Braat2008psf}
Braat, J.J.M., van Haver, S., Janssen, A.J.E.M., Dirksen, P.: Chapter 6
  Assessment of optical systems by means of point-spread functions, vol.~51,
  pp. 349--468. Elsevier (2008).
  \doi{https://doi.org/10.1016/S0079-6638(07)51006-1},
  \url{https://www.sciencedirect.com/science/article/pii/S0079663807510061}

\bibitem{mpi_sintel}
Butler, D.J., Wulff, J., Stanley, G.B., Black, M.J.: A naturalistic open source
  movie for optical flow evaluation. In: {A. Fitzgibbon et al. (Eds.)} (ed.)
  European Conf. on Computer Vision (ECCV). pp. 611--625. Part IV, LNCS 7577,
  Springer-Verlag (Oct 2012)

\bibitem{cao2024aberration}
Cao, Z., Li, N., Zhu, L., Wu, J., Dai, Q., Qiao, H.: Aberration-robust
  monocular passive depth sensing using a meta-imaging camera. Light: Science
  \& Applications  \textbf{13}(1), ~236 (2024)

\bibitem{chakravarthula2023thin}
Chakravarthula, P., Sun, J., Li, X., Lei, C., Chou, G., Bijelic, M., Froesch,
  J., Majumdar, A., Heide, F.: Thin on-sensor nanophotonic array cameras. ACM
  Transactions on Graphics (TOG)  \textbf{42}(6),  1--18 (2023)

\bibitem{chang2019deep}
Chang, J., Wetzstein, G.: Deep optics for monocular depth estimation and 3d
  object detection. In: Proceedings of the IEEE/CVF international conference on
  computer vision. pp. 10193--10202 (2019)

\bibitem{Chen2023intdepth}
Chen, M.K., Liu, X., Wu, Y., Zhang, J., Yuan, J., Zhang, Z., Tsai, D.P.: A
  meta-device for intelligent depth perception. Advanced Materials
  \textbf{35}(34),  2107465 (2023).
  \doi{https://doi.org/10.1002/adma.202107465},
  \url{https://onlinelibrary.wiley.com/doi/abs/10.1002/adma.202107465}

\bibitem{Chen2022semi}
Chen, R., Shao, Y., Zhou, Y., Dang, Y., Dong, H., Zhang, S., Wang, Y., Chen,
  J., Ju, B.F., Ma, Y.: A semisolid micromechanical beam steering system based
  on micrometa-lens arrays. Nano Letters  \textbf{22}(4),  1595--1603 (2022).
  \doi{10.1021/acs.nanolett.1c04493},
  \url{https://doi.org/10.1021/acs.nanolett.1c04493}, doi:
  10.1021/acs.nanolett.1c04493

\bibitem{Colburn2020paired}
Colburn, S., Majumdar, A.: Metasurface generation of paired accelerating and
  rotating optical beams for passive ranging and scene reconstruction. ACS
  Photonics  \textbf{7}(6),  1529--1536 (2020).
  \doi{10.1021/acsphotonics.0c00354},
  \url{https://doi.org/10.1021/acsphotonics.0c00354}, doi:
  10.1021/acsphotonics.0c00354

\bibitem{dosovitskiy2020image}
Dosovitskiy, A.: An image is worth 16x16 words: Transformers for image
  recognition at scale. arXiv preprint arXiv:2010.11929  (2020)

\bibitem{Fan2020arbitrary}
Fan, Q., Liu, M., Zhang, C., Zhu, W., Wang, Y., Lin, P., Yan, F., Chen, L.,
  Lezec, H.J., Lu, Y.: Independent amplitude control of arbitrary orthogonal
  states of polarization via dielectric metasurfaces. Physical Review Letters
  \textbf{125}(26),  267402 (2020)

\bibitem{fu2024geowizard}
Fu, X., Yin, W., Hu, M., Wang, K., Ma, Y., Tan, P., Shen, S., Lin, D., Long,
  X.: Geowizard: Unleashing the diffusion priors for 3d geometry estimation
  from a single image. In: ECCV (2024)

\bibitem{DualPixels}
Garg, R., Wadhwa, N., Ansari, S., Barron, J.T.: Learning single camera depth
  estimation using dual-pixels. In: Proceedings of the IEEE/CVF International
  Conference on Computer Vision (ICCV) (October 2019)

\bibitem{KITTI2012}
Geiger, A., Lenz, P., Urtasun, R.: Are we ready for autonomous driving? the
  kitti vision benchmark suite. In: Conference on Computer Vision and Pattern
  Recognition (CVPR) (2012)

\bibitem{ghanekar2024passive}
Ghanekar, B., Khan, S.S., Sharma, P., Singh, S., Boominathan, V., Mitra, K.,
  Veeraraghavan, A.: Passive snapshot coded aperture dual-pixel rgb-d imaging.
  In: Proceedings of the IEEE/CVF Conference on Computer Vision and Pattern
  Recognition. pp. 25348--25357 (2024)

\bibitem{ghanekar2022ps}
Ghanekar, B., Saragadam, V., Mehra, D., Gustavsson, A.K., Sankaranarayanan,
  A.C., Veeraraghavan, A.: Ps$^2$f: Polarized spiral point spread function for
  single-shot 3d sensing. IEEE Transactions on Pattern Analysis and Machine
  Intelligence  (2022)

\bibitem{gopakumar2024full}
Gopakumar, M., Lee, G.Y., Choi, S., Chao, B., Peng, Y., Kim, J., Wetzstein, G.:
  Full-colour 3d holographic augmented-reality displays with metasurface
  waveguides. Nature pp.~1--7 (2024)

\bibitem{rotationdepth}
Greengard, A., Schechner, Y.Y., Piestun, R.: Depth from diffracted rotation.
  Optics Letters  \textbf{31}(2),  181–183 (2006).
  \doi{10.1364/OL.31.000181},
  \url{https://opg.optica.org/ol/abstract.cfm?URI=ol-31-2-181}

\bibitem{Guizilini2023depth}
Guizilini, V., Vasiljevic, I., Chen, D., Ambruș, R., Gaidon, A.: Towards
  zero-shot scale-aware monocular depth estimation. In: 2023 IEEE/CVF
  International Conference on Computer Vision (ICCV). pp. 9199--9209 (2023).
  \doi{10.1109/ICCV51070.2023.00847}

\bibitem{Guo2019spidereye}
Guo, Q., Shi, Z., Huang, Y.W., Alexander, E., Qiu, C.W., Capasso, F., Zickler,
  T.: Compact single-shot metalens depth sensors inspired by eyes of jumping
  spiders. Proceedings of the National Academy of Sciences  \textbf{116}(46),
  22959--22965 (2019). \doi{doi:10.1073/pnas.1912154116},
  \url{https://www.pnas.org/doi/abs/10.1073/pnas.1912154116}

\bibitem{gur2019single}
Gur, S., Wolf, L.: Single image depth estimation trained via depth from defocus
  cues. In: Proceedings of the IEEE/CVF conference on computer vision and
  pattern recognition. pp. 7683--7692 (2019)

\bibitem{he2024lotus}
He, J., Li, H., Yin, W., Liang, Y., Li, L., Zhou, K., Zhang, H., Liu, B., Chen,
  Y.C.: Lotus: Diffusion-based visual foundation model for high-quality dense
  prediction. arXiv preprint arXiv:2409.18124  (2024)

\bibitem{hu2024metric3d}
Hu, M., Yin, W., Zhang, C., Cai, Z., Long, X., Chen, H., Wang, K., Yu, G.,
  Shen, C., Shen, S.: Metric3d v2: A versatile monocular geometric foundation
  model for zero-shot metric depth and surface normal estimation. IEEE
  Transactions on Pattern Analysis and Machine Intelligence  (2024)

\bibitem{huang2023int}
Huang, H., Overvig, A.C., Xu, Y., Malek, S.C., Tsai, C.C., Alù, A., Yu, N.:
  {Leaky-wave metasurfaces for integrated photonics}. Nat.~Nanotechnol.
  \textbf{18}(6),  580--588 (2023),
  \url{https://doi.org/10.1038/s41565-023-01360-z}

\bibitem{ikoma2021depth}
Ikoma, H., Nguyen, C.M., Metzler, C.A., Peng, Y., Wetzstein, G.: Depth from
  defocus with learned optics for imaging and occlusion-aware depth estimation.
  In: 2021 IEEE International Conference on Computational Photography (ICCP).
  pp. 1--12. IEEE (2021)

\bibitem{Jin2019dielectric}
Jin, C., Afsharnia, M., Berlich, R., Fasold, S., Zou, C., Arslan, D., Staude,
  I., Pertsch, T., Setzpfandt, F.: Dielectric metasurfaces for distance
  measurements and three-dimensional imaging. Advanced Photonics
  \textbf{1}(3),  036001 (2019), \url{https://doi.org/10.1117/1.AP.1.3.036001}

\bibitem{Jin2019dhpsf}
Jin, C., Zhang, J., Guo, C.: Metasurface integrated with double-helix point
  spread function and metalens for three-dimensional imaging. Nanophotonics
  \textbf{8}(3),  451--458 (2019). \doi{doi:10.1515/nanoph-2018-0216},
  \url{https://doi.org/10.1515/nanoph-2018-0216}

\bibitem{Martins2022metalidar}
Juliano~Martins, R., Marinov, E., Youssef, M.A.B., Kyrou, C., Joubert, M.,
  Colmagro, C., Gâté, V., Turbil, C., Coulon, P.M., Turover, D., Khadir, S.,
  Giudici, M., Klitis, C., Sorel, M., Genevet, P.: Metasurface-enhanced light
  detection and ranging technology. Nature Communications  \textbf{13}(1),
  ~5724 (2022). \doi{10.1038/s41467-022-33450-2},
  \url{https://doi.org/10.1038/s41467-022-33450-2}

\bibitem{ke2023repurposing}
Ke, B., Obukhov, A., Huang, S., Metzger, N., Daudt, R.C., Schindler, K.:
  Repurposing diffusion-based image generators for monocular depth estimation.
  In: Proceedings of the IEEE/CVF Conference on Computer Vision and Pattern
  Recognition (CVPR) (2024)

\bibitem{Kim2022fullspace}
Kim, G., Kim, Y., Yun, J., Moon, S.W., Kim, S., Kim, J., Park, J., Badloe, T.,
  Kim, I., Rho, J.: Metasurface-driven full-space structured light for
  three-dimensional imaging. Nature Communications  \textbf{13}(1), ~5920
  (2022). \doi{10.1038/s41467-022-32117-2},
  \url{https://doi.org/10.1038/s41467-022-32117-2}

\bibitem{Kim2021rev}
Kim, I., Martins, R.J., Jang, J., Badloe, T., Khadir, S., Jung, H.Y., Kim, H.,
  Kim, J., Genevet, P., Rho, J.: Nanophotonics for light detection and ranging
  technology. Nature Nanotechnology  \textbf{16}(5),  508--524 (2021).
  \doi{10.1038/s41565-021-00895-3},
  \url{https://doi.org/10.1038/s41565-021-00895-3}

\bibitem{Li2018cloud}
Li, Z., Dai, Q., Mehmood, M.Q., Hu, G., yanchuk, B.L., Tao, J., Hao, C., Kim,
  I., Jeong, H., Zheng, G., Yu, S., Alù, A., Rho, J., Qiu, C.W.: Full-space
  cloud of random points with a scrambling metasurface. Light: Science \&
  Applications  \textbf{7}(1), ~63 (2018). \doi{10.1038/s41377-018-0064-3},
  \url{https://doi.org/10.1038/s41377-018-0064-3}

\bibitem{liang2025distilling}
Liang, Y., Hu, Y., Shao, W., Fu, Y.: Distilling monocular foundation model for
  fine-grained depth completion. In: Proceedings of the Computer Vision and
  Pattern Recognition Conference. pp. 22254--22265 (2025)

\bibitem{depthanything3}
Lin, H., Chen, S., Liew, J.H., Chen, D.Y., Li, Z., Shi, G., Feng, J., Kang, B.:
  Depth anything 3: Recovering the visual space from any views. arXiv preprint
  arXiv:2511.10647  (2025)

\bibitem{lin2025prompting}
Lin, H., Peng, S., Chen, J., Peng, S., Sun, J., Liu, M., Bao, H., Feng, J.,
  Zhou, X., Kang, B.: Prompting depth anything for 4k resolution accurate
  metric depth estimation. In: Proceedings of the Computer Vision and Pattern
  Recognition Conference. pp. 17070--17080 (2025)

\bibitem{MIFDB16}
Mayer, N., Ilg, E., H{\"a}usser, P., Fischer, P., Cremers, D., Dosovitskiy, A.,
  Brox, T.: A large dataset to train convolutional networks for disparity,
  optical flow, and scene flow estimation. In: IEEE International Conference on
  Computer Vision and Pattern Recognition (CVPR) (2016),
  \url{http://lmb.informatik.uni-freiburg.de/Publications/2016/MIFDB16},
  arXiv:1512.02134

\bibitem{bire1}
Meuleman, A., Baek, S.H., Heide, F., Kim, M.H.: Single-shot monocular rgb-d
  imaging using uneven double refraction. In: Proceedings of the IEEE/CVF
  Conference on Computer Vision and Pattern Recognition. pp. 2465--2474 (2020)

\bibitem{nam2023depolarized}
Nam, S.W., Kim, Y., Kim, D., Jeong, Y.: Depolarized holography with
  polarization-multiplexing metasurface. ACM Transactions on Graphics (TOG)
  \textbf{42}(6),  1--16 (2023)

\bibitem{nyu_depth_v2}
Nathan~Silberman, Derek~Hoiem, P.K., Fergus, R.: Indoor segmentation and
  support inference from rgbd images. In: ECCV (2012)

\bibitem{ni2012broadband}
Ni, X., Emani, N.K., Kildishev, A.V., Boltasseva, A., Shalaev, V.M.: {Broadband
  light bending with plasmonic nanoantennas}. Science  \textbf{335}(6067),
  427--427 (Jan 2012), \url{http://dx.doi.org/10.1126/science.1214686}

\bibitem{Ni2020fov}
Ni, Y., Chen, S., Wang, Y., Tan, Q., Xiao, S., Yang, Y.: Metasurface for
  structured light projection over 120° field of view. Nano Letters
  \textbf{20}(9),  6719--6724 (2020). \doi{10.1021/acs.nanolett.0c02586},
  \url{https://doi.org/10.1021/acs.nanolett.0c02586}, doi:
  10.1021/acs.nanolett.0c02586

\bibitem{DualPixels2}
Pan, L., Chowdhury, S., Hartley, R., Liu, M., Zhang, H., Li, H.: Dual pixel
  exploration: Simultaneous depth estimation and image restoration. In:
  Proceedings of the IEEE/CVF Conference on Computer Vision and Pattern
  Recognition (CVPR). pp. 4340--4349 (June 2021)

\bibitem{park2024depth}
Park, J.H., Jeong, C., Lee, J., Jeon, H.G.: Depth prompting for sensor-agnostic
  depth estimation. In: Proceedings of the IEEE/CVF Conference on Computer
  Vision and Pattern Recognition. pp. 9859--9869 (2024)

\bibitem{Park2021slm}
Park, J., Jeong, B.G., Kim, S.I., Lee, D., Kim, J., Shin, C., Lee, C.B.,
  Otsuka, T., Kyoung, J., Kim, S., Yang, K.Y., Park, Y.Y., Lee, J., Hwang, I.,
  Jang, J., Song, S.H., Brongersma, M.L., Ha, K., Hwang, S.W., Choo, H., Choi,
  B.L.: All-solid-state spatial light modulator with independent phase and
  amplitude control for three-dimensional lidar applications. Nature
  Nanotechnology  \textbf{16}(1),  69--76 (2021).
  \doi{10.1038/s41565-020-00787-y},
  \url{https://doi.org/10.1038/s41565-020-00787-y}

\bibitem{rotationdepth2}
Pavani, S.R.P., Piestun, R.: Three dimensional tracking of fluorescent
  microparticles using a photon-limited double-helix response system. Optics
  express  \textbf{16}(26),  22048–22057 (2008)

\bibitem{piccinelli2025unidepthv2}
Piccinelli, L., Sakaridis, C., Yang, Y.H., Segu, M., Li, S., Abbeloos, W.,
  Van~Gool, L.: Unidepthv2: Universal monocular metric depth estimation made
  simpler. arXiv preprint arXiv:2502.20110  (2025)

\bibitem{piccinelli2024unidepth}
Piccinelli, L., Yang, Y.H., Sakaridis, C., Segu, M., Li, S., Van~Gool, L., Yu,
  F.: {U}ni{D}epth: Universal monocular metric depth estimation. In:
  Proceedings of the IEEE/CVF Conference on Computer Vision and Pattern
  Recognition (CVPR) (2024)

\bibitem{Prasad2013psf}
Prasad, S.: Rotating point spread function via pupil-phase engineering. Optics
  Letters  \textbf{38}(4),  585--587 (2013). \doi{10.1364/OL.38.000585},
  \url{https://opg.optica.org/ol/abstract.cfm?URI=ol-38-4-585}

\bibitem{ranftl2021vision}
Ranftl, R., Bochkovskiy, A., Koltun, V.: Vision transformers for dense
  prediction. In: Proceedings of the IEEE/CVF international conference on
  computer vision. pp. 12179--12188 (2021)

\bibitem{hypersim}
Roberts, M., Ramapuram, J., Ranjan, A., Kumar, A., Bautista, M.A., Paczan, N.,
  Webb, R., Susskind, J.M.: {Hypersim}: {A} photorealistic synthetic dataset
  for holistic indoor scene understanding. In: International Conference on
  Computer Vision (ICCV) 2021 (2021)

\bibitem{shen2023monocular}
Shen, Z., Zhao, F., Jin, C., Wang, S., Cao, L., Yang, Y.: Monocular metasurface
  camera for passive single-shot 4d imaging. Nature Communications
  \textbf{14}(1), ~1035 (2023)

\bibitem{shi2021towards}
Shi, L., Li, B., Kim, C., Kellnhofer, P., Matusik, W.: Towards real-time
  photorealistic 3d holography with deep neural networks. Nature
  \textbf{591}(7849),  234--239 (2021)

\bibitem{shi2022end}
Shi, L., Li, B., Matusik, W.: End-to-end learning of 3d phase-only holograms
  for holographic display. Light: Science \& Applications  \textbf{11}(1), ~247
  (2022)

\bibitem{splitaperturecameras}
Shi, Z., Chugunov, I., Bijelic, M., C\^{o}t\'{e}, G., Yeom, J., Fu, Q., Amata,
  H., Heidrich, W., Heide, F.: Split-aperture 2-in-1 computational cameras. ACM
  Trans. Graph.  \textbf{43}(4) (jul 2024). \doi{10.1145/3658225},
  \url{https://doi.org/10.1145/3658225}

\bibitem{sun}
Sun, J.Q., Weng, H., Xing, X., Yeum, C.M., Crowley, M.: View invariant learning
  for vision-language navigation in continuous environments. IEEE Robotics and
  Automation Letters  \textbf{11}(5),  5861--5868 (2026).
  \doi{10.1109/LRA.2026.3669785}

\bibitem{talegaonkar2025repurposingmarigoldzeroshotmetric}
Talegaonkar, C., Suresh, N.G., Novack, Z., Belhe, Y., Nagasamudra, P., Antipa,
  N.: Repurposing marigold for zero-shot metric depth estimation via defocus
  blur cues (2025), \url{https://arxiv.org/abs/2505.17358}

\bibitem{Tan2021dispersion}
Tan, S., Yang, F., Boominathan, V., Veeraraghavan, A., Naik, G.V.: 3d imaging
  using extreme dispersion in optical metasurfaces. ACS Photonics
  \textbf{8}(5),  1421--1429 (2021). \doi{10.1021/acsphotonics.1c00110},
  \url{https://doi.org/10.1021/acsphotonics.1c00110}, doi:
  10.1021/acsphotonics.1c00110

\bibitem{tang2017depth}
Tang, H., Cohen, S., Price, B., Schiller, S., Kutulakos, K.N.: Depth from
  defocus in the wild. In: Proceedings of the IEEE conference on computer
  vision and pattern recognition. pp. 2740--2748 (2017)

\bibitem{tseng2021neural}
Tseng, E., Colburn, S., Whitehead, J., Huang, L., Baek, S.H., Majumdar, A.,
  Heide, F.: Neural nano-optics for high-quality thin lens imaging. Nature
  communications  \textbf{12}(1), ~6493 (2021)

\bibitem{wang2025moge2accuratemonoculargeometry}
Wang, R., Xu, S., Dong, Y., Deng, Y., Xiang, J., Lv, Z., Sun, G., Tong, X.,
  Yang, J.: Moge-2: Accurate monocular geometry with metric scale and sharp
  details (2025), \url{https://arxiv.org/abs/2507.02546}

\bibitem{wei2024spatially}
Wei, K., Li, X., Froech, J., Chakravarthula, P., Whitehead, J., Tseng, E.,
  Majumdar, A., Heide, F.: Spatially varying nanophotonic neural networks.
  Science Advances  \textbf{10}(45),  eadp0391 (2024)

\bibitem{Wijayasingha_2024_WACV}
Wijayasingha, L., Alemzadeh, H., Stankovic, J.A.: Camera-independent single
  image depth estimation from defocus blur. In: Proceedings of the IEEE/CVF
  Winter Conference on Applications of Computer Vision (WACV). pp. 3749--3758
  (January 2024)

\bibitem{Phasecam3d}
Wu, Y., Boominathan, V., Chen, H., Sankaranarayanan, A., Veeraraghavan, A.:
  Phasecam3d—learning phase masks for passive single view depth estimation.
  In: 2019 IEEE International Conference on Computational Photography (ICCP).
  p. 1–12. IEEE (2019)

\bibitem{Yan2024nano}
Yan, T., Zhou, T., Guo, Y., Zhao, Y., Shao, G., Wu, J., Huang, R., Dai, Q.,
  Fang, L.: Nanowatt all-optical 3d perception for mobile robotics. Science
  Advances  \textbf{10}(27),  eadn2031 (2024).
  \doi{doi:10.1126/sciadv.adn2031},
  \url{https://www.science.org/doi/abs/10.1126/sciadv.adn2031}

\bibitem{depth_anything_v1}
Yang, L., Kang, B., Huang, Z., Xu, X., Feng, J., Zhao, H.: Depth anything:
  Unleashing the power of large-scale unlabeled data. In: Proceedings of the
  IEEE/CVF conference on computer vision and pattern recognition. pp.
  10371--10381 (2024)

\bibitem{depth_anything_v2}
Yang, L., Kang, B., Huang, Z., Zhao, Z., Xu, X., Feng, J., Zhao, H.: Depth
  anything v2. Advances in Neural Information Processing Systems  \textbf{37},
  21875--21911 (2024)

\bibitem{yin2023metric}
Yin, W., Zhang, C., Chen, H., Cai, Z., Yu, G., Wang, K., Chen, X., Shen, C.:
  Metric3d: Towards zero-shot metric 3d prediction from a single image. In:
  Proceedings of the IEEE/CVF international conference on computer vision. pp.
  9043--9053 (2023)

\bibitem{yu2011light}
Yu, N., Genevet, P., Kats, M.A., Aieta, F., Tetienne, J.P., Capasso, F.,
  Gaburro, Z.: Light propagation with phase discontinuities: generalized laws
  of reflection and refraction. science  \textbf{334}(6054),  333--337 (2011)

\bibitem{zheng2023close}
Zheng, C., Zhao, G., So, P.: Close the design-to-manufacturing gap in
  computational optics with a'real2sim'learned two-photon neural lithography
  simulator. In: SIGGRAPH Asia 2023 Conference Papers. pp.~1--9 (2023)

\bibitem{LenslessCameras}
Zheng, Y., Salman~Asif, M.: Joint image and depth estimation with mask-based
  lensless cameras. IEEE Transactions on Computational Imaging  \textbf{6},
  1167--1178 (2020). \doi{10.1109/TCI.2020.3010360}

\end{thebibliography}

\clearpage
\appendix
\setcounter{page}{1}



\noindent In the supplementary material, we provide additional results, analyses, and implementation details. As an overview:
\begin{enumerate}
    \item \textbf{Experiments.} We show our framework's generalization to other DFMs (\cref{sec:supp:generalization}), qualitative analysis of depth prior (\cref{sec:supp:qualitative_ablation}) and depth consistency (\cref{sec:supp:depth_consistency}). We also extend our simulated (\cref{sec:supp:sim}) and physical (\cref{sec:supp:physical}) experiments, including additional results, experiment details, and discussion (\cref{sec:supp:discussion}). 
    \item \textbf{Software.} We discuss additional details for the neural network training (\cref{sec:supp:nn}) and simulator (\cref{sec:supp:simulator}). 
    \item \textbf{Hardware.} We compare our PSF with alternative designs (\cref{sec:supp:compare}), and further present the physical principles underlying the metalens (\cref{sec:supp:principles}), as well as the material design and fabrication procedures (\cref{sec:supp:hw_design_and_fab}).
\end{enumerate}


\section{Generalization to Other DFMs}
\label{sec:supp:generalization}
We examine whether the proposed input adaptation generalizes beyond Depth Anything V2 \cite{depth_anything_v2}. Specifically, we fine-tune UniDepth V2 \cite{piccinelli2025unidepthv2} with the same adapted three-channel input and training setup. The results in \cref{tab:supp:generalization} show that UniDepth V2 likewise adapts well to our polarization measurements. Remarkably, the fine-tuned UniDepth small model (137 MB) surpasses the original large model (1.42 GB). This suggests that our adaptation strategy can be model-agnostic, enabling different monocular depth foundation models to leverage polarization cues while retaining their pretrained visual priors.
\begin{table}[]
\centering
\scriptsize
\setlength{\tabcolsep}{2.5pt} 
\renewcommand{\arraystretch}{1.15}
\begin{tabular}{l|cccc|cccc}
\toprule
\multirow{2}{*}{\makecell{Dataset}} & \multicolumn{4}{c|}{Finetuned-Small} & \multicolumn{4}{c}{{Original-Large}}\\
& {MAE$\downarrow$} & {\makecell{RMSE$\downarrow$}} & {\makecell{AbsRel$\downarrow$}} & {\makecell{$\delta_{0.5}\uparrow$}} & {\makecell{MAE$\downarrow$}} & {\makecell{RMSE$\downarrow$}} & {\makecell{AbsRel$\downarrow$}} & {\makecell{$\delta_{0.5}\uparrow$}} \\
\hline
{NYU Depth V2} & 0.0217 & 0.0404 &  0.0375 & 0.9573
& 0.0338 & 0.0591 & 0.0633 & 0.8653 \\
{MIT-CGH-4k}            & 0.0708 & 0.1307 & 0.1159 & 0.7521
& 0.1273 & 0.1635 &  0.2588&  0.3599 \\
{Real Data}                 & 0.0348 & 0.0096 & 0.0619 & 0.8636
& 0.1068 & 0.1446 & 0.1957 & 0.5917 \\

\bottomrule
\end{tabular}
\vspace{6pt}
\Caption{Generalization of the proposed input adaptation to UniDepth V2.} {We test on NYU Depth V2, MIT-CGH-4k and our real dataset captured in physical experiment.}
\label{tab:supp:generalization}

\vspace{-15pt}
\end{table}

\section{Qualitative Analysis of Depth Prior}
\label{sec:supp:qualitative_ablation}
\begin{figure*}[]
\centering
\includegraphics[width=1.0\linewidth]{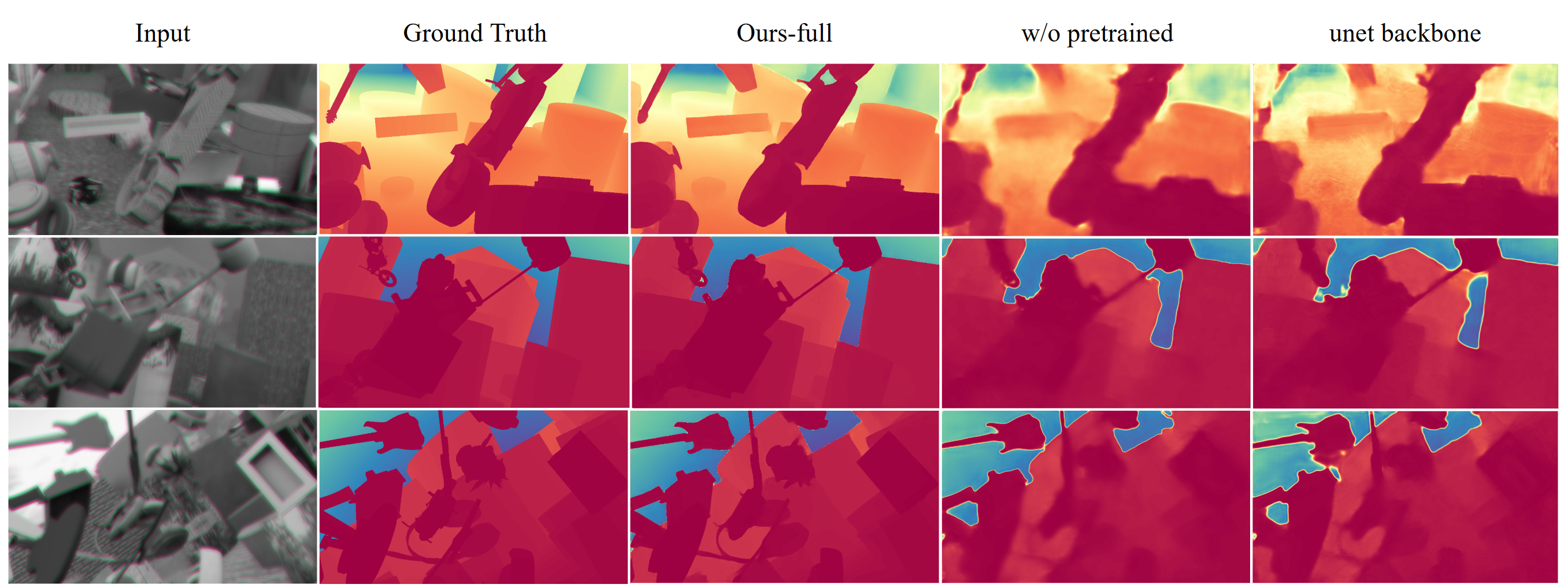}
\Caption{Qualitative Analysis of Depth Prior.}{Pretrained initialization clearly improves fine structures, edge sharpness, and shape consistency, leading to more accurate and visually coherent depth predictions. }
\label{fig:supp:ablation}
\vspace{-12pt}
\end{figure*}

To further illustrate the role of pretrained initialization, we present qualitative comparisons in \cref{fig:supp:ablation} following the ablation study in \cref{tab:ablation_compare_with_dfd}. From left to right, we show the simulated input generated from the FlyingThings3D dataset, the ground-truth depth, the output of our full model, the output without pretrained initialization, and the output with a U-Net backbone. Pretrained initialization clearly improves fine structures, edge sharpness, and shape consistency, leading to more accurate and visually coherent depth predictions. In contrast, removing pretrained initialization results in degraded details and less reliable geometry. For reference, replacing the ViT backbone with a U-Net also yields inferior results, with less smooth depth maps and more visible artifacts.

\begin{figure*}[b!]
\vspace{-10pt}
\centering
\includegraphics[width=0.98\linewidth, trim={0cm 0.0cm 0cm 0.0cm}]{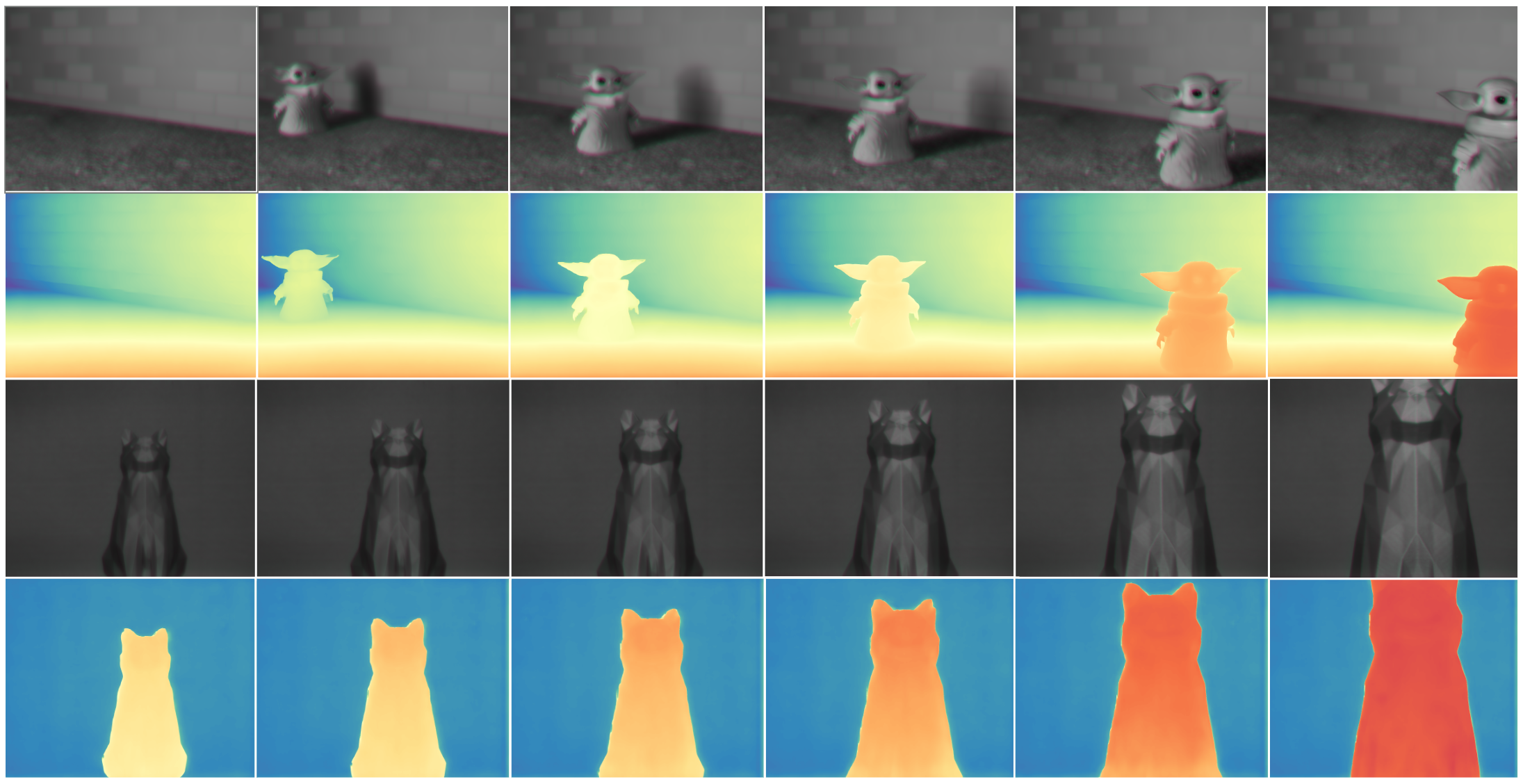}

\Caption{Qualitative results on moving objects.}{The upper two rows show results on a rendered video in simulation. The lower two rows show results from a physical experiment in which an object moves toward the camera. Our method produces temporally consistent depth predictions with smoothly varying estimates as the object approaches.}

\label{fig:supp:video}
\vspace{-10pt}
\end{figure*}

\section{Depth Consistency}
\vspace{-10pt}
\label{sec:supp:depth_consistency}
We evaluate the consistency of our depth predictions under object motion in both simulation and real experiments. As shown in \cref{fig:supp:video}, the predicted depth evolves smoothly as the object moves toward the camera, with little temporal fluctuation or abrupt frame-to-frame changes. In the rendered simulation sequence, our method accurately reflects the continuous decrease in object distance. In the physical experiment, despite additional real-world noise and optical imperfections, the predictions remain stable and follow the same monotonic depth trend. This demonstrates that our method produces temporally consistent depth estimates and robustly captures physically grounded depth cues for moving objects.

\section{Simulated Experiments}
\label{sec:supp:sim}

\subsection{Evaluation on Additional Datasets}
\label{sec:supp:additional_quantative}

We present quantitative comparisons against all baselines on MPI Sintel \cite{mpi_sintel} and Hypersim \cite{hypersim}. For each dataset, we report the \textit{mean} and \textit{standard deviation} of all depth metrics (defined in \cref{tab:metrics}) to provide a more comprehensive evaluation. We additionally include a fine-tuned and post-aligned DepthAnything v2 \cite{depth_anything_v2} baseline to illustrate the effect of our post-alignment procedure; this baseline is highlighted using the diagonal color cell in each table. For completeness, we also provide extended evaluations on NYU Depth V2 \cite{nyu_depth_v2} and MIT-CGH-4K \cite{shi2021towards,shi2022end} in the supplementary tables. We evaluate on 100 uniformly sampled data samples from each dataset’s test split. All datasets are evaluated at their original image resolutions.
Taken together, these experiments demonstrate that our approach generalizes reliably across diverse domains and effectively exploits polarization-encoded physical depth cues.

\begin{table}[]

\footnotesize
\centering
\renewcommand{\arraystretch}{1.4}  
\begin{tabular}{c c}
\toprule
\textbf{Metric} & \textbf{Definition} \\
\midrule
$\text{MAE}$ 
& $\frac{1}{N}\sum_{i} |d_i - \hat{d}_i|$ \\
$\text{RMSE}$ 
& $\sqrt{\frac{1}{N}\sum_{i} (d_i - \hat{d}_i)^2}$ \\
$\text{AbsRel}$ 
& $\frac{1}{N}\sum_{i} \frac{|d_i - \hat{d}_i|}{d_i}$ \\
$\delta_{0.5}$ 
& $\frac{1}{N}\sum_{i} \mathbf{1}\!\left(\max\!\left(\frac{d_i}{\hat{d}_i}, \frac{\hat{d}_i}{d_i}\right) < 1.25^{0.5}\right)$ \\

\bottomrule
\end{tabular}
\Caption{
Definitions of depth evaluation metrics.}{Here, $d_i$ and $\hat{d}_i$ denote the ground-truth and predicted depth at pixel $i$,  
$N$ is the number of valid pixels, and $\mathbf{1}(\cdot)$ is the indicator function.
}
\label{tab:metrics}
\end{table}

\begin{table*}[]
\centering
\scriptsize

{
\setlength{\tabcolsep}{0pt}          
\renewcommand{\arraystretch}{1.15}      

\begin{tabular}{l|cccc}
\toprule
\multirow{2}{*}{\tinyspace \makecell{\tiny \colorbox{\colorTableGreen}{Train}/ \colorbox{\colorTableYellow}{Post.}/ \\ \tiny \colorbox{\colorTableGray}{w/ LiDAR}}}&
\multicolumn{4}{c}{\textbf{MPI Sintel}} \\
& MAE$\downarrow$ & RMSE$\downarrow$ & AbsRel$\downarrow$ & $\delta_{0.5}\uparrow$ \\
\hline
 \tableGreen{\tinyspace \textbf{Ours-Large}} 
 & \rankFirst{\tinyspace \textbf{0.0310 $\pm$ 0.0120}\tinyspace } & \rankFirst{\tinyspace \textbf{0.0600 $\pm$ 0.0259}\tinyspace } & \rankFirst{\tinyspace \textbf{0.0460 $\pm$ 0.0154}\tinyspace } & \rankFirst{\tinyspace \textbf{0.9288 $\pm$ 0.0433}\tinyspace } \\
 \tableGreen{\tinyspace \textbf{Ours-Base}} 
 & \rankSecond{0.0367 $\pm$ 0.0128} & \rankThird{0.0681 $\pm$ 0.0236} & \rankSecond{0.0516 $\pm$ 0.0165} & \rankSecond{0.8908 $\pm$ 0.0481} \\ 
 \tableGreen{\tinyspace \textbf{Ours-Small}} 
 & \rankThird{0.0418 $\pm$ 0.0179} & {0.0742 $\pm$ 0.0290} & \rankThird{0.0630 $\pm$ 0.0244} & \rankThird{0.8681 $\pm$ 0.0727} \\ 
 \tableGreen{\tinyspace DepthAny. v2$^\ast$ }
 & 0.1399 $\pm$ 0.0604 & 0.1711 $\pm$ 0.0690 & 0.2243 $\pm$ 0.0772 & 0.2743 $\pm$ 0.1846  \\ 
 
 \triangcellWH{7.1em}{1.2em}{DepthAny. v2$^\ast$}
 & 0.0782 $\pm$ 0.0332 & 0.0999 $\pm$ 0.0327 & 0.1527 $\pm$ 0.0720 & 0.5178 $\pm$ 0.1970 \\ 
 
 \tableYellow{\tinyspace DepthAny. v2} 
 & {0.0681 $\pm$ 0.0254} & {0.0896 $\pm$ 0.0282} & {0.1308 $\pm$ 0.0489} & {0.5704 $\pm$ 0.1912} \\ 
 \tableYellow{\tinyspace Depth Pro} 
 & {0.0482 $\pm$ 0.0192} & \rankThird{0.0675 $\pm$ 0.0212} & {0.0865 $\pm$ 0.0353} & {0.7363 $\pm$ 0.1708} \\ 
 \tableYellow{\tinyspace Lotus} 
 & 0.0937 $\pm$ 0.0242 & 0.1271 $\pm$ 0.0316 & 0.1908 $\pm$ 0.0578 & 0.4819 $\pm$ 0.1432 \\ 
 \tableYellow{\tinyspace Marigold}
 & 0.0908 $\pm$ 0.0300 & 0.1234 $\pm$ 0.0368 & 0.1765 $\pm$ 0.0596 & 0.4693 $\pm$ 0.1269 \\ 
 \tableYellow{\tinyspace Metric3D v2} 
 & 0.0812 $\pm$ 0.0230 & 0.1080 $\pm$ 0.0281 & 0.1576 $\pm$ 0.0500 & 0.4784 $\pm$ 0.1886 \\ 
 \tableYellow{\tinyspace UniDepth} v2  
 & {0.0577 $\pm$ 0.0195} & {0.0804 $\pm$ 0.0231} & {0.1114 $\pm$ 0.0392} & {0.6330 $\pm$ 0.1945} \\
 \tableYellow{\tinyspace ZoeDepth}
 & 0.0792 $\pm$ 0.0250 & 0.1094 $\pm$ 0.0332 & 0.1494 $\pm$ 0.0470 & 0.5511 $\pm$ 0.1221  \\ 
\hline
 \tableGray{\tinyspace PromptDA} 
 & \tableLightGray{0.0206 $\pm$ 0.0106} & \tableLightGray{0.0385 $\pm$ 0.0196} & \tableLightGray{0.0355 $\pm$ 0.0148} & \tableLightGray{0.9678 $\pm$ 0.0481} \\
\bottomrule
\end{tabular}
}

\Caption{Quantitative comparisons on MPI Sintel.}{
\colorbox{\colorTableGreen}{Train}: fine-tuned on our dataset; \colorbox{\colorTableYellow}{Post.}: post-aligned with GT using least-square fitting; \colorbox{\colorTableGray}{w/ LiDAR}: with additional simulated LiDAR input. Method with * is finetuned on our dataset. We highlight the top three metrics among LiDAR-free methods. Post-alignment can artificially improve baseline scores by fitting their scale and shift to each test sample.
}

\label{tab:supp:sintel}
\vspace{6pt}
\end{table*}

\paragraph{NYU Depth V2} This indoor dataset provides dense LiDAR ground truth and serves as a standard benchmark for indoor metric depth estimation. As shown in \cref{tab:supp:nyu-depth-v2}, our method demonstrates a clear advantage in the zero-shot setting, outperforming all LiDAR-free baselines and even surpassing several non–zero-shot baselines. Remarkbly, Compared to the LiDAR-assisted model, our approach achieves comparable performance, with lower RMSE, AbsRel, and $\delta_{0.5}$, and a similarly competitive MAE error.

\paragraph{MPI Sintel} MPI Sintel is a rendered animation movie dataset originally designed for optical-flow evaluation, and also widely used for depth estimation. In \cref{tab:supp:sintel}, our method achieves the strongest performance among all LiDAR-free baselines, demonstrating strong results without relying on domain-specific training data.

\paragraph{Hypersim} In \cref{tab:supp:hypersim}, the fine-tuned and post-aligned DepthAnything v2 baseline reports substantially improved results compared to its non–post-aligned version, illustrating how post alignment can artificially inflate monocular baseline performance. Even with this advantage, our method still achieves the best or near-best performance across most metrics among all LiDAR-free approaches. This highlights the effectiveness of our physical prompting in leveraging metric cues beyond what can be recovered through fine-tuning and post alignment alone.

\begin{table*}[]
\centering
\scriptsize

{
\setlength{\tabcolsep}{0pt}          
\renewcommand{\arraystretch}{1.15}      

\begin{tabular}{l|cccc}
\toprule
\multirow{2}{*}{\tinyspace \makecell{\tiny \colorbox{\colorTableGreen}{Train}/ \colorbox{\colorTableYellow}{Post.}/ \\ \tiny \colorbox{\colorTableGray}{w/ LiDAR}}}&
\multicolumn{4}{c}{\textbf{Hypersim-Test}} \\
& MAE$\downarrow$ & RMSE$\downarrow$ & AbsRel$\downarrow$ & $\delta_{0.5}\uparrow$ \\
\hline  
 \tableGreen{\tinyspace \textbf{Ours-Large}} 
 & \rankFirst{\tinyspace \textbf{{0.0258 $\pm$ 0.0249}}\tinyspace } & \rankSecond{\tinyspace {0.0438 $\pm$ 0.0274}\tinyspace } & \rankFirst{\tinyspace \textbf{0.0447 $\pm$ 0.0654}\tinyspace } & \rankFirst{\tinyspace \textbf{0.9482 $\pm$ 0.0971}\tinyspace } \\
 \tableGreen{\tinyspace \textbf{Ours-Base}} 
 & \rankThird{0.0300 $\pm$ 0.0226} & \rankThird{0.0495 $\pm$ 0.0256} & \rankSecond{0.0495 $\pm$ 0.0553} & \rankSecond{0.9333 $\pm$ 0.0930} \\ 
 \tableGreen{\tinyspace \textbf{Ours-Small}} 
 & {0.0347 $\pm$ 0.0301} & {0.0563 $\pm$ 0.0331} & {0.0625 $\pm$ 0.0850} & \rankThird{0.8989 $\pm$ 0.1016} \\ 

 \tableGreen{\tinyspace DepthAny. v2$^\ast$}
 & {0.0739 $\pm$ 0.0396} & {0.0860 $\pm$ 0.0401} & {0.1534 $\pm$ 0.0891} & {0.5065 $\pm$ 0.2611} \\ 
 
 \triangcellWH{7.1em}{1.2em}{DepthAny. v2$^\ast$} 
 & \rankSecond{0.0289 $\pm$ 0.0149} & \rankFirst{\textbf{0.0430 $\pm$ 0.0190}} & \rankThird{0.0538 $\pm$ 0.0270} & {0.8957 $\pm$ 0.1125} \\ 
 
 \tableYellow{\tinyspace DepthAny}. v2 
 & {0.0383 $\pm$ 0.0256} & {0.0559 $\pm$ 0.0347} & {0.0698 $\pm$ 0.0502} & {0.8429 $\pm$ 0.1948} \\ 
 \tableYellow{\tinyspace Depth Pro} 
 & {0.0398 $\pm$ 0.0302} & {0.0568 $\pm$ 0.0401} & {0.0729 $\pm$ 0.0627} & {0.8395 $\pm$ 0.1770} \\ 
 \tableYellow{\tinyspace Lotus} 
 & {0.0793 $\pm$ 0.0332} & {0.1032 $\pm$ 0.0408} & {0.1508 $\pm$ 0.0747} & {0.4932 $\pm$ 0.1990} \\ 
 \tableYellow{\tinyspace Marigold}
 & {0.0437 $\pm$ 0.0311} & {0.0628 $\pm$ 0.0397} & {0.0827 $\pm$ 0.0653} & {0.7985 $\pm$ 0.1914} \\ 
 \tableYellow{\tinyspace Metric3D v2} 
 & {0.0468 $\pm$ 0.0367} & {0.0667 $\pm$ 0.0475} & {0.0896 $\pm$ 0.0782} & {0.7809 $\pm$ 0.2144} \\ 
\tableYellow{\tinyspace UniDepth} v2 
 & {0.0442 $\pm$ 0.0374} & {0.0642 $\pm$ 0.0462} & {0.0813 $\pm$ 0.0758} & {0.8179 $\pm$ 0.2032} \\ 
 \tableYellow{\tinyspace ZoeDepth} 
 & {0.0746 $\pm$ 0.0400} & {0.1033 $\pm$ 0.0477} & {0.1427 $\pm$ 0.0813} & {0.5775 $\pm$ 0.1980} \\ 
 
\hline
 \tableGray{\tinyspace PromptDA} 
 & \tableLightGray{0.0121 $\pm$ 0.0056} & \tableLightGray{0.0275 $\pm$ 0.0167} & \tableLightGray{0.0210 $\pm$ 0.0076} & \tableLightGray{0.9845 $\pm$ 0.0190} \\
\bottomrule

\end{tabular}
}

\Caption{Quantitative comparisons on Hypersim.}{
}

\label{tab:supp:hypersim}
\vspace{6pt}
\end{table*}

\begin{table*}[]
\centering
\scriptsize

{
\setlength{\tabcolsep}{0pt}          
\renewcommand{\arraystretch}{1.15}      

\begin{tabular}{l|cccc}
\toprule
\multirow{2}{*}{\tinyspace \makecell{\tiny \colorbox{\colorTableGreen}{Train}/ \colorbox{\colorTableYellow}{Post.}/ \\ \tiny \colorbox{\colorTableGray}{w/ LiDAR}}}&
\multicolumn{4}{c}{\textbf{NYU Depth V2}} \\
& MAE$\downarrow$ & RMSE$\downarrow$ & AbsRel$\downarrow$ & $\delta_{0.5}\uparrow$ \\
\hline
 \tableGreen{\tinyspace\textbf{Ours-Large}\tinyspace} & \rankSecond{0.0228 $\pm$ 0.0058} & \rankSecond{0.0396 $\pm$ 0.0095} & \rankSecond{0.0387 $\pm$ 0.0085} & \rankSecond{0.9513 $\pm$ 0.0275} \\
 \tableGreen{ \textbf{Ours-Base} } & \rankFirst{ \tinyspace\textbf{0.0215 $\pm$ 0.0053}\tinyspace} & \rankFirst{ \tinyspace\textbf{0.0392 $\pm$ 0.0099}\tinyspace} & \rankFirst{ \tinyspace\textbf{0.0356 $\pm$ 0.0075}\tinyspace} & \rankFirst{\tinyspace\textbf{0.9571 $\pm$ 0.0237}\tinyspace} \\ 
 \tableGreen{\tinyspace\textbf{Ours-Small} \tinyspace} & \rankThird{0.0249 $\pm$ 0.0062} & \rankThird{0.0430 $\pm$ 0.0098} & \rankThird{0.0430 $\pm$ 0.0096} & \rankThird{0.9359 $\pm$ 0.0357} \\ 

 \tableGreen{\tinyspace DepthAny. v2$^\ast$} 
 & 0.1277 $\pm$ 0.0720 & 0.1483 $\pm$ 0.0731 & 0.2666 $\pm$ 0.1841 & 0.3412 $\pm$ 0.2540  \\ 
 \triangcellWH{7.1em}{1.2em}{DepthAny. v2$^\ast$}& 0.0543 $\pm$ 0.0247 & 0.0788 $\pm$ 0.0328 & 0.0975 $\pm$ 0.0520 & 0.7309 $\pm$ 0.1598  \\ 
 \tableYellow{\tinyspace DepthAny. v2} 
 & {0.0431 $\pm$ 0.0268} & {0.0666 $\pm$ 0.0358} & {0.0790 $\pm$ 0.0557} & {0.8049 $\pm$ 0.1880} \\ 
 \tableYellow{\tinyspace Depth Pro}  
 & {0.0383 $\pm$ 0.0267} & {0.0610 $\pm$ 0.0350} & {0.0709 $\pm$ 0.0566} & {0.8412 $\pm$ 0.1696} \\ 
 \tableYellow{\tinyspace Lotus} 
 & 0.0687 $\pm$ 0.0235 & 0.0927 $\pm$ 0.0287 & 0.1267 $\pm$ 0.0496 & 0.5748 $\pm$ 0.1833  \\ 
 \tableYellow{\tinyspace Marigold}
 & 0.0453 $\pm$ 0.0271 & 0.0696 $\pm$ 0.0350 & 0.0847 $\pm$ 0.0570 & 0.7845 $\pm$ 0.1798     \\ 
 \tableYellow{\tinyspace Metric3D v2} 
 & 0.0561 $\pm$ 0.0563 & 0.0816 $\pm$ 0.0615 & 0.1099 $\pm$ 0.1197 & 0.7658 $\pm$ 0.2570\\ 
 \tableYellow{\tinyspace UniDepth} v2 
 & 0.0338 $\pm$ 0.0269 & 0.0591 $\pm$ 0.0360 & 0.0633 $\pm$ 0.0581 & 0.8653 $\pm$ 0.1829 \\ 
  \tableYellow{\tinyspace ZoeDepth} 
 & 0.0413 $\pm$ 0.0188 & 0.0619 $\pm$ 0.0243 & 0.0759 $\pm$ 0.0373 & 0.8045 $\pm$ 0.1527 \\ 
 
\hline
\tableGray{\tinyspace PromptDA} & \tableLightGray{0.0205 $\pm$ 0.0064} & \tableLightGray{0.0424 $\pm$ 0.0185} & \tableLightGray{0.0358 $\pm$ 0.0105} & \tableLightGray{0.9549 $\pm$ 0.0332} \\
\bottomrule

\end{tabular}
}

\Caption{Quantitative comparisons on NYU Depth V2.}{}
\label{tab:supp:nyu-depth-v2}
\vspace{-6pt}
\end{table*}

\begin{table*}[]
\centering
\scriptsize

{
\setlength{\tabcolsep}{0pt}          
\renewcommand{\arraystretch}{1.15}      

\begin{tabular}{l|cccc}
\toprule
\multirow{2}{*}{\tinyspace \makecell{\tiny \colorbox{\colorTableGreen}{Train}/ \colorbox{\colorTableYellow}{Post.}/ \\ \tiny \colorbox{\colorTableGray}{w/ LiDAR}}}&
\multicolumn{4}{c}{\textbf{MIT-CGH-4k}} \\

& MAE$\downarrow$ & RMSE$\downarrow$ & AbsRel$\downarrow$ & $\delta_{0.5}\uparrow$ \\
\hline 
\tableGreen{\tinyspace \textbf{Ours-Large}} 
 & \rankFirst{\textbf{\tinyspace 0.0673 $\pm$ 0.0097\tinyspace }} & \rankFirst{\textbf{\tinyspace 0.1258 $\pm$ 0.0178\tinyspace }} & \rankSecond{{\tinyspace 0.1053 $\pm$ 0.0162\tinyspace }} & \rankSecond{{\tinyspace 0.7644 $\pm$ 0.0412\tinyspace }} \\
 \tableGreen{\tinyspace \textbf{Ours-Base}} 
  & \rankSecond{\tinyspace 0.0678 $\pm$ 0.0101 \tinyspace } & \rankSecond{\tinyspace 0.1291 $\pm$ 0.0183 \tinyspace } & \rankFirst{\textbf{\tinyspace 0.1023 $\pm$ 0.0159\tinyspace }} & \rankFirst{\tinyspace \textbf{0.7717 $\pm$ 0.0402} \tinyspace } \\ 
 \tableGreen{\tinyspace \textbf{Ours-Small}} 
 & \rankThird{0.0756 $\pm$ 0.0100} & \rankThird{0.1365 $\pm$ 0.0176} & \rankThird{0.1249 $\pm$ 0.0211} & \rankThird{0.7241 $\pm$ 0.0435} \\ 

 \tableGreen{\tinyspace DepthAny. v2$^\ast$}
 & {0.3014 $\pm$ 0.0524} & {0.3711 $\pm$ 0.0571} & {0.4104 $\pm$ 0.0473} & {0.1000 $\pm$ 0.0499} \\ 
 
 \triangcellWH{7.1em}{1.2em}{DepthAny. v2$^\ast$} 
 & {0.1800 $\pm$ 0.0338} & {0.2200 $\pm$ 0.0336} & {0.3711 $\pm$ 0.0849} & {0.2412 $\pm$ 0.0901} \\ 
 
 \tableYellow{\tinyspace DepthAny}. v2 
 & {0.1510 $\pm$ 0.0264} & {0.1899 $\pm$ 0.0289} & {0.3077 $\pm$ 0.0659} & {0.3004 $\pm$ 0.0896} \\ 
 \tableYellow{\tinyspace Depth Pro} 
 & {0.1437 $\pm$ 0.0247} & {0.1812 $\pm$ 0.0275} & {0.2917 $\pm$ 0.0569} & {0.3094 $\pm$ 0.0887} \\ 
 \tableYellow{\tinyspace Lotus} 
 & {0.1624 $\pm$ 0.0264} & {0.2008 $\pm$ 0.0277} & {0.3304 $\pm$ 0.0659} & {0.2680 $\pm$ 0.0741} \\ 
 \tableYellow{\tinyspace Marigold}
 & {0.1736 $\pm$ 0.0334} & {0.2127 $\pm$ 0.0324} & {0.3548 $\pm$ 0.0808} & {0.2427 $\pm$ 0.0748} \\ 
 \tableYellow{\tinyspace Metric3D v2} 
 & {0.2124 $\pm$ 0.0330} & {0.2517 $\pm$ 0.0306} & {0.4418 $\pm$ 0.0870} & {0.1834 $\pm$ 0.0661} \\ 
 \tableYellow{\tinyspace UniDepth} v2 
 & {0.1273 $\pm$ 0.0225} & {0.1635 $\pm$ 0.0246} & {0.2588 $\pm$ 0.0557} & {0.3599 $\pm$ 0.0994} \\ 
\tableYellow{\tinyspace ZoeDepth} 
 & {0.2051 $\pm$ 0.0298} & {0.2467 $\pm$ 0.0274} & {0.4280 $\pm$ 0.0798} & {0.2038 $\pm$ 0.0866} \\ 
 
\hline
 \tableGray{\tinyspace PromptDA} 
 & \tableLightGray{0.0575 $\pm$ 0.0079} & \tableLightGray{0.1132 $\pm$ 0.0147} & \tableLightGray{0.0974 $\pm$ 0.0158} & \tableLightGray{0.8044 $\pm$ 0.0314} \\
\bottomrule

\end{tabular}
}

\Caption{Quantitative comparisons on MIT-CGH-4k.}{
}

\label{tab:supp:mit}
\end{table*}

\begin{figure}[t!]
\centering
\includegraphics[width=0.98\linewidth]{images/qual_sim.pdf}
\vspace{-10pt}

\Caption{Qualitative results of simulated experiments.}{Bottom-right insets show the error map where dark colors indicate low error. Baseline results are post aligned to the ground truth while our results are directly visualized.}
\label{fig:supp:results_sim}
\end{figure}

\paragraph{MIT-CGH-4K}
MIT-CGH-4K contains scenes with extremely limited semantics, making it especially challenging for monocular models that rely on learned visual priors. In \cref{tab:supp:mit}, our method and the LiDAR-assisted baseline significantly outperform all other approaches, demonstrating that our polarization-encoded physical cues remain effective even when semantic information is scarce. These results underscore the robustness of our system in settings where purely data-driven monocular depth estimation methods typically fail.

\subsection{Qualitative Results}
Additional qualitative results from simulated experiments are shown in \cref{fig:supp:results_sim}. Compared to baselines, our method provides the most reliable metric scale and preserves crisp, well-defined object boundaries.

\section{Physical Experiments}
\label{sec:supp:physical}

\subsection{Hardware Prototype Details}
\label{sec:supp:hw_details}
We built a prototype depth camera with a fabricated metalens, and the hardware specifications are listed in \cref{tab:hardware}. The hardware includes four main components: the \matName metalens, a 1-inch tube for optical alignment, a 590-nm optical bandpass filter, and a monochrome CMOS sensor.

\begin{table}[]
    \begin{center}
    \begin{tabular}{c|c}
        \toprule
         Operation Wavelength & $\approx$590 nm \\
         \hline
         Metasurface & 1.5 mm radius, 700 nm thick \matName\\
         \hline
         Substrate & 500 $\mathrm{\mu m}$ thick glass\\
         \hline
         Focal length & 34 mm\\
        \bottomrule
    \end{tabular}
    \Caption{Specifications of our metasurface imaging hardware.}{}
    \label{tab:hardware}
    \end{center}
\end{table}

\paragraph{Metasurface Imaging Setup} 

We set the focal length of the metalens as $\focus=34$ mm. We mount the metalens 37.6 mm away from the monochrome CMOS sensor to set an in-focus depth $\infocus=$ of 35 cm. A separation of $2\sepy=6.5$ mm ensures that the two images occupy the CMOS sensor without overlapping.

\paragraph{Camera and optical filter}
We employ a FLIR Blackfly\textsuperscript{S} BFS-U3-200S6M-C USB\,3.1 camera, equipped with a 1-inch Sony IMX183 CMOS sensor providing $5472\times3648$ pixels at 2.4-\textmu m pitch. To suppress out-of-band light and enhance image contrast, we place a 10-nm optical bandpass filter centered at a wavelength of 590 nm before the CMOS sensor. This preserves the single-wavelength assumption central to our rotating-PSF design.

\paragraph{Apertures and mounting}
For stray-light suppression and to prevent overlap of the image pair, we installed a custom-made aperture in front of the metasurface. The aperture is sized to match the design field of view so that the deflected $x$- and $y$-polarized images occupy non-overlapping halves on the sensor. A standard 1-inch lens tube holds the metasurface, filter, and aperture in rigid alignment with the camera housing.

\paragraph{Optical rail setup}
We perform experimental validations on a 1.8-m optical rail, where the metasurface–camera assembly is fixed at one end, and a platform carrying the test objects slides along the $z$-axis. Fine translations in $x$, $y$, and $z$ allow precise measurement of object positions relative to the metasurface. The focal distance is adjusted so that the in-focus plane lies approximately 35 cm from the metasurface, matching the design for our single-helix PSF. This arrangement enables controlled data acquisition for a range of real-world scenes, which are then processed by our neural network for dense depth reconstruction.

\subsection{Data Processing}
\begin{figure*}[t]
\centering
    \subfloat[raw sensor capture from our prototype]{
        \includegraphics[width=0.28\textwidth, angle=90, trim={0cm 0cm 0cm 0},clip]{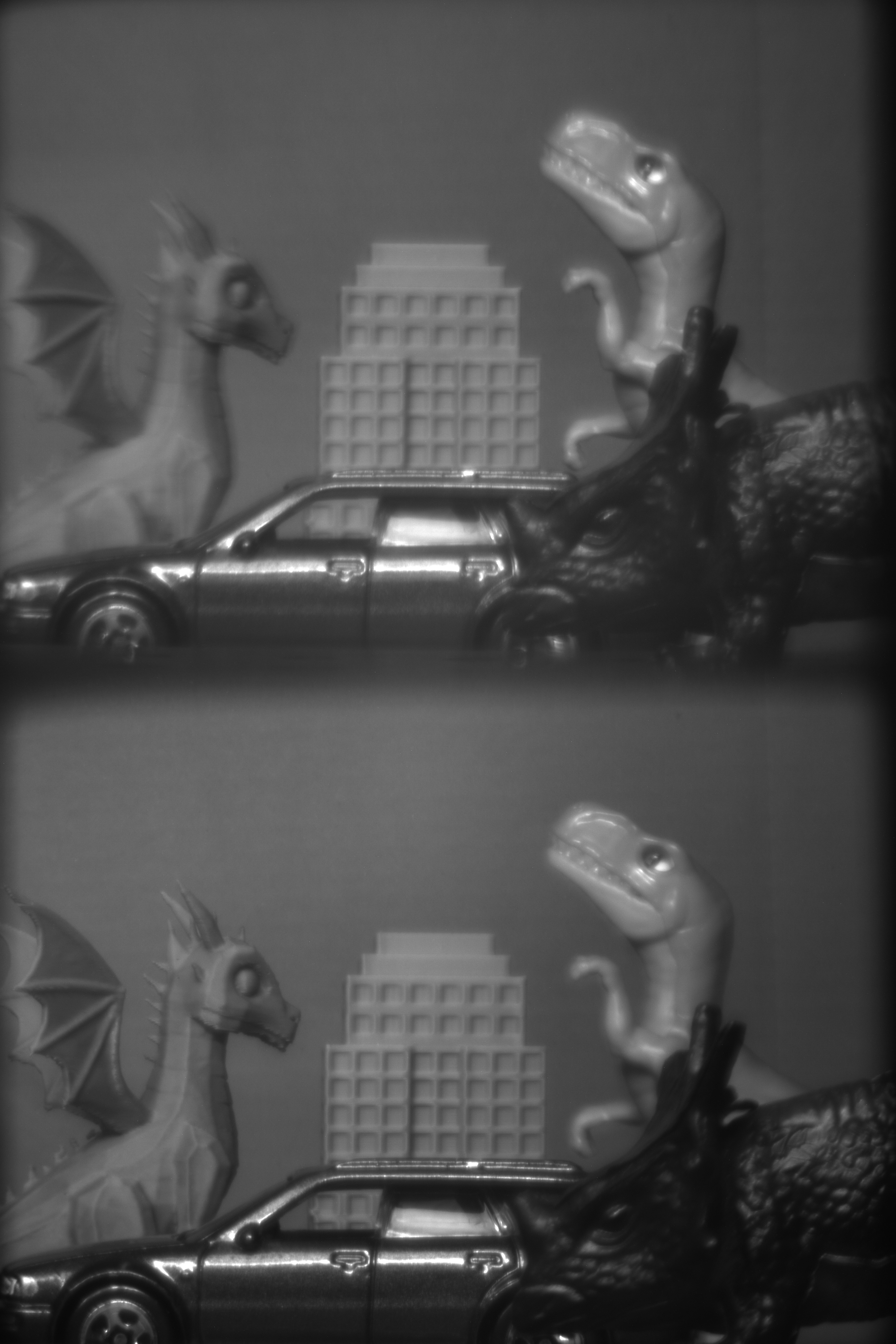}
        \label{fig:eval:hw_setting}
    } 
    \hspace{12pt}
    \subfloat[polarization channels, model input, and approximate label]{
        \includegraphics[width=.41\linewidth, trim={0cm 0.0cm 0cm 0.0cm},clip]{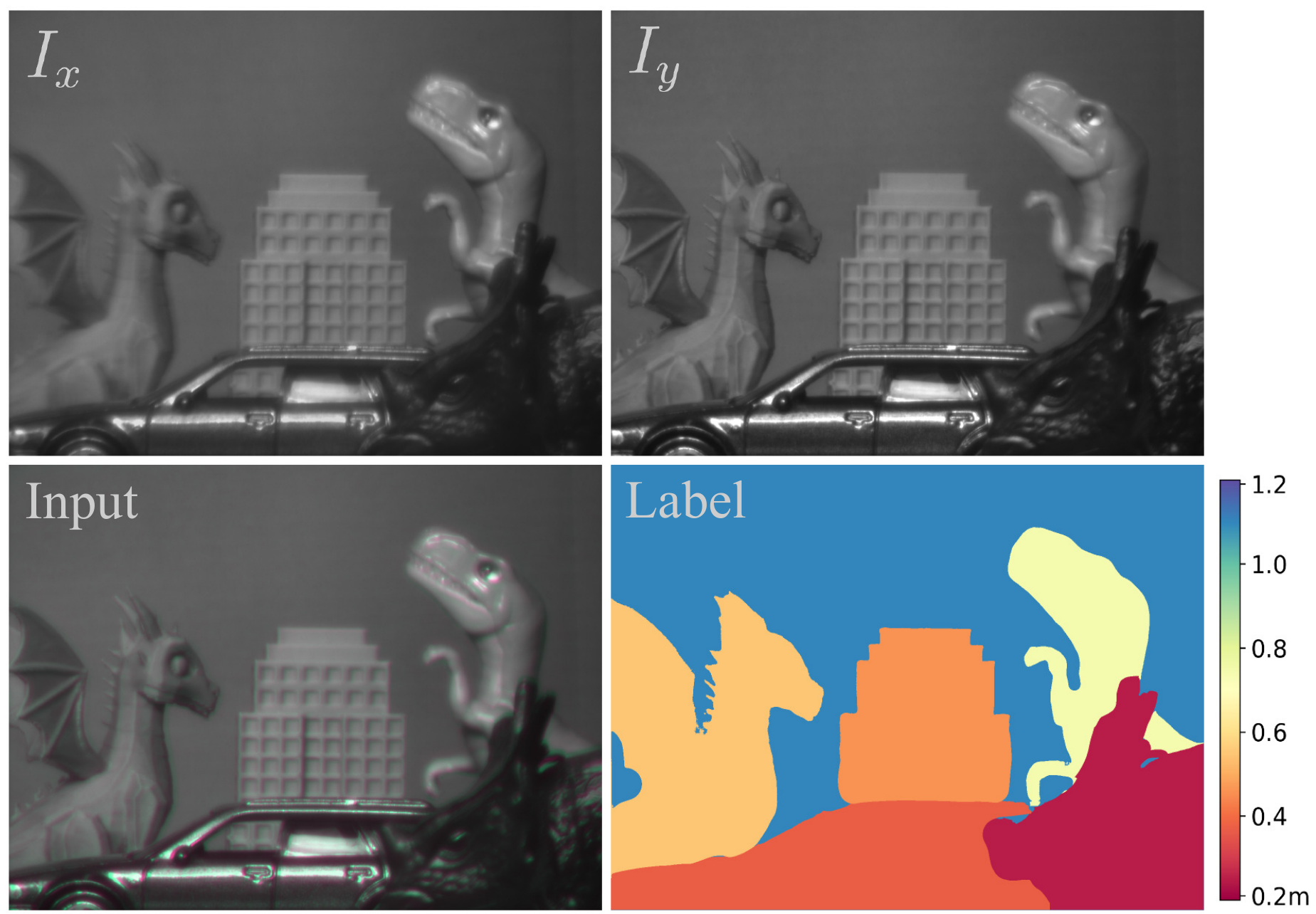}
        \label{fig:eval:label}
    } 
    
    \vspace{-5pt}
    \Caption{Data processing in physical experiments.}{
An example of our data processing workflow. We first crop the raw sensor capture to extract the two polarization channels and compose them into a pseudo-RGB image for model input. We then segment each object and assign its corresponding depth value to generate approximate ground-truth labels.
}
\vspace{-6pt}
    \label{fig:supp:label}
\end{figure*}

 As illustrated in \cref{fig:supp:label}, we begin by cropping the raw sensor capture to isolate the two polarized sub-images and compose them into a pseudo-RGB input for our model. We then manually segment individual objects and assign each region its corresponding depth value to form approximate ground-truth depth maps. Although these annotations are not perfectly precise, they provide sufficiently consistent supervision for validating the stability of our encoded depth cues.

\begin{figure*}[h!]
\centering
\includegraphics[width=0.94\linewidth]{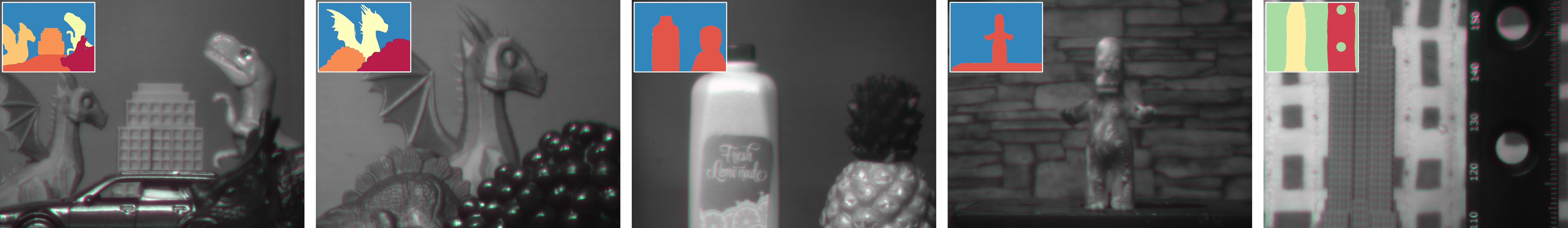}
\vspace{-5pt}
\Caption{Five real samples incorporated into the training set.}{The top-left image shows the corresponding annotated ground-truth depth.} 
\label{fig:supp:real_for_train}
\vspace{-10pt}
\end{figure*}
\begin{figure}[b!]
\centering
\includegraphics[width=0.98\linewidth, trim={0cm 0.0cm 0cm 10.0cm}]{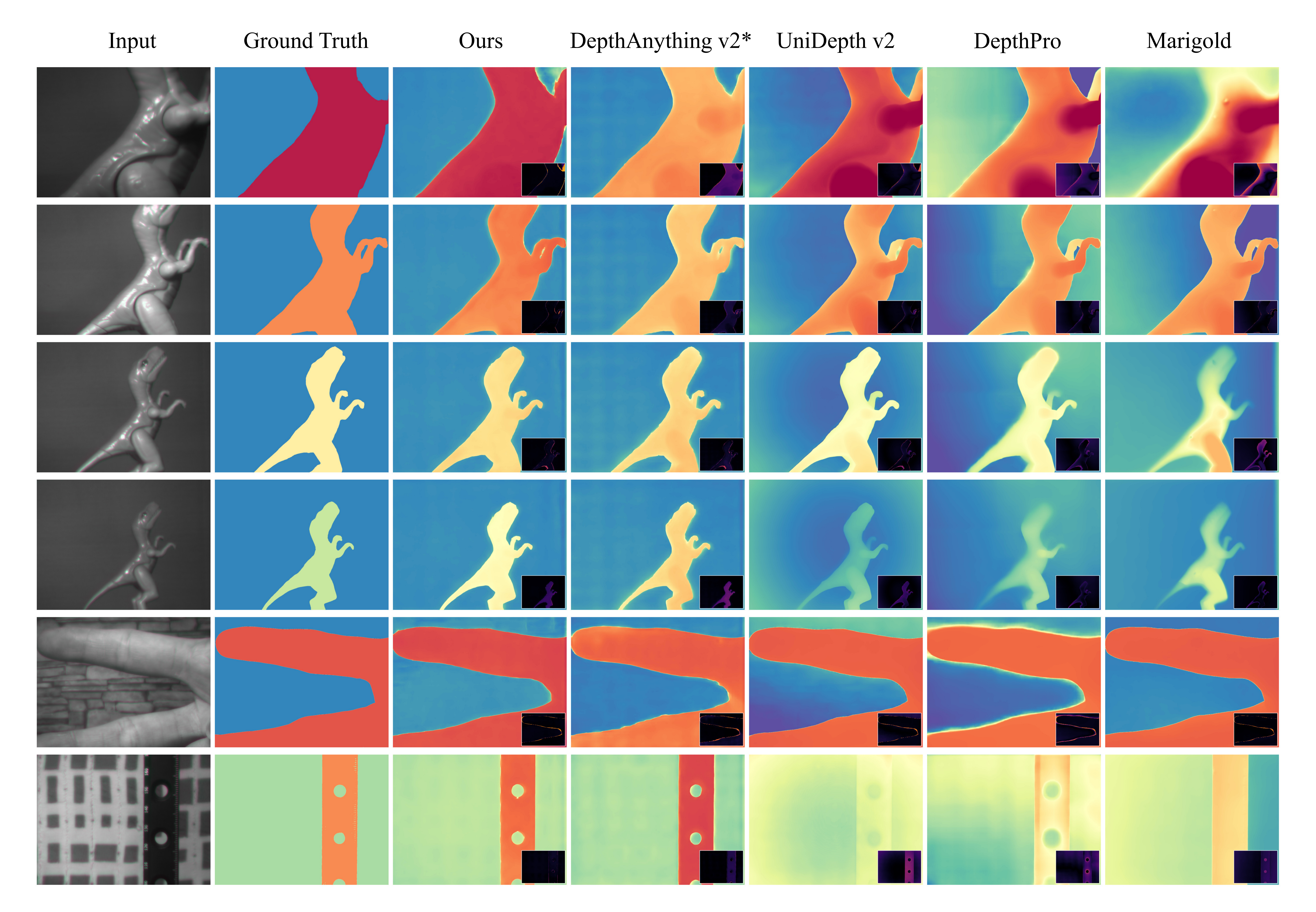}

\Caption{Qualitative results of physical experiment.}{
Bottom-right insets show error maps, where darker colors indicate lower error. Our results and DepthAnything~V2* (fine-tuned on our dataset) are visualized directly, while other methods are post-aligned to ground truth.
}

\label{fig:supp:results_physical}
\end{figure}
\begin{figure*}[]
\centering
\includegraphics[width=1.0\linewidth]{images/qual_phy_2.pdf}
\Caption{Additional qualitative results of physical experiments.}{
We show extended results complementing those shown in \cref{fig:supp:results_physical}.}
\label{fig:supp:results_physical_more}
\end{figure*}
\subsection{Qualitative Results}
We show qualitative results from physical experiments in \cref{fig:supp:results_physical} and \cref{fig:supp:results_physical_more}, comparing our method with fine-tuned DepthAnything~V2 and other baselines. After fine-tuning on the same dataset including five real data, DepthAnything~V2 produces clean relative depth but remains inaccurate in metric scale. Other baselines are post-aligned to ground truth, so their visualizations reflect only relative depth. In contrast, our method recovers metric depth directly without alignment and preserves sharp object boundaries. It also generalizes well to unseen objects and unseen depth ranges, demonstrating the strength of the physically encoded cues provided by our metalens.

\section{Discussion on Experiments}
\label{sec:supp:discussion}

\paragraph{Post Alignment.}
To eliminate the inherent scale--shift ambiguity in monocular prediction and more fairly evaluate the baselines' ability to estimate \emph{relative} scene geometry (e.g., object-to-object distance ratios and object size consistency), we apply a per-sample least-squares scale-and-shift alignment to all monocular baselines. The optimal scale and shift $\{s,t\}$ is found to aligns predictions $\hat{D}$ with the ground-truth $D$:
\begin{equation}    
(s^{*}, t^{*}) = \arg\min_{s, t} \|\, s \hat{D} + t - D \,\|_2^2.
\end{equation}

This post alignment significantly improves their reported performance compared with directly computing metrics on raw outputs. As shown in \cref{tab:supp:nyu-depth-v2} to \cref{tab:supp:mit}, fine-tuned DepthAnything v2 exhibits a large gap between its aligned and non-aligned results, illustrating how post alignment can inflate the accuracy of monocular methods. We highlight that our approach is not only accurate in estimating global scale, but also inherently superior in relative depth structure.

\paragraph{Physical experiments.}
The ground-truth annotations in our physical experiments are approximate and contain errors from two sources: (1) many objects are not planar, and (2) the object segmentation is not perfectly accurate. As a result, the quantitative numbers should be interpreted as approximate indicators rather than absolute measurements. Their primary purpose is to validate the correctness and stability of our physically encoded depth cues. Additionally, our current hardware prototype has a limited field of view and F-number, constraining the diversity and scale of physical scenes we can capture. We plan to improve the optical design, refine the calibration and labeling pipeline, and evaluate on a richer range of indoor scenes in future work.

\paragraph{Simulated experiments.}
Our simulated training data currently includes only indoor scenes (\cref{tab:supp:nyu-depth-v2}, 
\cref{tab:supp:hypersim}) and rendered scenes (\cref{tab:supp:sintel}, \cref{tab:supp:hypersim}, \cref{tab:supp:mit}), primarily due to the limited depth range supported by the current metalens design. Consequently, we do not evaluate on large-range outdoor datasets such as KITTI \cite{KITTI2012}. We plan to extend the depth range of our optical system and generate outdoor-scale training data in future work, enabling validation on outdoor datasets and broader real-world scenarios.

\begin{figure*}[b!]
\centering
\includegraphics[width=1.0\linewidth]{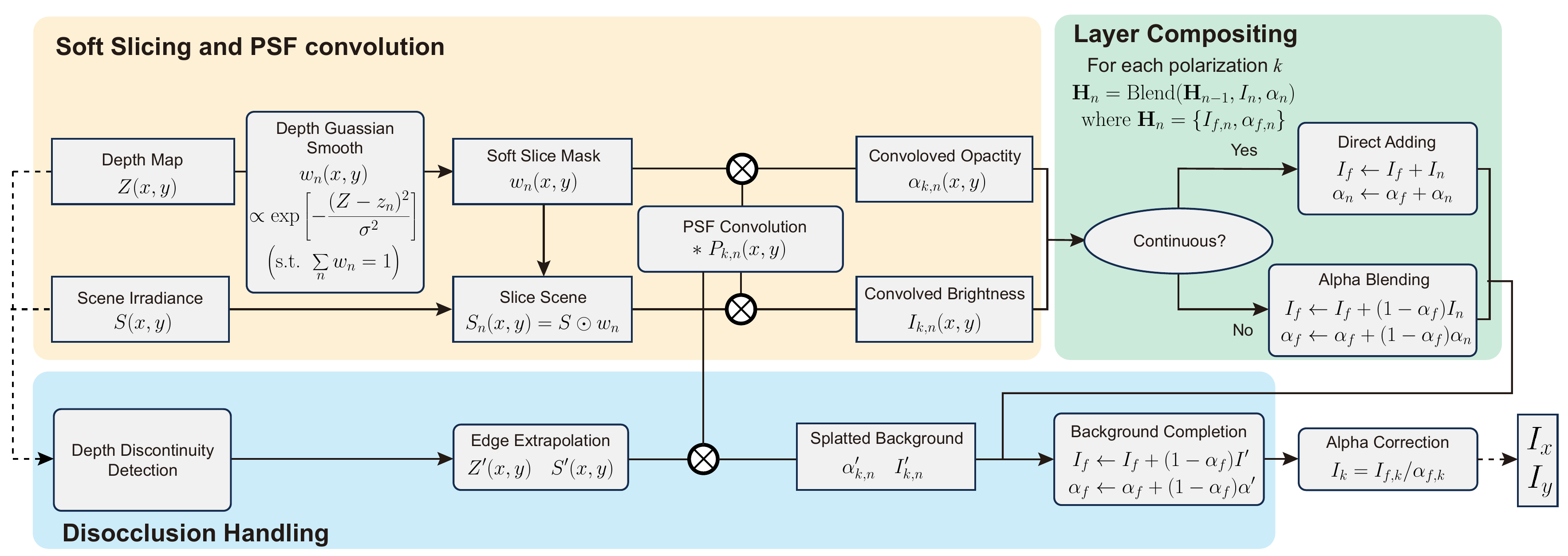}
\Caption{Illustration of our wave-propagation simulator.}
{
}
\label{fig:supp:simulator}
\end{figure*}

\section{Training the Neural Network}
\label{sec:supp:nn}

We use DepthAnything v2 (ViT-Small/Base/Large) as our backbone. Starting from the metric-pretrained weights, we fine-tune on the Hypersim dataset with depth linearly mapped to 0.2–1.2 m, followed by our data-preparation pipeline. The training loss is a combination of $L_1$ and gradient loss, $L = L_1 + 0.5,L_{\mathrm{grad}}$. During training, each input is randomly cropped to 518×518, while inference uses the original sensor resolution; we find the model to be robust to this change in resolution. We additionally incorporate five manually annotated real scenes (see \cref{fig:supp:real_for_train}) into the training set. Each provides ground-truth depth, serving as a small but effective set of real anchors that improves sim-to-real generalization when mixed into training with probability 0.05. We train for 80k steps with a learning rate of $4\times10^{-6}$, using a step learning-rate scheduler that reduces the learning rate by a factor of 0.8 every 10k iterations. Batch sizes are 2 for ViT-Large and 8 for ViT-Small/Base. For our largest model, training requires roughly 20 hours on a single A100 GPU. 

\section{Wave Propagation Simulator}
\label{sec:supp:simulator}
We provide a detailed illustration of our wave propagation simulator in \cref{fig:supp:simulator}. To build the simulator, we first numerically compute the depth-dependent point spread function (PSF) (\cref{sec:supp:numerical}). Subsequently, we develop an optical forward model featuring soft slicing, PSF convolution, and layer compositing to accurately render the image formation and mitigate simulation artifacts (\cref{sec:supp:rendering}). Finally, we address disocclusion through explicit edge extrapolation and background completion (\cref{sec:supp:disocclusion}).

\subsection{Numerical Computation of 3D PSF}
\label{sec:supp:numerical}
Here, we present the method for numerically computing the depth-dependent PSF. Please refer to \cref{sec:meta:psf} for the physical and analytical details.\paragraph{Free-Space Propagator}Light propagation between the metasurface and the sensor is governed by the diffraction formula in \cref{eq:meta:diffraction}. This diffraction integral can be formulated as a convolution between the complex transmission field of the metalens, $\outE(x_m,y_m)$, and the free-space impulse response $h(x_m,y_m)$:
\begin{align}
&U(x_i, y_i) \nonumber \\ 
&= \iint_{-\infty}^{\infty} \outE(x_m, y_m) h(x_i-x_m, y_i-y_m) \, dx_m \, dy_m \nonumber \\
&= \outE(x_m, y_m) \ast h(x_m, y_m),
\end{align}
where $(x_i,y_i)$ denotes the coordinates on the sensor plane, $(x_m, y_m)$ represents the coordinates on the metalens plane, and $h(x,y)$ is given by:
\begin{equation}
h(x, y) = \frac{z}{i\lambda} \frac{\exp\left(ik\sqrt{x^2 + y^2 + \Delta z^2}\right)}{x^2 + y^2 + \Delta z^2}
\end{equation}
where $\lambda$ is the wavelength, $k = 2\pi/\lambda$ is the wavenumber, and $\Delta z$ is the distance between the metasurface and the sensor. To reduce the computational complexity, we apply the convolution theorem to evaluate this integral in the frequency domain:
\begin{equation}
U(x_i, y_i) = \mathcal{F}^{-1} \{ \mathcal{F} \{ \outE(x_m, y_m) \} \cdot \mathcal{F} \{ h(x_m, y_m) \} \}
\end{equation}
where $\mathcal{F}$ and $\mathcal{F}^{-1}$ denote the forward and inverse Fast Fourier Transforms (FFT), respectively. Here, $H = \mathcal{F} \{ h \}$ denotes the Transfer Function (or Angular Spectrum propagator) of free space. Since $H$ is independent of the input field, it can be pre-computed to improve efficiency. Given the wavelength dependence of $H$, we sample the operating spectrum using 5 discrete wavelengths centered at 590 nm.

\paragraph{Depth-Dependent PSF Library} We discretize the depth range of 20--120 cm into 400 steps to compute the depth-dependent PSF. Specifically, for each depth $z$, we model a point source located on the optical axis. The field transmitted through the metasurface is calculated as the product of the incident spherical wavefront and the metasurface phase modulation, $\exp(i\phi_{m,x})$. We focus exclusively on the x-polarization channel, as the y-polarized PSF is simply a $180^\circ$ rotation of the x-polarized counterpart. A detailed comparison between the simulated and experimentally measured PSFs is shown in \cref{fig:psfdetail}.

\subsection{Optical Forward Model}
\label{sec:supp:rendering}
To accurately model the imaging process using depth-dependent PSFs, our optical forward model transforms the input depth map $Z(x,y)$ and scene irradiance $S(x,y)$ into sensor-plane measurements. This process consists of three key stages: depth soft slicing, PSF convolution, and a hybrid layer compositing strategy.
\paragraph{Depth Gaussian Smoothing} To mitigate discretization artifacts arising from hard depth binning, we first apply Gaussian smoothing along the $z$-direction. For a given pixel $(x,y)$, its contribution to the $n$-th depth slice at distance $z_n$ is determined by a soft slice mask $w_n(x,y)$. This weight is computed using a normalized Gaussian function centered at the pixel's true depth $Z(x,y)$:
\begin{align}
w_n(x, y) = &\frac{1}{\mathcal{N}} \exp \left[ - \frac{(Z(x,y) - z_n)^2}{\sigma^2} \right],\nonumber \\ &\quad \text{s.t.} \sum_{n} w_n(x,y) = 1
\end{align}
where $\sigma$ controls the smoothness of the distribution, and $\mathcal{N}$ is the normalization factor ensuring energy conservation.
\paragraph{Scene Slicing and PSF Convolution} Next, we define the sliced scene irradiance $S_n(x,y)$ by modulating the total irradiance with the soft mask: $S_n(x,y) = S(x,y) \odot w_n(x,y)$. We then project these slice contributions onto the image plane by convolving them with the depth-dependent PSF, $P_{k,n}(x,y)$, for polarization channel $k$. This yields the projected brightness $I_{k,n}$ and the layer opacity $\alpha_{k,n}$ for each slice:
\begin{align}I_{k,n}(x, y) &= S_n(x, y) \ast P_{k,n}(x, y) \\ \alpha_{k,n}(x, y) &= w_n(x, y) \ast P_{k,n}(x, y)\end{align}
Here, the opacity map $\alpha_{k,n}$ represents the blur kernel's footprint, which is crucial for correctly handling occlusions.
\paragraph{Hybrid Layer Compositing} Finally, we accumulate the convolved slices front-to-back to form the final image. We maintain an accumulation state $\mathbf{H}_n = \{I_{f,n}, \alpha_{f,n}\}$ representing the foreground brightness and opacity. To address artifacts at surface boundaries, we employ a hybrid compositing strategy based on surface continuity:
\begin{equation}
\begin{bmatrix} I_f \\ \alpha_f \end{bmatrix} \leftarrow
\begin{cases} 
\begin{bmatrix} I_f + I_n \\ \alpha_f + \alpha_n \end{bmatrix} & \text{if Cont.} \\ \\ 
\begin{bmatrix} I_f + (1-\alpha_f)I_n \\ \alpha_f + (1-\alpha_f)\alpha_n \end{bmatrix} & \text{if Discont.}
\end{cases}
\end{equation}
To robustly distinguish between surface continuity and occlusion, we maintain a record of the last updated depth, $z_{\text{last}}$, for each pixel. We introduce a depth threshold $\tau$ (e.g., 3 cm) as the decision criterion. For the current slice at depth $z_n$, we calculate the depth interval $\Delta z = |z_n - z_{\text{last}}|$. If $\Delta z < \tau$, the current slice is considered part of the same continuous surface as the previous accumulation. In this case, we use direct adding to integrate the energy spread across adjacent bins. Conversely, if $\Delta z \ge \tau$, it indicates a significant depth jump, implying a discontinuity or a new object entering the line of sight. Here, we switch to alpha blending to correctly handle the occlusion relationships.

\subsection{Disocclusion Solution}
\label{sec:supp:disocclusion}

Our strategy for handling disocclusion involves identifying pixels along \textbf{depth discontinuities} and extrapolating their \textbf{background properties} into the occluded regions. The complete procedure is detailed in \cref{algorithm}. We begin by normalizing the input depth map $\mathbf{D}$ and extracting the edge map $\mathbf{E}$ alongside gradient orientations $\mathbf{\Theta}$ using Sobel operators (with threshold $\tau_{\text{edge}}$). For each edge pixel, we perform an outward trace along the direction derived from $\mathbf{\Theta}$ to sample the local background depth and intensity ($\mathbf{D}_{\text{bg}}, \mathbf{G}_{\text{bg}}$). To generate spatially coherent dense maps ($\mathbf{D}_{\text{fill}}, \mathbf{G}_{\text{fill}}$), we propagate these sparse samples into a surrounding band $\mathbf{M}_{\text{band}}$ via masked Gaussian convolution. The final extension mask $\mathbf{M}_{\text{ext}}$ is derived by verifying that the filled depth $\mathbf{D}_{\text{fill}}$ is significantly \textit{farther} than the original depth $\mathbf{D}_{\text{norm}}$ (controlled by $\tau_{\text{depth}}$), thereby isolating valid disocclusion areas. Finally, these regions are convolved with the PSF and blended with the accumulated rendering to complete the background.

\begin{algorithm}[t]
\caption{Depth Edge-Based Background Extension}
\label{algorithm}
\begin{algorithmic}[1]

\Require Gray $\mathbf{G}$, Depth $\mathbf{D}$, Radius $N$,
\Statex \hphantom{\textbf{Require:} } Thresholds $\tau_{\text{edge}}, \tau_{\text{depth}}$
\Ensure Mask $\mathbf{M}_{\text{ext}}$, Gray $\mathbf{G}_{\text{ext}}$, Depth $\mathbf{D}_{\text{ext}}$

\Statex
\State \textbf{1. Gradient-Based Edge Extraction}
\State $\mathbf{D}_{\text{norm}} \gets \text{NORMALIZE}(\mathbf{D})$
\State $(\mathbf{g}_x, \mathbf{g}_y) \gets \text{SOBEL}(\mathbf{D}_{\text{norm}})$
\State $\mathbf{M}_{\text{mag}} \gets \sqrt{\mathbf{g}_x^2 + \mathbf{g}_y^2}$
\State $\mathbf{E} \gets \text{MORPH\_CLOSE}(\mathbf{M}_{\text{mag}} > \tau_{\text{edge}}, 2)$
\State $\mathbf{\Theta} \gets \text{ARCTAN2}(\mathbf{g}_y, \mathbf{g}_x)$

\Statex
\State \textbf{2. Background Sampling via Tracing}
\State $\mathbf{D}_{\text{bg}}, \mathbf{G}_{\text{bg}} \gets \text{INIT\_NAN}(H, W)$
\ForAll{pixel $\mathbf{p}$ where $\mathbf{E}(\mathbf{p})$ is True}
    \State $\mathbf{v} \gets (\cos(\mathbf{\Theta}_\mathbf{p}), \sin(\mathbf{\Theta}_\mathbf{p}))$
    \State $(d^*, g^*) \gets \text{TRACE}(\mathbf{p}, \mathbf{v}, \mathbf{D}_{\text{norm}}, \mathbf{G}, N)$
    \State $\mathbf{D}_{\text{bg}}(\mathbf{p}) \gets d^*$; $\quad \mathbf{G}_{\text{bg}}(\mathbf{p}) \gets g^*$
\EndFor

\Statex
\State \textbf{3. Sparse-to-Dense Propagation}
\State $\mathbf{M}_{\text{band}} \gets \text{DIST\_TRANS}(\neg \mathbf{E}) \leq N$
\State $\mathbf{M}_{\text{valid}} \gets \neg \text{IS\_NAN}(\mathbf{D}_{\text{bg}})$
\State $\mathbf{D}_{\text{fill}} \gets \text{MASKED\_GAUSS}(\mathbf{D}_{\text{bg}}, \mathbf{M}_{\text{valid}})$
\State $\mathbf{G}_{\text{fill}} \gets \text{MASKED\_GAUSS}(\mathbf{G}_{\text{bg}}, \mathbf{M}_{\text{valid}})$

\Statex
\State \textbf{4. Disocclusion Masking}
\State $\mathbf{M}_{\text{depth}} \gets \mathbf{D}_{\text{fill}} > \mathbf{D}_{\text{norm}} \cdot (1 + \tau_{\text{depth}})$
\State $\mathbf{M}_{\text{ext}} \gets \mathbf{M}_{\text{band}} \wedge \mathbf{M}_{\text{depth}}$
\State $\mathbf{tmp} \gets \text{APPLY}(\mathbf{D}_{\text{fill}}, \mathbf{M}_{\text{ext}})$
\State $\mathbf{D}_{\text{ext}} \gets \text{RESCALE}(\mathbf{tmp}, \mathbf{D})$
\State $\mathbf{G}_{\text{ext}} \gets \text{APPLY}(\mathbf{G}_{\text{fill}}, \mathbf{M}_{\text{ext}})$

\State \textbf{return} $\mathbf{M}_{\text{ext}}, \mathbf{G}_{\text{ext}}, \mathbf{D}_{\text{ext}}$

\end{algorithmic}
\end{algorithm}

\section{Comparison of PSF Designs}
\label{sec:supp:compare}

In this section, we compare our polarization-multiplexed single-helix PSF with several alternative designs to analyze its advantages and limitations. This comparison also helps explain why our PSF provides effective depth-encoding cues for a depth foundation model, leading to better performance in the ablation study in \cref{tab:ablation_compare_with_dfd} of the main text.

We consider three representative baselines: (1) a deep-learning-optimized depth-from-defocus PSF (DeepDfD)~\cite{ikoma2021depth}, (2) an ideal double-helix PSF (DH-PSF) with a total rotation of \(120^\circ\) over the target depth range ~\cite{rotationdepth,rotationdepth2}, and (3) the PSF of a conventional lens. These baselines represent learned, engineered, and standard optical designs, respectively.

For a fair comparison, all PSFs are evaluated over the same depth range from 1~m to 5~m. Our PSF is generated using the rotation phase in \cref{eq:method:rotationphase}, with a diameter of 5~mm, a focal length of 50~mm, and a focus distance of 1.7~m. This same parameter setting is also used in the comparisons reported in \cref{tab:compare_with_dfd} and \cref{tab:ablation_compare_with_dfd} of the main text. These parameters are chosen to be close to those of DeepDfD, which uses a focal length of 50~mm and a DOE aperture of 5.6~mm. The conventional lens baseline uses the same optical parameters.

We organize the comparison into three parts. First, we compare the PSFs directly using an on-axis point source to evaluate their intrinsic depth encoding. Second, we use a step edge with various orientations to represent more general extended boundaries and textures. Third, we compare the global similarity of the PSFs across depth to evaluate ambiguity in the encoded depth cues.

\begin{figure*}[h]
\centering
\includegraphics[width=1.0\linewidth]{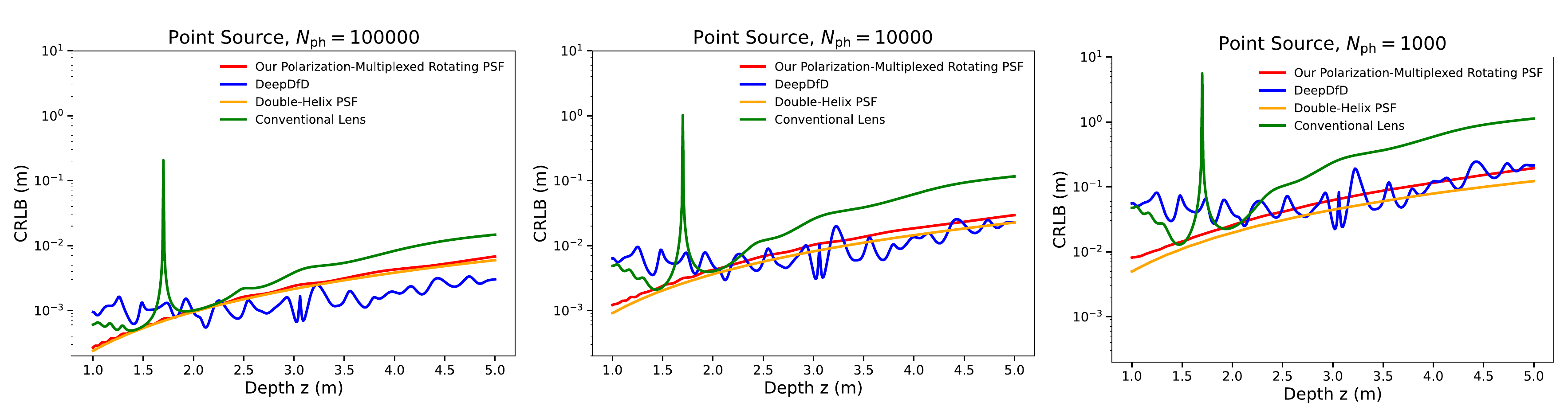}
\Caption{CRLB comparison of different PSF designs under varying photon budgets.}{Lower photon budgets correspond to higher relative noise levels. Our rotating PSF exhibits more robust depth encoding, as evidenced by its smoother CRLB trend and reduced sensitivity to photon-budget variation.}

\label{fig:supp:CRLB}
\end{figure*}

\subsection{Point-source Comparison of PSF designs}
To directly compare the depth-encoding capability of different PSFs, we first consider the Fisher information of the image formed by an on-axis point source. Let the unknown parameter vector be \(\boldsymbol{\theta} = (\theta_1,\theta_2,\dots)\), which may include depth and other variables of interest. Under a Poisson image-formation model with background, the Fisher information matrix (FIM) is given by~\cite{ghanekar2022ps}
\begin{equation}
\mathbf{I}(\boldsymbol{\theta})_{ij}
=
\sum_{k=1}^{N_{\mathrm{pix}}}
\frac{1}{\mu_{\boldsymbol{\theta}}(k)+\beta}
\frac{\partial \mu_{\boldsymbol{\theta}}(k)}{\partial \theta_i}
\frac{\partial \mu_{\boldsymbol{\theta}}(k)}{\partial \theta_j},
\label{eq:supp:fim_general}
\end{equation}
where \(N_{\mathrm{pix}}\) is the total number of sensor pixels, \(k\) indexes the pixels, \(\mu_{\boldsymbol{\theta}}(k)\) is the expected photon count at pixel \(k\) for a point source under parameter \(\boldsymbol{\theta}\), and \(\beta\) is the background noise level. More specifically, \(\mu_{\boldsymbol{\theta}}(k)\) is the normalized PSF or image scaled by the total photon number \(N_{\mathrm{ph}}\), i.e.,
\begin{equation}
\mu_{\boldsymbol{\theta}}(k)=N_{\mathrm{ph}}\,h_{\boldsymbol{\theta}}(k),
\end{equation}
where \(h_{\boldsymbol{\theta}}(k)\) denotes the normalized PSF or image intensity at pixel \(k\), satisfying \(\sum_{k=1}^{N_{\mathrm{pix}}} h_{\boldsymbol{\theta}}(k)=1\). Physically, \(\partial \mu_{\boldsymbol{\theta}}(k)/\partial \theta_i\) describes how sensitively the image intensity at each pixel changes with respect to parameter \(\theta_i\). Therefore, the FIM quantifies how strongly the recorded image responds to changes in the underlying parameters.

The Cram\'er--Rao lower bound (CRLB) is obtained from the inverse of the FIM,
\begin{equation}
\mathrm{Cov}(\hat{\boldsymbol{\theta}})
\succeq
\mathbf{I}(\boldsymbol{\theta})^{-1},
\label{eq:supp:crlb_general}
\end{equation}
where \(\hat{\boldsymbol{\theta}}\) is an unbiased estimator of \(\boldsymbol{\theta}\). The diagonal elements of \(\mathbf{I}^{-1}\) give the minimum achievable variances of the corresponding parameters,
\begin{equation}
\mathrm{Var}(\hat{\theta}_i)\ge
\left[\mathbf{I}(\boldsymbol{\theta})^{-1}\right]_{ii},
\qquad
\mathrm{CRLB}_{\theta_i}
=
\sqrt{
\left[\mathbf{I}(\boldsymbol{\theta})^{-1}\right]_{ii}
}.
\label{eq:supp:crlb_diag}
\end{equation}

In the special case where only a single parameter is estimated, such as the depth \(z\), the FIM reduces to a scalar,
\begin{equation}
I(z)
=
\sum_{k=1}^{N_p}
\frac{1}{\mu_z(k)+\beta}
\left(
\frac{\partial \mu_z(k)}{\partial z}
\right)^2,
\label{eq:supp:fim_depth}
\end{equation}
and the corresponding CRLB becomes
\begin{equation}
\mathrm{CRLB}_{z}
=
\frac{1}{\sqrt{I(z)}}.
\label{eq:supp:crlb_depth}
\end{equation}
This scalar form is useful for directly comparing the intrinsic depth sensitivity of different PSF designs in the point-source setting.

For the point-source comparison, we fix the background noise level to \(\sigma = 10\) and set the background Poisson level as \(\beta=\sigma^2\), and then compute the depth CRLB under different photon budgets \(N_{\mathrm{ph}}\). As shown in \cref{fig:supp:CRLB}, all engineered PSFs outperform the conventional lens over most of the evaluated depth range. The conventional lens only provides good depth sensitivity near the focal region, but performs poorly both exactly at focus and away from focus, where the image changes little with depth. This leads to large CRLB values and even singular behavior around the focus distance. For this reason, we do not include the conventional lens in the following comparisons. 

Both rotating PSFs, namely our PSF and the DH-PSF, perform better at shorter distances, with the CRLB gradually increasing as \(z\) increases. This trend is expected because the rotation angle is intrinsically linear with \(1/z\), so the PSF changes more rapidly with depth in the near range than in the far range. In other words, these PSFs are naturally more depth-sensitive at short distances. This also explains why, in our physical prototype, we focus on the range of \(0.2\)--\(1.2\)~m, where the rotating-PSF design is better matched to the target depth range. 

DeepDfD performs reasonably well when the photon budget is high, e.g., at \(N_{\mathrm{ph}}=10^5\). However, its performance degrades quickly as the photon budget decreases to \(10^3\), corresponding to a lower signal-to-noise ratio (SNR). In contrast, the two rotating PSFs are less sensitive to the photon budget and show a more stable CRLB trend across noise levels. This robustness comes from the fact that they encode depth mainly through PSF position change and rotation, rather than through weaker blur variations. As a result, the encoded depth cue remains more distinguishable under relative noise, which is an important advantage of our design.

\begin{figure*}[h]
\centering
\includegraphics[width=1.0\linewidth]{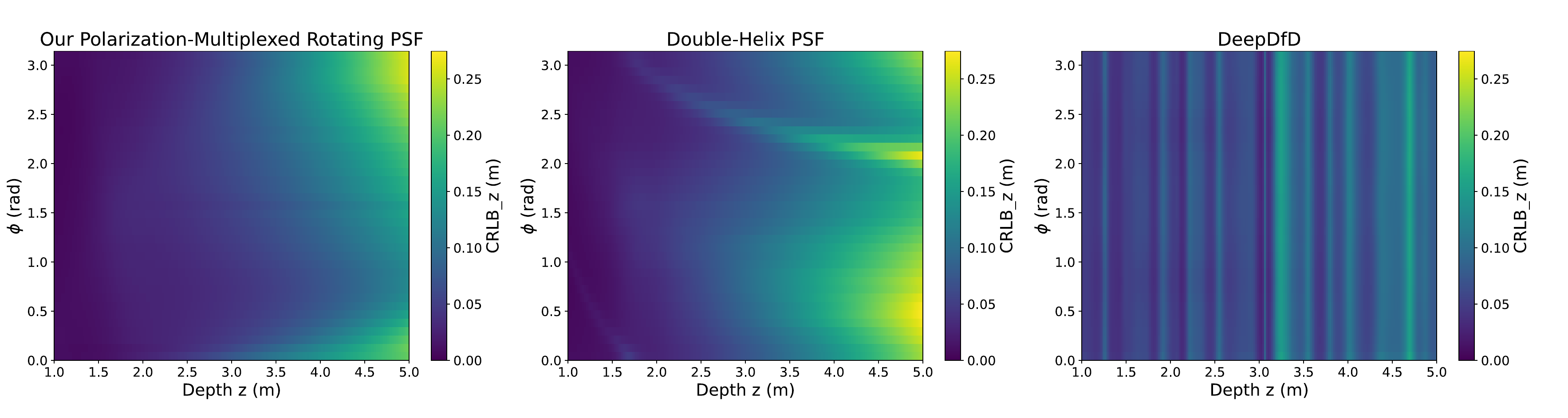}
\Caption{CRLB comparison of different PSF designs for a step edge with respect to depth $z$ and orientation $\phi$}

\label{fig:supp:stepedge}
\end{figure*}
\begin{figure*}[h]
\centering
\includegraphics[width=1.0\linewidth]{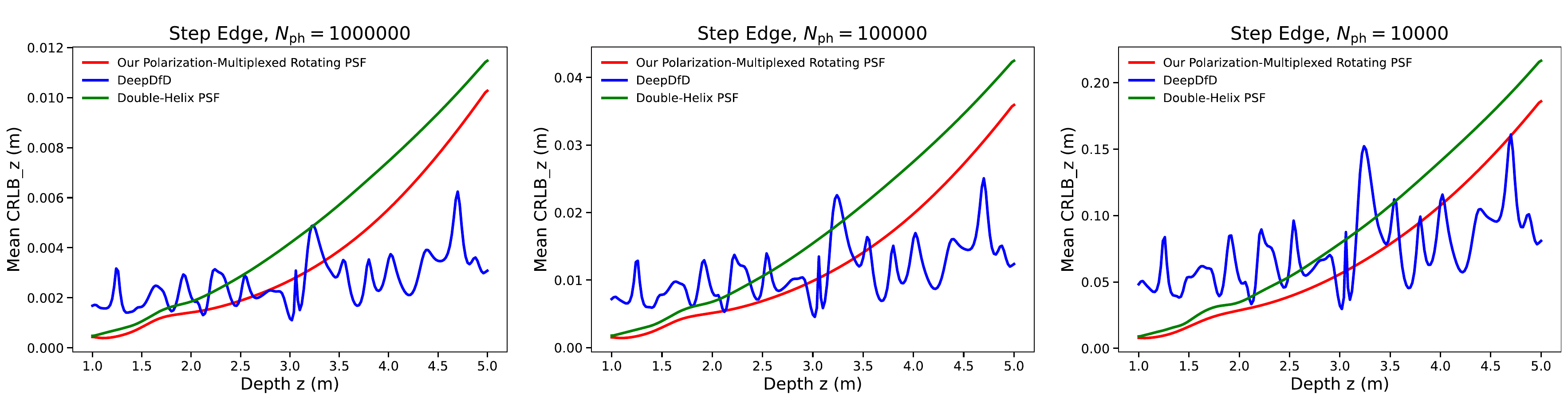}
\Caption{Mean depth CRLB comparison of different PSF designs for a step edge under varying photon budgets }

\label{fig:supp:stepmean}
\end{figure*}

\subsection{Step-edge Comparison of PSF Designs}
To move beyond the point-source setting, we next consider a step edge as a simple model for general scene boundaries and local texture transitions. Let \((x,y)\) denote the image-plane coordinates. A step edge with orientation \(\phi\) can be defined as
\begin{equation}
s_{\phi}(x,y)=H(x\cos\phi+y\sin\phi),
\label{eq:supp:step_edge}
\end{equation}
where \(H(\cdot)\) is the Heaviside step function. This represents a straight edge passing through the origin, with one side bright and the other side dark. Although simple, this model captures the local structure of many boundaries and texture changes in real scenes.

In this case, the unknown parameters are \((z,\phi)\), and the CRLB is computed in the two-dimensional parameter space of depth and edge orientation. Let \(\mathbf{I}(z,\phi)\) denote the Fisher information matrix defined in \cref{eq:supp:fim_general}. The depth CRLB is then given by
\begin{equation}
\mathrm{CRLB}_{z}(z,\phi)
=
\sqrt{\left[\mathbf{I}(z,\phi)^{-1}\right]_{11}},
\label{eq:supp:crlb_edge}
\end{equation}
which explicitly depends on both \(z\) and \(\phi\). This quantity measures the best achievable depth precision when the edge orientation is also unknown and jointly estimated.

The results in \cref{fig:supp:stepedge} show how \(\mathrm{CRLB}_{z}\) varies with both depth and edge orientation for the three engineered PSFs. Our polarization-multiplexed single-helix PSF exhibits a relatively smooth dependence on \((z,\phi)\), indicating stable depth encoding across different edge orientations. DeepDfD is nearly independent of \(\phi\), which is expected because its PSF is radially symmetric. By contrast, the DH-PSF shows clear degradation at specific depths and orientations. This behavior arises because the DH-PSF forms two separated lobes, and when the edge orientation aligns with the direction of lobe motion, the encoded depth cue becomes less distinguishable, resulting in a larger depth CRLB. 

To summarize the overall trend, we further average \(\mathrm{CRLB}_{z}(z,\phi)\) over \(\phi\), and plot the mean depth CRLB as a function of \(z\) in \cref{fig:supp:stepmean}. Our polarization-multiplexed single-helix PSF consistently outperforms the DH-PSF after angular averaging, mainly because it avoids the ghosting-like ambiguity associated with the two-lobe structure. As in the point-source case, the rotating PSFs perform better in the near range and gradually degrade with increasing distance, while remaining relatively insensitive to the photon budget. This again shows that depth cues encoded through PSF motion are more robust to relative noise than those based primarily on blur variation.

\begin{figure*}[h]
\centering
\includegraphics[width=1.0\linewidth]{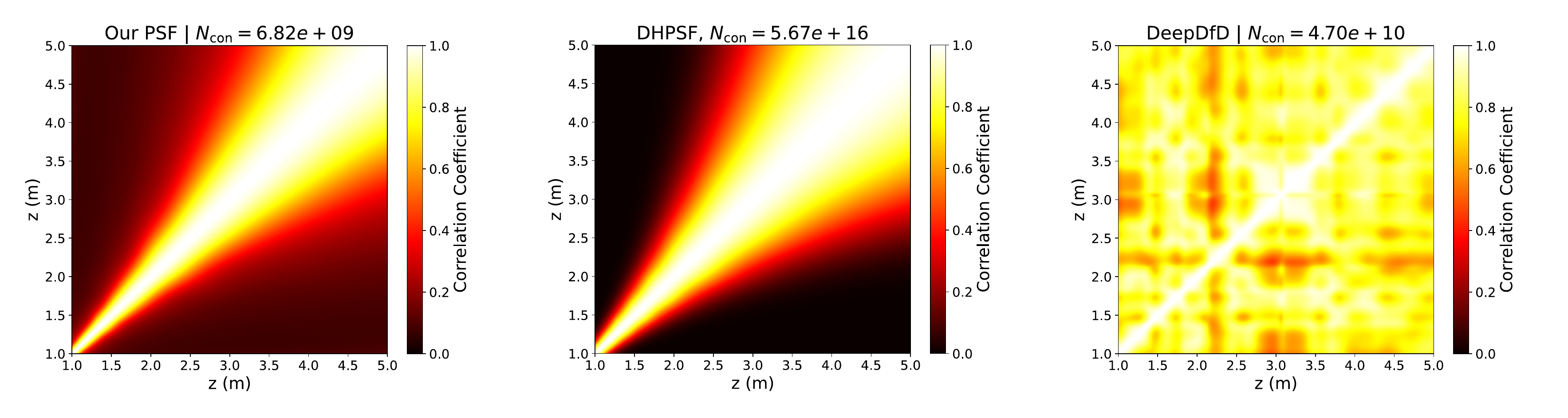}
\Caption{Cross-depth correlation comparison of different PSF designs.}{Our PSF exhibits the lowest cross-depth correlation, indicating the least depth ambiguity and, consequently, the smallest condition number among the compared designs.}

\label{fig:supp:correlation}
\end{figure*}

\subsection{Global Similarity Analysis of PSF Designs}

The CRLB measures the local sensitivity of a PSF to depth at a given depth value. However, depth estimation from a PSF can also be viewed as a Tikhonov-regularized least-squares inverse problem, as discussed in ~\cite{ikoma2021depth}. From this perspective, it is also important to consider the global similarity of the PSF across different depths. Lower similarity between PSFs at different depths reduces ambiguity in the inverse problem and improves its conditioning.

To quantify this effect, we compute the pairwise correlation between PSFs at different depths for the three engineered designs, as shown in \cref{fig:supp:correlation}. Compared with the DH-PSF and DeepDfD, our PSF exhibits the lowest overall cross-depth correlation. This behavior arises because the bright center of our PSF rotates continuously with depth, so the most informative high-intensity region moves to substantially different image locations at different depths. As a result, PSFs from different depths are less similar to each other, which reduces depth ambiguity. By comparison, the DH-PSF still retains stronger structural similarity across depth despite its rotation, while DeepDfD shows even higher global similarity due to its largely blur-based encoding.

This difference is also reflected in the condition number. The condition number of our PSF is only about one-seventh that of the DeepDfD PSF, indicating a substantially better-conditioned inverse problem. In other words, our PSF provides lower-ambiguity depth encoding, which is more favorable for a depth foundation model to extract reliable and less ambiguous depth information.

\begin{figure*}[]
\newcommand{\subW}{0.1}
\newcommand{\subH}{35pt}
\newcommand{\RowSpacing}{\vspace{10pt}}
\newcommand{\mysideCaption}{
  \hspace{1pt}
  \raisebox{35pt}{\parbox[c]{10pt}{\rotatebox{-90}{\scriptsize Measured \hspace{3.5pt} Simulated }}}
}
\captionsetup[subfloat]{labelformat=empty}
\centering
  \subfloat[depth = 20 cm]{
  \begin{tabular}[b]{c}
  \includegraphics[height=\subH]{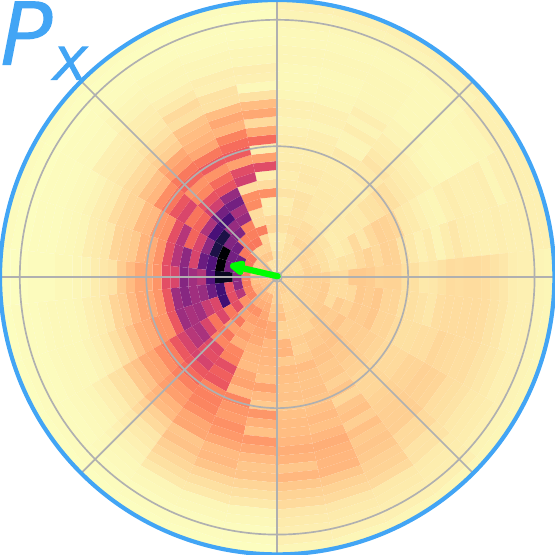}
  \includegraphics[height=\subH]{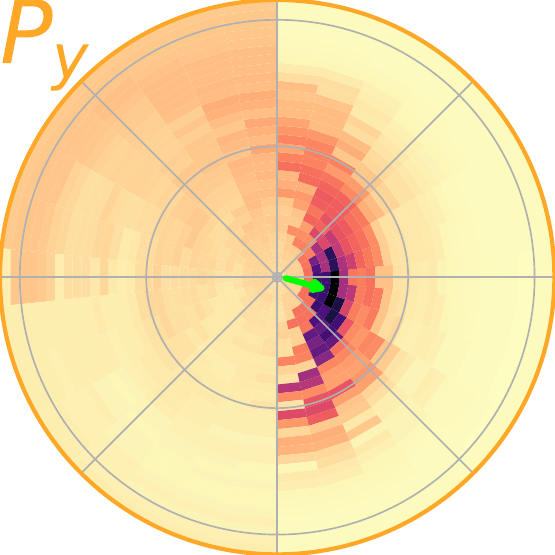}\\            
  \includegraphics[height=\subH]{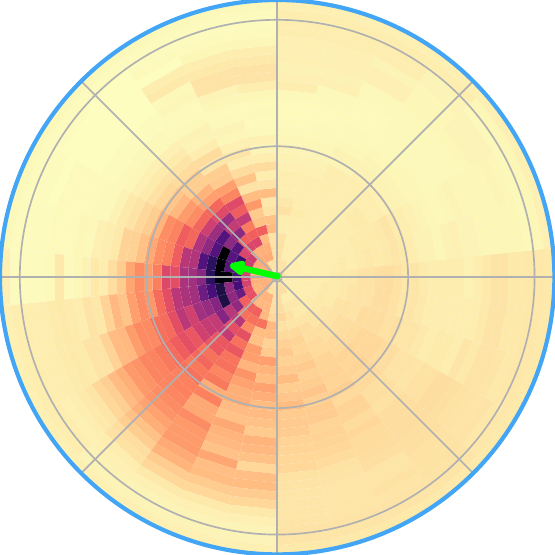}
  \includegraphics[height=\subH]{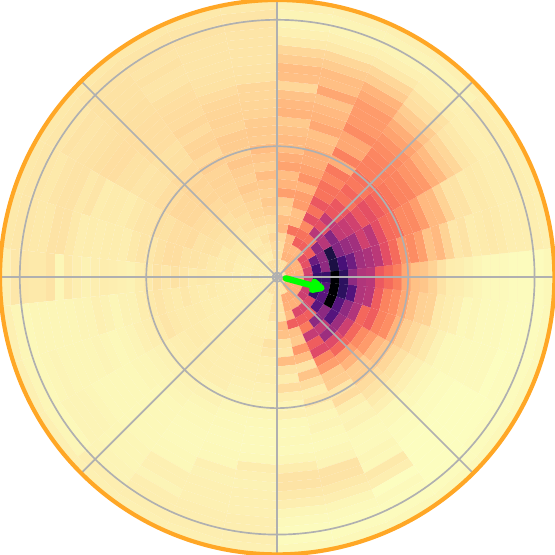}
  \end{tabular}
  }
  \subfloat[depth = 30 cm]{
  \begin{tabular}[b]{c}
  \includegraphics[height=\subH]{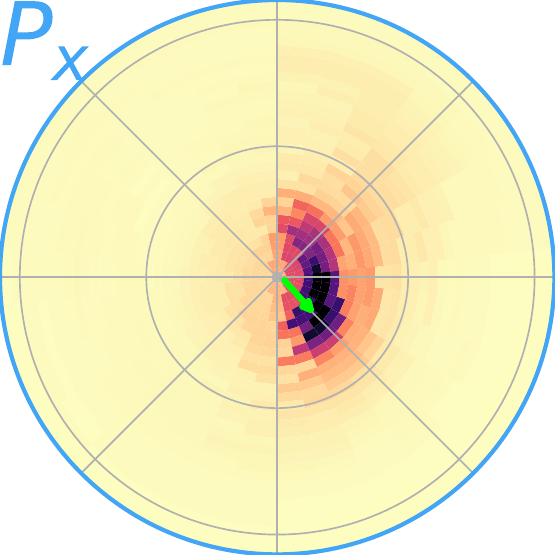}
  \includegraphics[height=\subH]{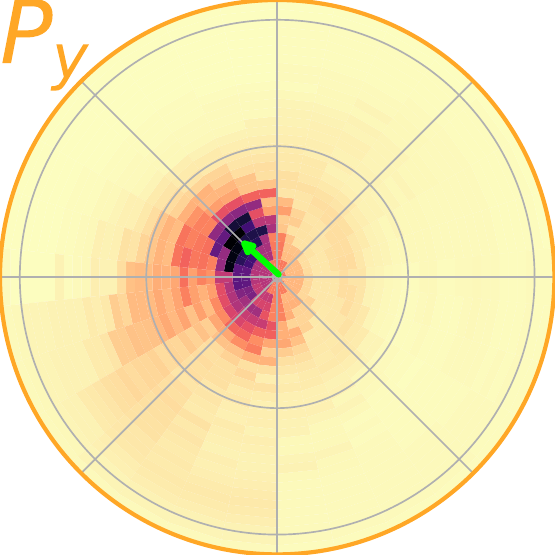}\\            
  \includegraphics[height=\subH]{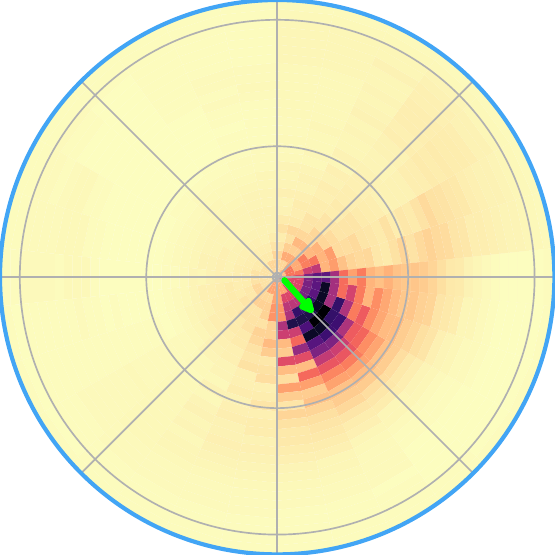}
  \includegraphics[height=\subH]{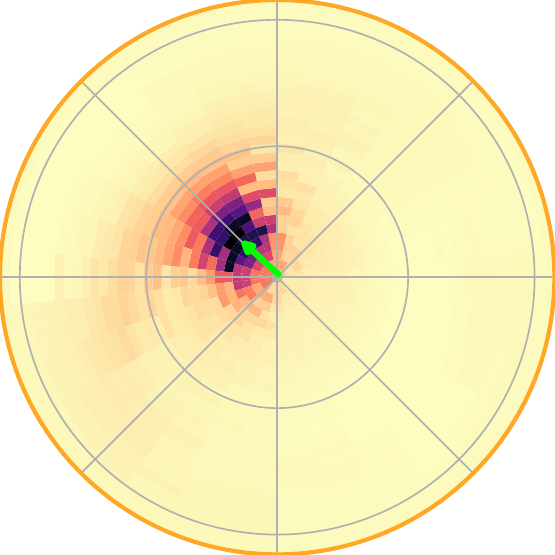}
  \end{tabular}
  }
  \subfloat[depth = 40 cm]{
  \begin{tabular}[b]{c}
  \includegraphics[height=\subH]{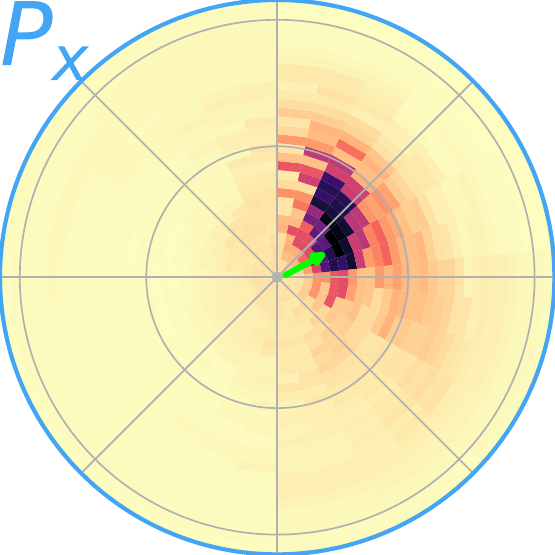}
  \includegraphics[height=\subH]{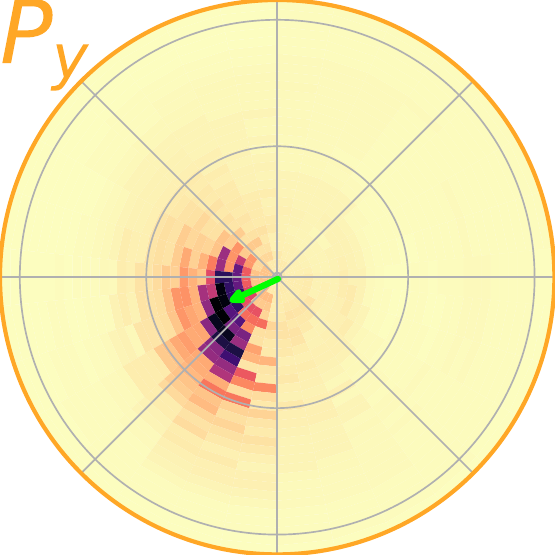}\\            
  \includegraphics[height=\subH]{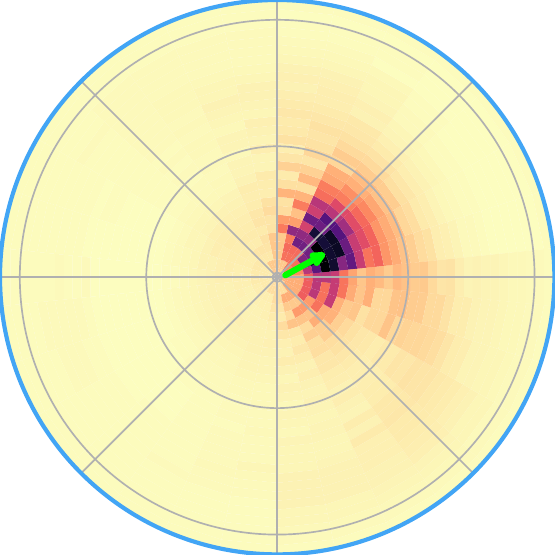}
  \includegraphics[height=\subH]{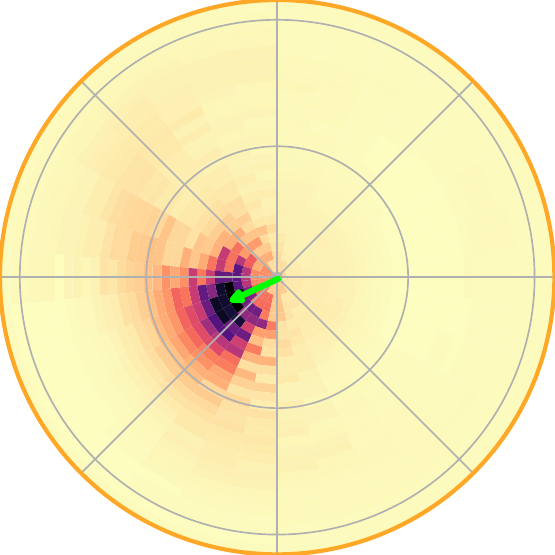}
  \end{tabular}
  }%
  \subfloat[depth = 50 cm]{
  \begin{tabular}[b]{c}
  \includegraphics[height=\subH]{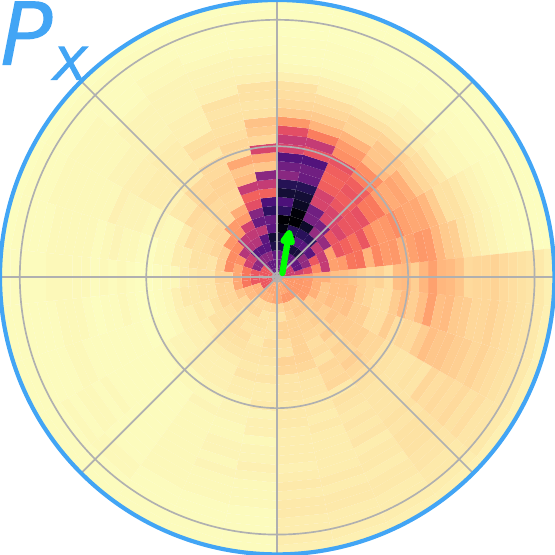}
  \includegraphics[height=\subH]{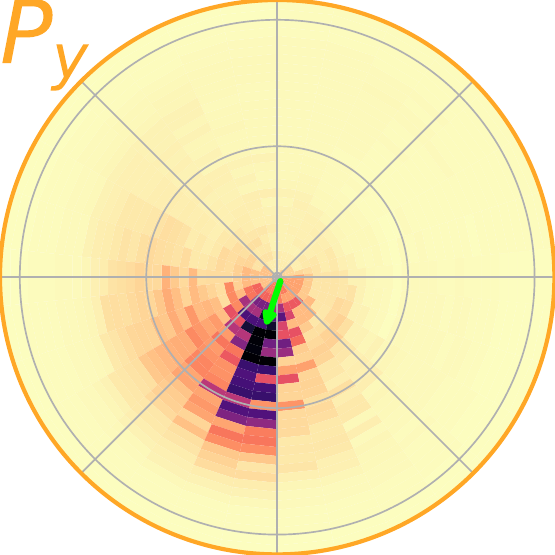}\\            
  \includegraphics[height=\subH]{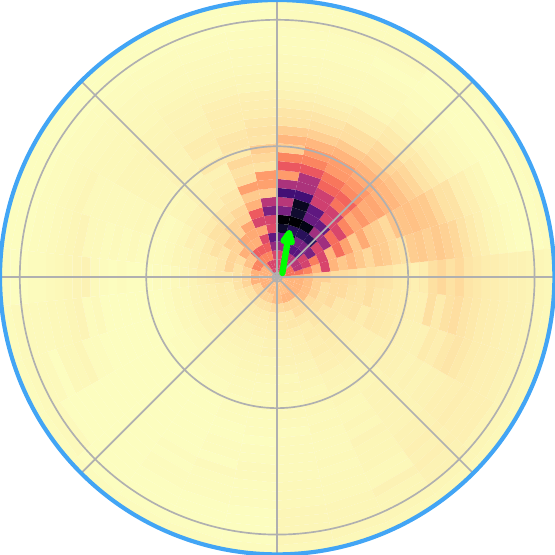}
  \includegraphics[height=\subH]{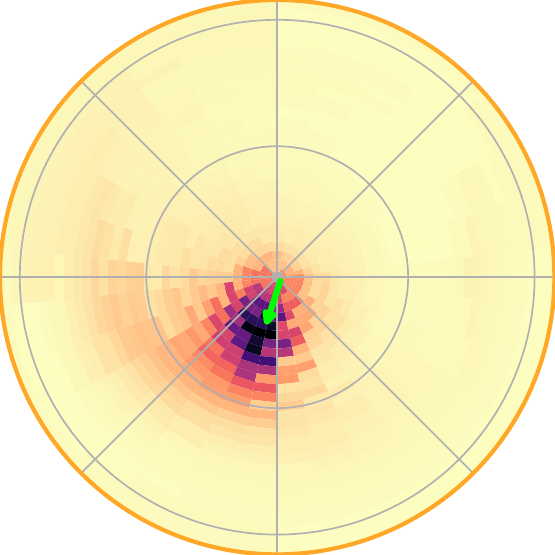}
  \end{tabular}
  }%
  \mysideCaption
  \RowSpacing
  \subfloat[depth = 60 cm]{
  \begin{tabular}[b]{c}
  \includegraphics[height=\subH]{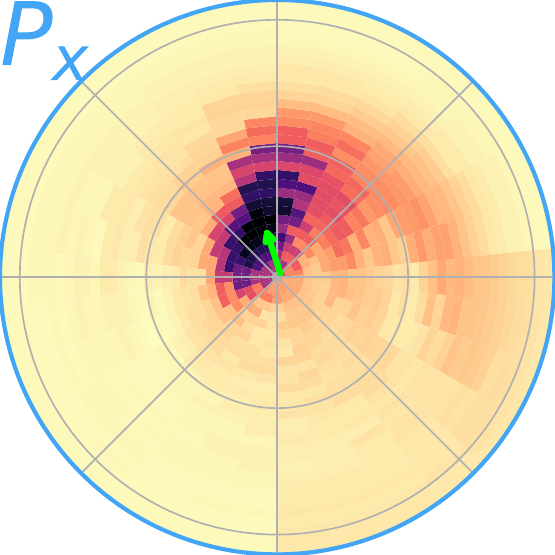}
  \includegraphics[height=\subH]{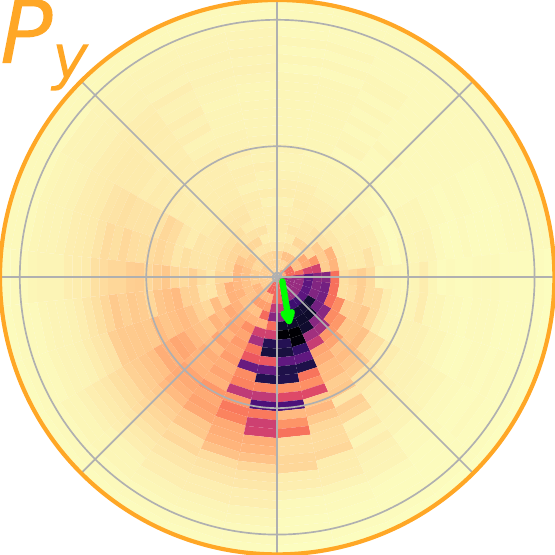}\\            
  \includegraphics[height=\subH]{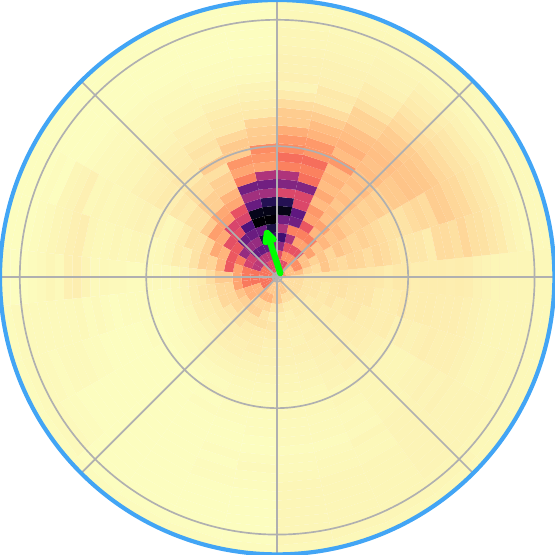}
  \includegraphics[height=\subH]{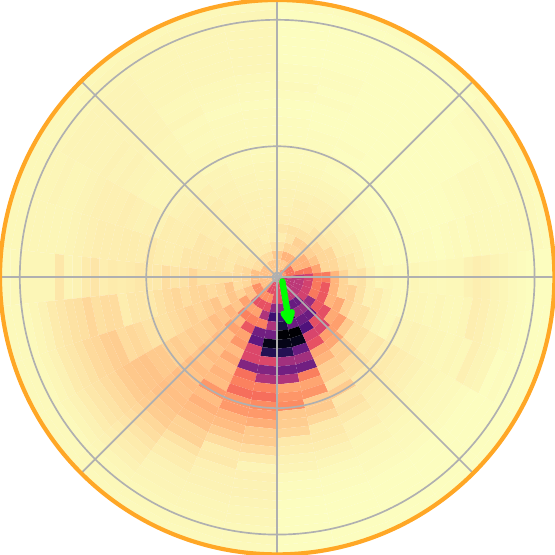}
  \end{tabular}
  }
  \subfloat[depth = 70 cm]{
  \begin{tabular}[b]{c}
  \includegraphics[height=\subH]{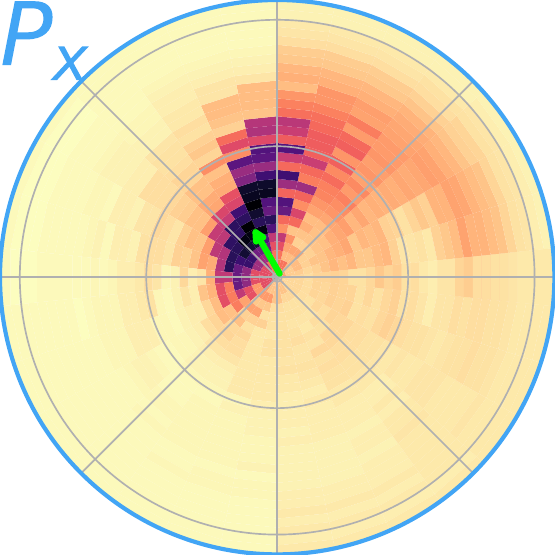}
  \includegraphics[height=\subH]{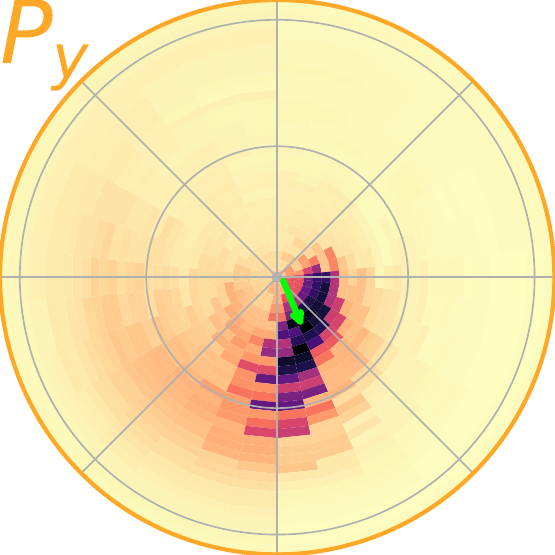}\\            
  \includegraphics[height=\subH]{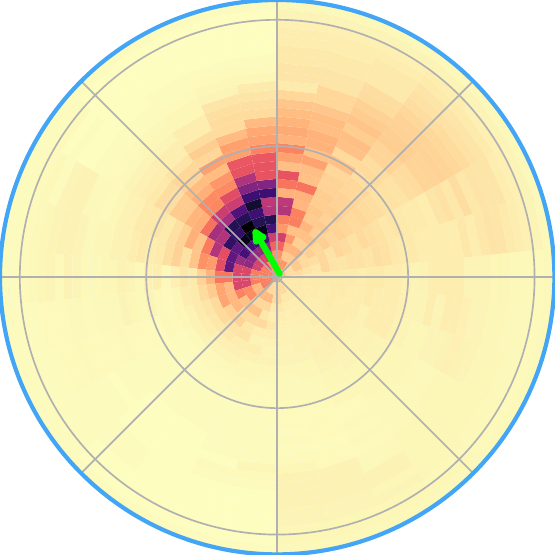}
  \includegraphics[height=\subH]{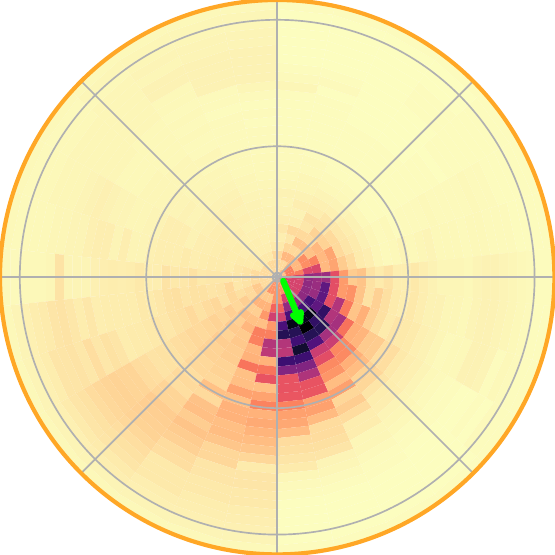}
  \end{tabular}
  }
  \subfloat[depth = 80 cm]{
  \begin{tabular}[b]{c}
  \includegraphics[height=\subH]{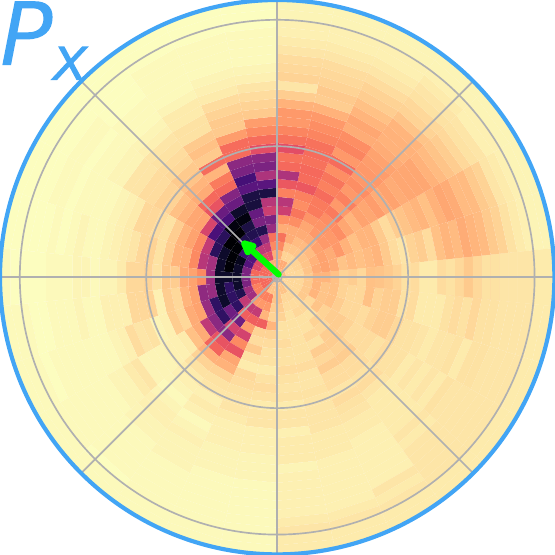}
  \includegraphics[height=\subH]{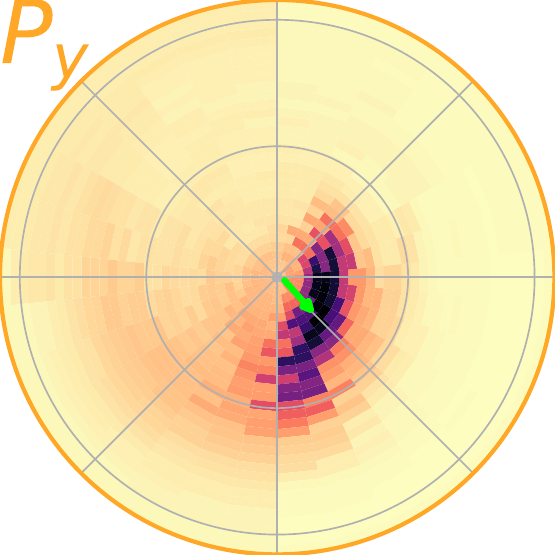}\\            
  \includegraphics[height=\subH]{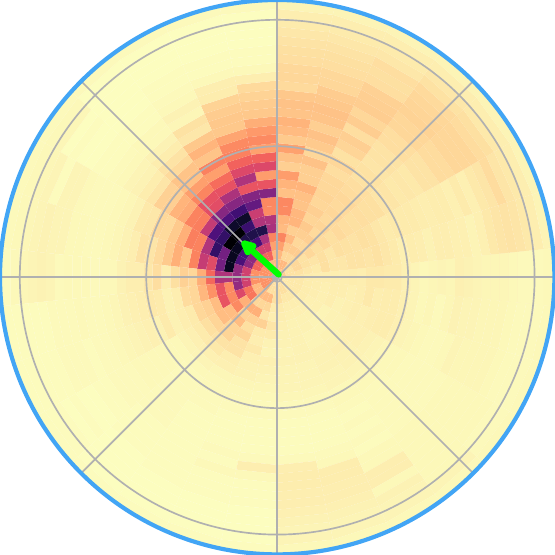}
  \includegraphics[height=\subH]{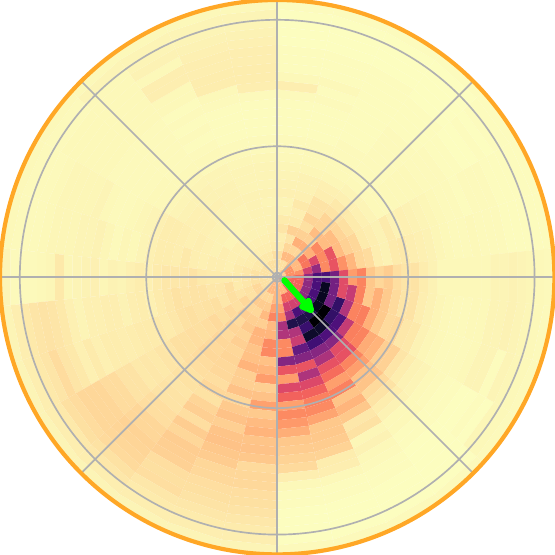}
  \end{tabular}
  }%
  \subfloat[depth = 90 cm]{
  \begin{tabular}[b]{c}
  \includegraphics[height=\subH]{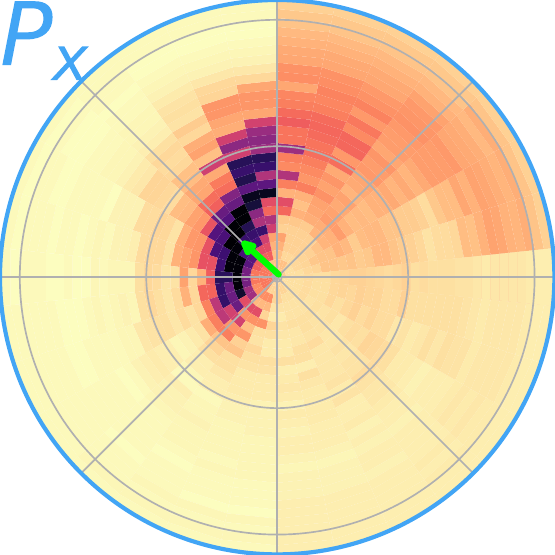}
  \includegraphics[height=\subH]{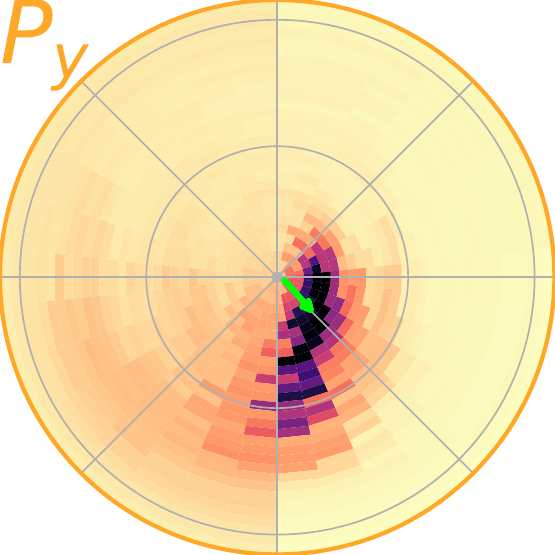}\\            
  \includegraphics[height=\subH]{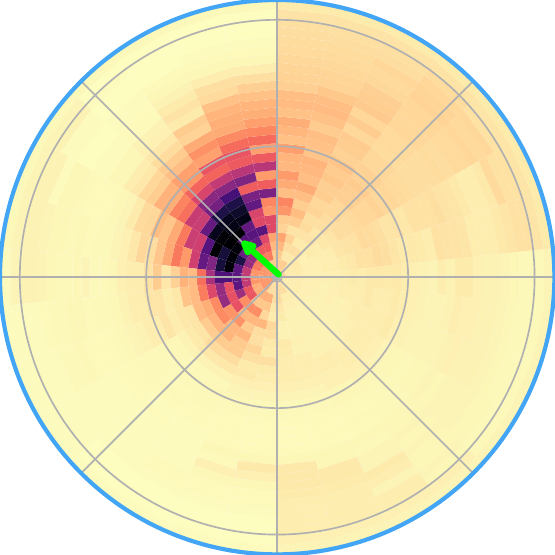}
  \includegraphics[height=\subH]{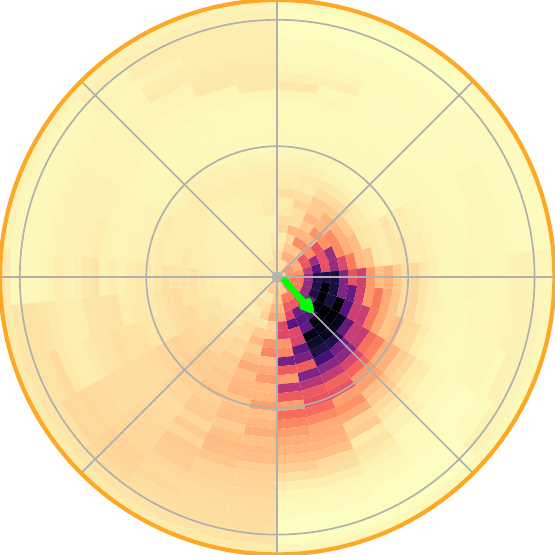}
  \end{tabular}
  }%
  \mysideCaption
  \subfloat[depth = 100 cm]{
  \begin{tabular}[b]{c}
  \includegraphics[height=\subH]{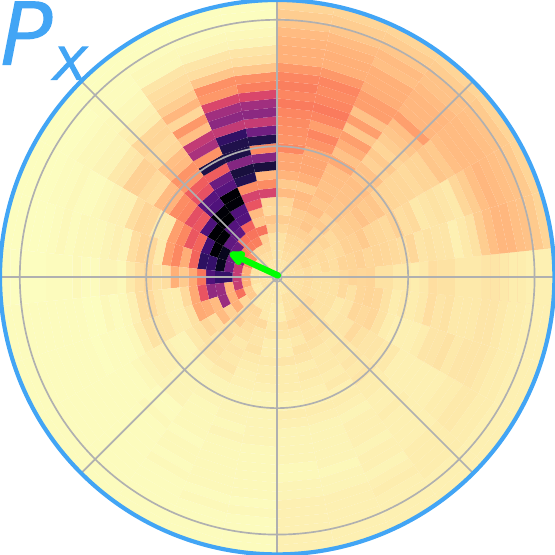}
  \includegraphics[height=\subH]{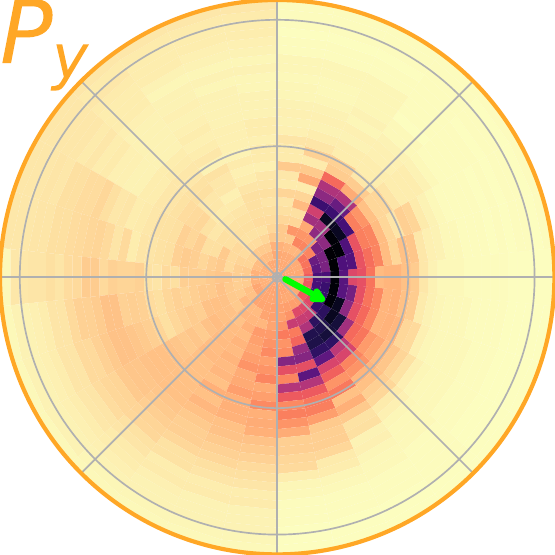}\\            
  \includegraphics[height=\subH]{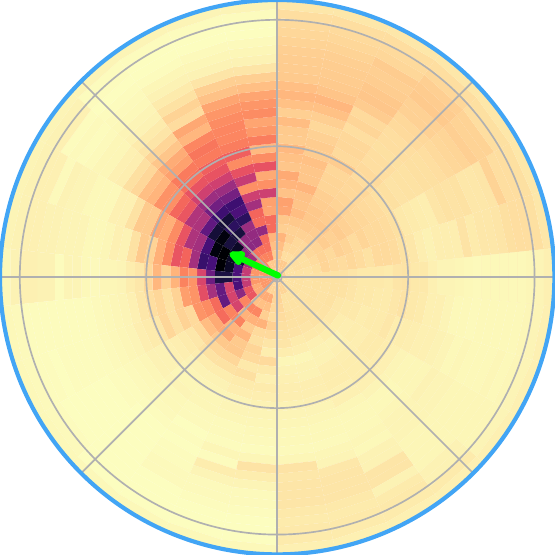}
  \includegraphics[height=\subH]{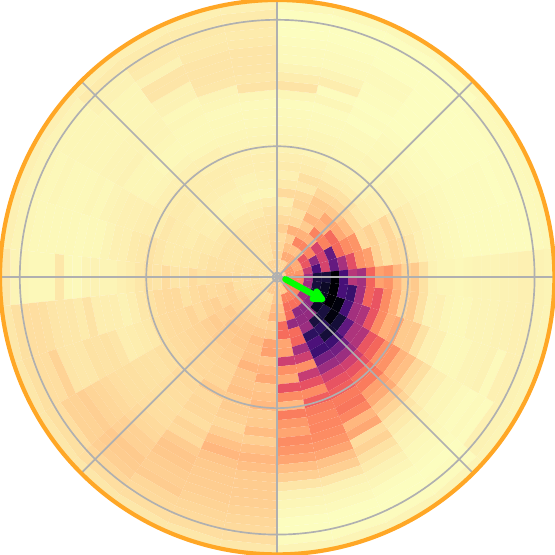}
  \end{tabular}
  }
  \subfloat[depth = 110 cm]{
  \begin{tabular}[b]{c}
  \includegraphics[height=\subH]{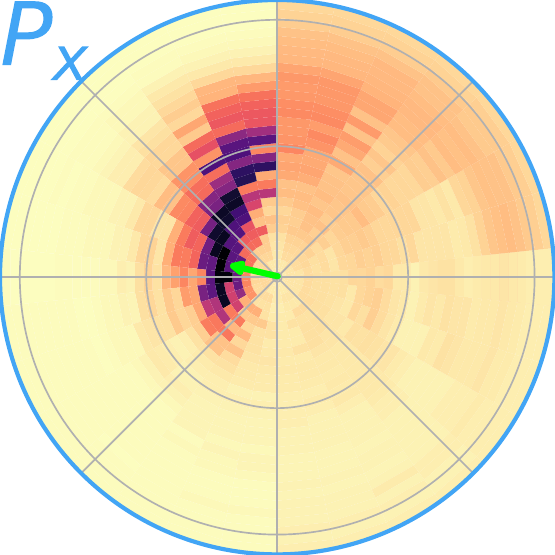}
  \includegraphics[height=\subH]{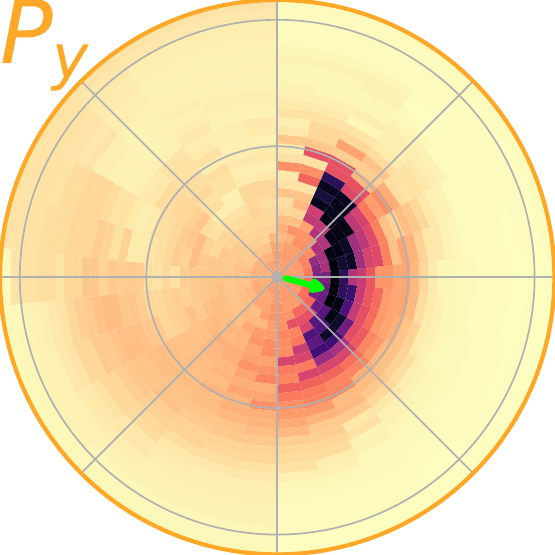}\\            
  \includegraphics[height=\subH]{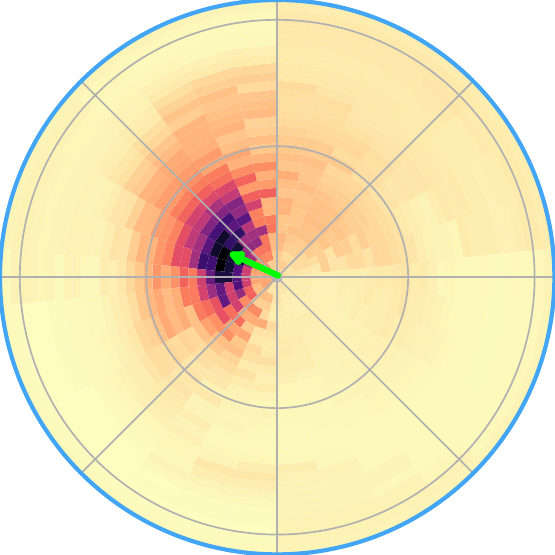}
  \includegraphics[height=\subH]{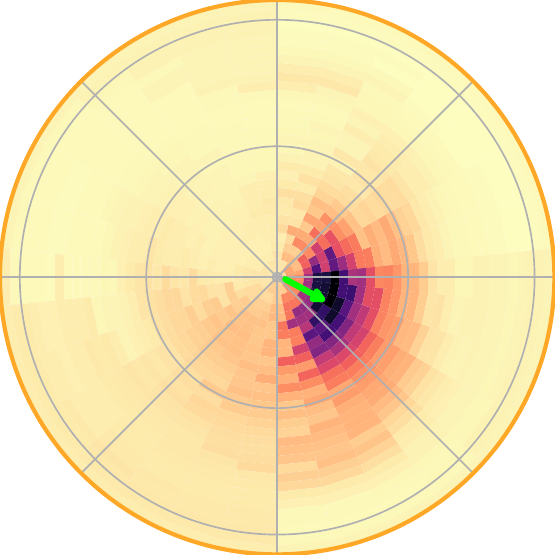}
  \end{tabular}
  }
  \subfloat[depth = 120 cm]{
  \begin{tabular}[b]{c}
  \includegraphics[height=\subH]{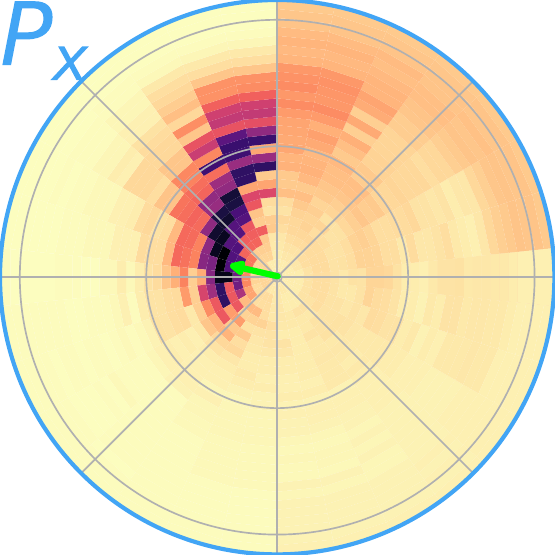}
  \includegraphics[height=\subH]{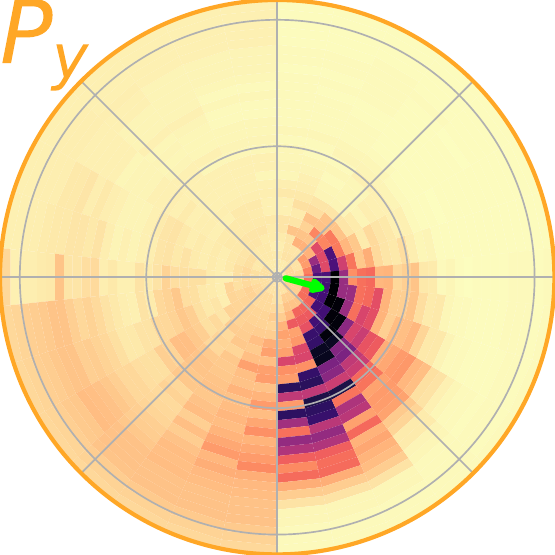}\\            
  \includegraphics[height=\subH]{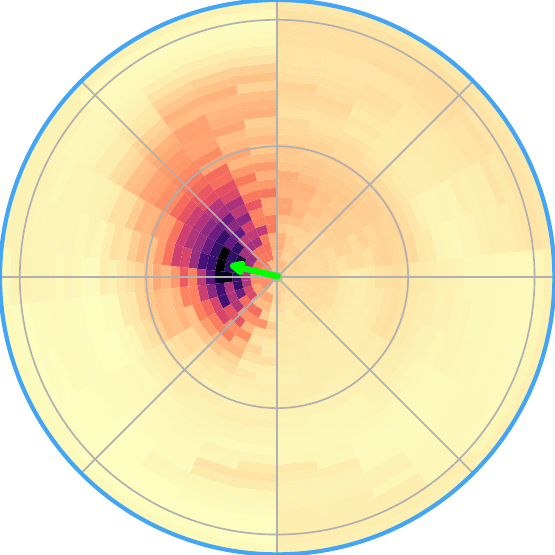}
  \includegraphics[height=\subH]{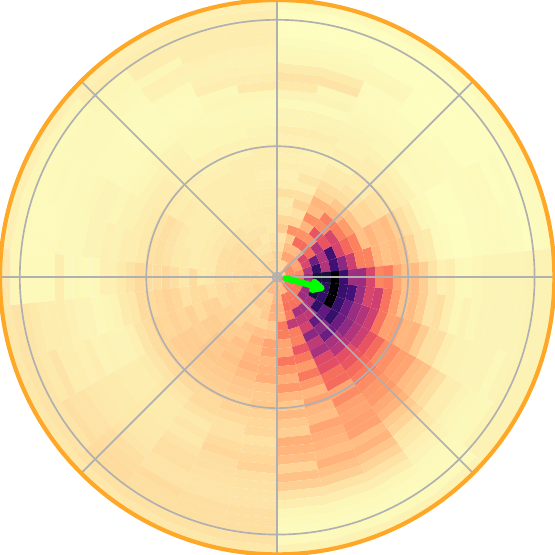}
  \end{tabular}
  }%
  \subfloat[depth = 130 cm]{
  \begin{tabular}[b]{c}
  \includegraphics[height=\subH]{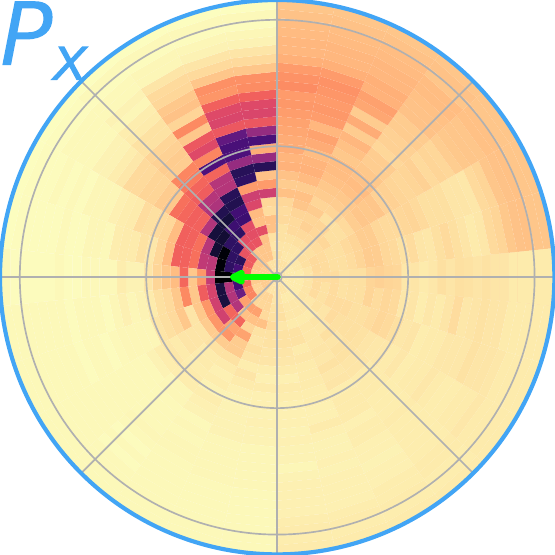}
  \includegraphics[height=\subH]{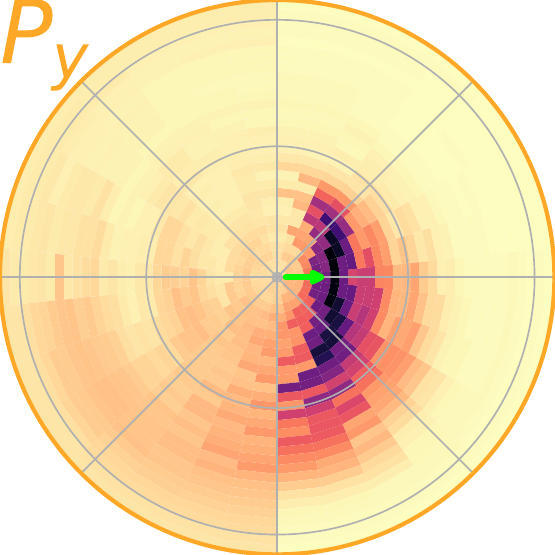}\\            
  \includegraphics[height=\subH]{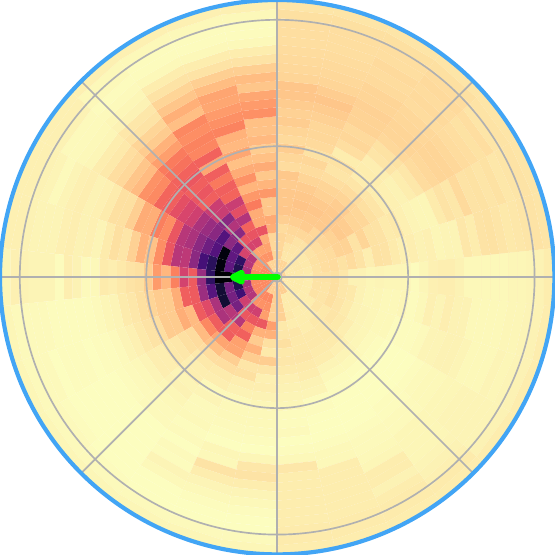}
  \includegraphics[height=\subH]{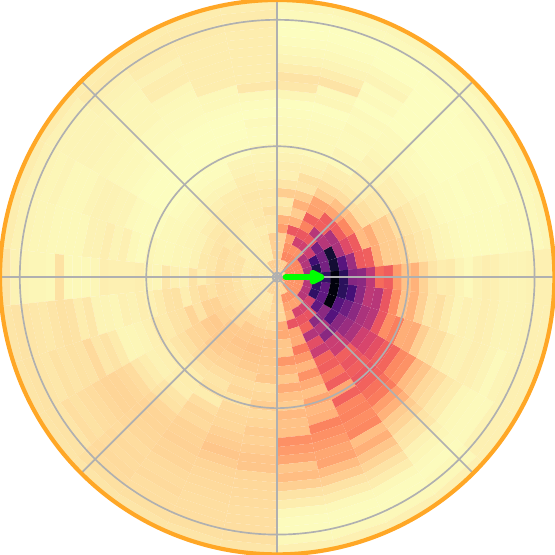}
  \end{tabular}
  }%
  \mysideCaption
  \RowSpacing

\begin{minipage}{0.439\textwidth}
\vspace{10pt}
  \subfloat[]{
    \begin{tabular}[b]{c}
      \includegraphics[width=0.95\linewidth]{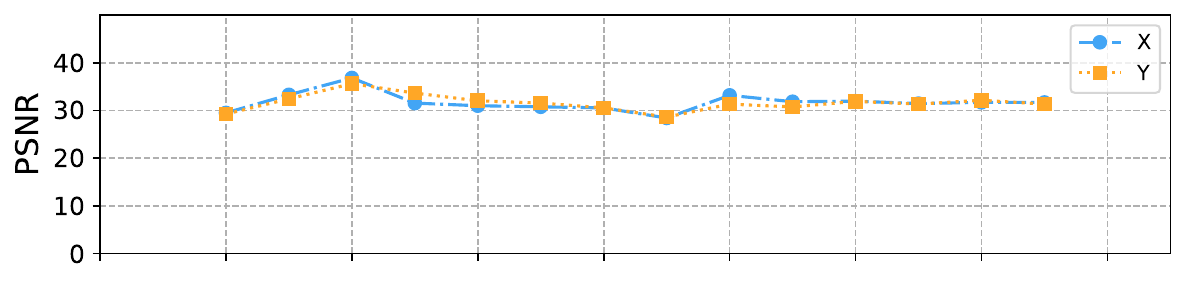}\\
      \includegraphics[width=0.95\linewidth]{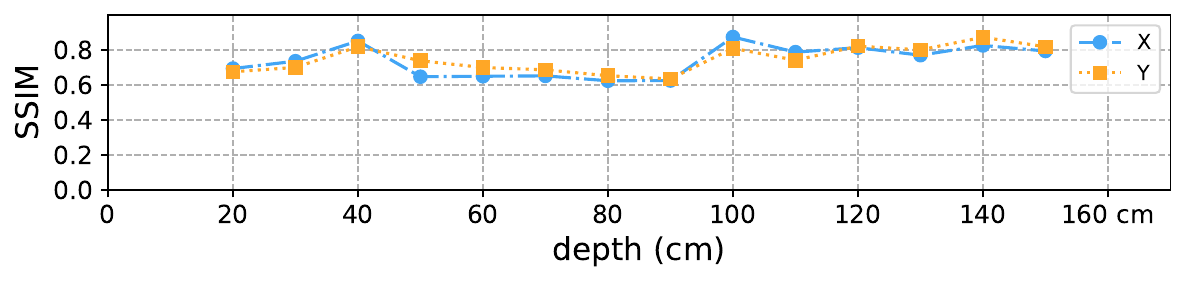}
    \end{tabular}
  }
\end{minipage}
\begin{minipage}{0.50\textwidth}
  \subfloat[depth = 140 cm]{
    \begin{tabular}[b]{c}
      \includegraphics[height=\subH]{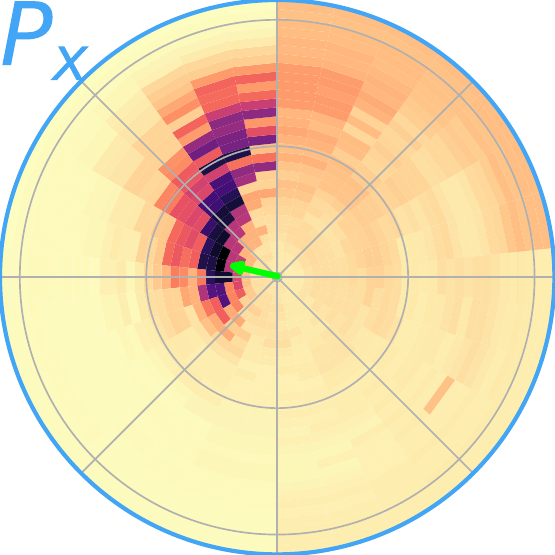}
      \includegraphics[height=\subH]{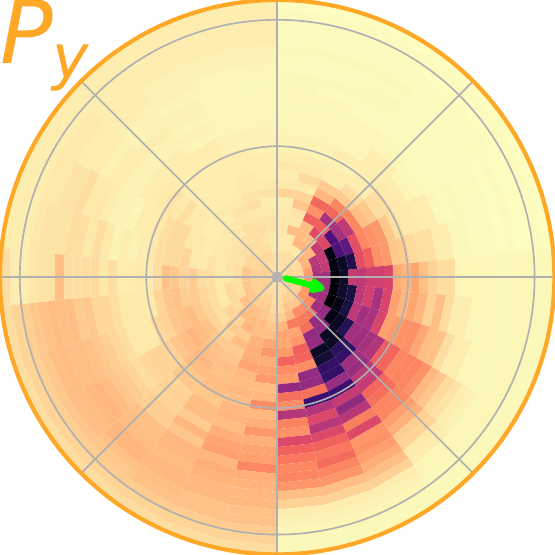}\\
      \includegraphics[height=\subH]{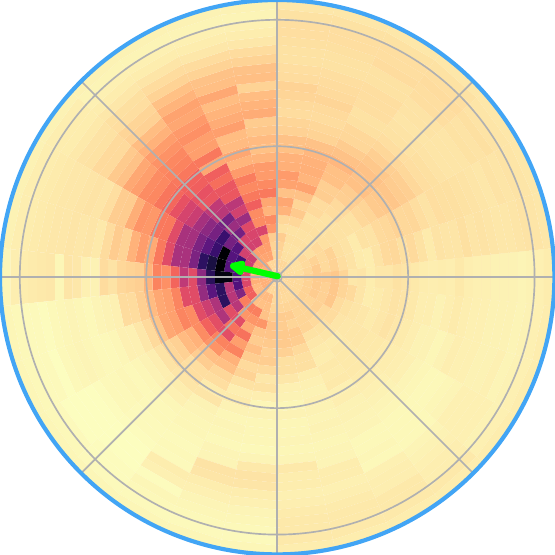}
      \includegraphics[height=\subH]{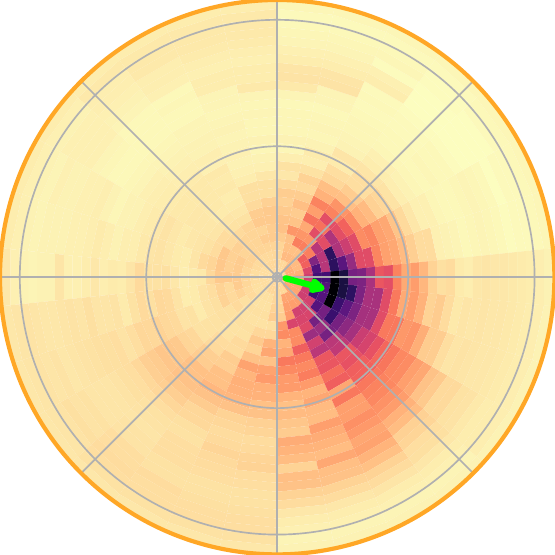}
    \end{tabular}
  }
  \subfloat[depth = 150 cm]{
    \begin{tabular}[b]{c}
      \includegraphics[height=\subH]{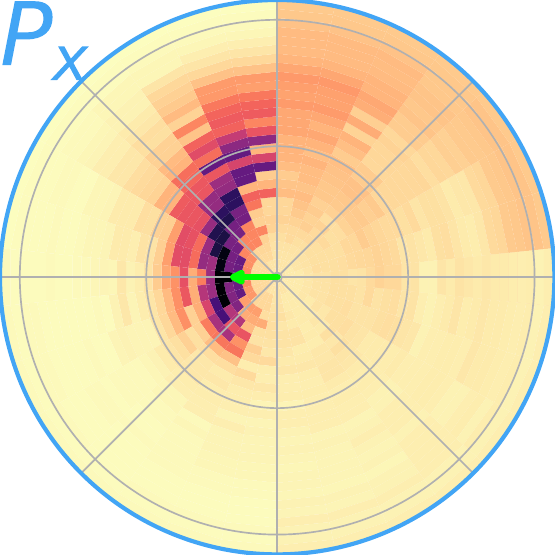}
      \includegraphics[height=\subH]{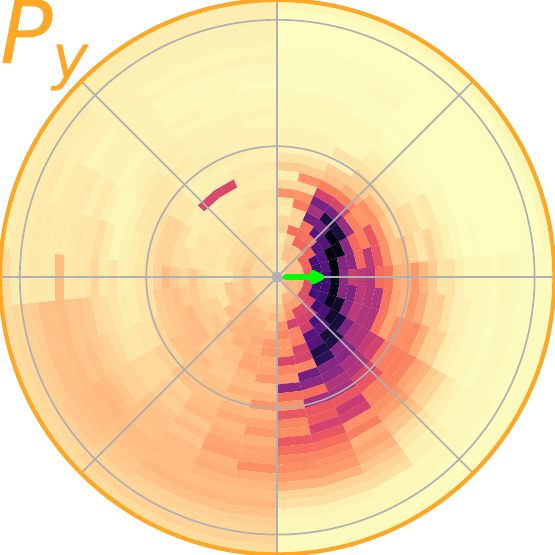}\\
      \includegraphics[height=\subH]{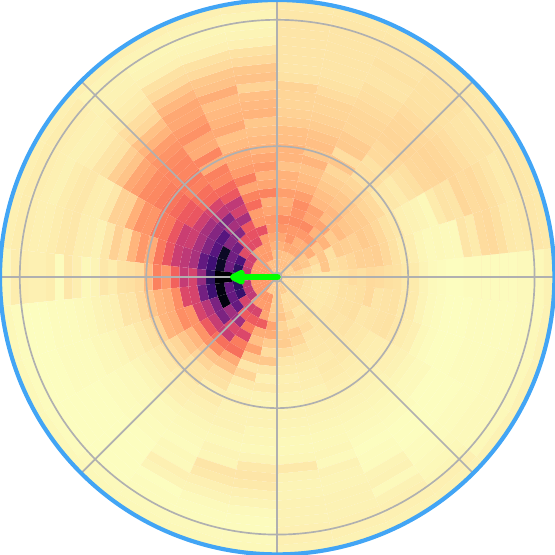}
      \includegraphics[height=\subH]{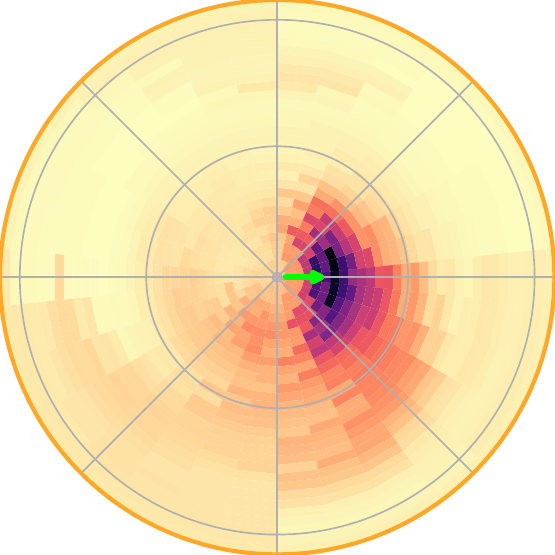}
    \end{tabular}
    \hspace{1pt}
    \raisebox{33pt}{\parbox[c]{10pt}{\rotatebox{-90}{\scriptsize Measured \hspace{3.5pt} Simulated }}}
  }
\end{minipage}

\Caption{ Measured and simulated metasurface's responses to a point light source.}
{We visualize PSFs from 20 cm to 150 cm, although the metalens is designed for a 20–120-cm depth range (\cref{fig:metasurface-psf}). For each depth, the left/right sub-figures show X-/Y-polarized images, and the top/bottom rows show measured/simulated PSFs. Green arrows denote the PSF shift vectors. The bottom-left plots report peak signal-to-noise ratio (PSNR) of the measured PSFs and structural similarity index measure (SSIM) between the measured and simulated PSFs. The simulated PSFs closely match the measured ones across the full depth range, both quantitatively and qualitatively.}
\label{fig:psfdetail}
\end{figure*}


\section{Birefringent Metalens for Polarization-Multiplexing Depth Encoding}
\label{sec:supp:principles}

We first introduce the operating principles of birefringent metasurfaces in \cref{sec:meta:basics}. We then describe how these metasurfaces engineer the Point Spread Function (PSF) in \cref{sec:meta:psf}. Specifically, we demonstrate how depth information is encoded into the rotation of the PSF in \cref{sec:meta:rotating}. Finally, we show how polarization multiplexing is employed to generate two images on a single sensor, encoding depth within the disparity of the polarized image pair (\cref{sec:meta:pair}).

\subsection{Birefringent Metasurface}
\label{sec:meta:basics}
A metasurface is an ultra-thin optical film, typically only hundreds of nanometers in thickness, that can fully modulate electromagnetic waves with subwavelength spatial resolution. Unlike traditional refractive optics that rely on bulk curvature, a metasurface is constructed from a dense, two-dimensional array of microscopic structures like nanopillars, which are called meta-units~\cite{yu2011light, ni2012broadband}. Each meta-unit can independently manipulate the local amplitude, phase, and polarization of the transmitted light. In this context, phase refers to the time delay of the light wave that dictates the wavefront shape, while polarization describes the geometric orientation of the oscillation direction of the electric field component of the light wave. This capability allows the metasurface to achieve complex optical functions within a planar form factor.

\paragraph{Phase Modulation.}We focus on phase-only metasurfaces that impose a spatially-varying phase delay over the incident wavefront while preserving amplitude and polarization. Let $\inE(\metaVector)$ be the incident field at metasurface coordinate $\metaVector$. The transmitted field $\outE(\metaVector)$ is
\begin{align}
\outE\left(\metaVector\right)=
\trans\left(\metaVector\right)\exp\left[i \metaPhase\left(\metaVector\right)\right]\inE\left(\metaVector\right),
\label{eq:meta:basic}
\end{align}
where $\trans(\metaVector)\approx1$ is the near-unity transmission coefficient; $\metaPhase(\metaVector)$ is the designed spatially-varying phase profile. We decompose 
\begin{align}
\metaPhase\left(\metaVector\right)=
\phaseLens\left(\metaVector\right)+
\phaseFresnel\left(\metaVector\right),
\label{eq:meta:phasesplit}
\end{align}
where $\phaseLens$ provides focusing power for the metalens, and $\phaseFresnel$ encodes an additional function (i.e., a helical PSF for depth encoding).

\paragraph{Birefringent Phase Modulation.}
Birefringence is an optical property where a material exhibits different responses depending on the polarization state of the light wave. By designing meta-units with different dimensions in the $x$ and $y$ directions, such as rectangular or cross-shaped pillars, the metasurface can impart distinct phase delays to $x$- and $y$-polarized light. Once the birefringent meta-units are assembled into a metasurface, each unit cell at spatial position $\metaVector$ imparts independent phase shifts, $\metaPhaseX(\metaVector)$ and $\metaPhaseY(\metaVector)$, on the $x$- and $y$-polarized components of the incident electric field, respectively. For a light wave under near-normal incidence with an electric field
\begin{equation}
\inE(\metaVector) 
=
\begin{pmatrix}
\inEx(\metaVector)\\[3pt]
\inEy(\metaVector)
\end{pmatrix},
\end{equation}
the transmitted fields are given by:
\begin{equation}
\outEx(\metaVector)
=
\exp[{\,i\,\metaPhaseX(\metaVector)}]
\,
\inEx(\metaVector).
\end{equation}
\begin{equation}
\outEy(\metaVector)
=
\exp[{\,i\,\metaPhaseY(\metaVector)}]
\,
\inEy(\metaVector).
\end{equation}
In the subsequent text, we use the index $k \in \{x, y\}$ to represent an arbitrary polarization channel when a specific direction is not specified. Accordingly, notation such as $\metaPhaseboth$ denotes the birefringent phase modulation for the $k$-polarized component.

\subsection{PSF Engineered by Metalens}
\label{sec:meta:psf}
\paragraph{Point Spread Function.} The imaging performance of a metasurface is characterized by its amplitude Point Spread Function (PSF), $\amplitudePSF(\imageVector;\physicalPoint)$, which defines the complex field amplitude at the image-plane coordinate $\imageVector=(\imageX,\imageY)$ resulting from a point source $\point$ at $\physicalPoint=(\pointX,\pointY,\pointZ)$. It is important to note that the sensor records intensity; therefore, the observable blur kernel is given by the intensity PSF, $\psf=|\amplitudePSF|^2$. For an extended scene under incoherent illumination, the final captured image is formed by the superposition of these intensity point responses across the field of view.

\paragraph{Kirchhoff's Diffraction for PSF Calculation.} Once the phase profile $\metaPhase$ of the metalens is defined, we can derive the PSF of the metasurface using Kirchhoff's diffraction theory~\cite{born2013principles,Braat2008psf}. Each meta-unit acts as a secondary emitter that imparts a phase delay $\metaPhase$ to the spherical wave originating from a point source at $\physicalPoint$. Integrating these secondary waves across the entire metasurface yields $\amplitudePSF(\imageVector;\physicalPoint)$: 
\begin{align}
\amplitudePSF\left(\imageVector;\physicalPoint\right)
&=
-\frac{i}{\wavelength}
\iint_{\mathrm{MS}}
  \frac{
    \exp\left[
      i k\left|\metaVector-\physicalPoint\right|
    \right]
  }{
    \left|\metaVector-\physicalPoint\right|
  }
  \exp\left[
    i\metaPhase\left(\metaVector\right)
  \right]
  \nonumber \\
&\quad\times
  \frac{
    \exp\left[
      i\,k\,\left|\imageVector-\metaVector + \imageDistance\right|
    \right]
  }{
    \left|\imageVector-\metaVector+\imageDistance\right|
  }
  \dd^2\metaVector,
\label{eq:meta:diffraction}
\end{align}
where the integral is over the 2D metasurface aperture $\mathrm{MS}$, $\wavelength$ is the wavelength, $k=2\pi/\wavelength$, and $\imageDistance$ is the distance from the metasurface to the image plane along the optical axis.  The exponential term $\exp\left[i\,\metaPhase\left(\metaVector\right)\right]$ accounts for the metasurface-imposed phase, while the remaining exponential terms model free-space propagation from $\physicalPoint$ to $\metaVector$ and from $\metaVector$ to $\imageVector$.

\paragraph{Depth-Dependent PSF.}
By evaluating this integral for point sources $\physicalPoint$ across a range of depths, we construct the system’s depth-dependent PSF. To simplify Equation \cref{eq:meta:diffraction}, we introduce the defocus term $\defocus(\metaVector;\depth)$, which arises when the object depth $\depth$ deviates from the designed in-focus plane $\infocus$\cite{Prasad2013psf}: 
\begin{align}
\defocus(\metaVector;\depth)
\;=\;
\frac{\pi\,\pupilR^{2}}{\lambda}(\frac{1}{\depth}-\frac{1}{\infocus}),
\label{eq:meta:defocus}
\end{align}
where $\pupilR=|\metaVector|$. If we assume the focusing phase $\phaseLens$ renders the optical setup an ideal imaging system, we can approximate the depth-dependent PSF using the 2D Fourier Transform $\Fourier$:
\begin{equation}
\psf(z)=|\Fourier\{\mathrm{exp}[i(\phaseFresnel(\metaVector)-\defocus(\metaVector;z))]\}|^2.
\label{eq:meta:psf}
\end{equation}

\subsection{Depth Encoding with Rotating PSFs}
\label{sec:meta:rotating}
Following \cite{Prasad2013psf,shen2023monocular}, we design the phase $\Fresnelphaseboth$ to encode depth $\depth$ as a PSF rotation. In the imaging plane's polar coordinates $(\imageR,\imageAngle)$, the engineered PSF for both polarizations, $\psfboth$, rotates by the same depth-dependent angle $\Delta\imageAngle(\depth)$: 
\begin{equation}
\psfboth(\imageR,\imageAngle;\depth) \approx \psfboth(\imageR,\imageAngle-\Delta\imageAngle(\depth);\infocus),
\end{equation}
where $\infocus$ is the in-focus depth. We set the two polarized patterns $180^\circ$ apart:
\begin{equation}
\psf_x(\imageR,\imageAngle;\depth) =\psf_y(\imageR,\imageAngle-\pi;\depth),
\end{equation}
so their relative disparity vector's angle directly tracks their co-rotation $\Delta\imageAngle(\depth)$, enabling robust depth estimation.

\paragraph{Rotating Phase Profile.}
To realize the PSF rotation, we partition the metalens at the pupil (radius $\radius$) into $\RingNumber=8$ concentric rings, each with a topological charge of $\RingIndex$ ($\RingIndex=1,\dots,\RingNumber$) \cite{Prasad2013psf}. In the pupil polar coordinates $(\pupilR,\pupilAngle)$, the x-polarized phase profile is:
\begin{align}
\Fresnelphasex\left(\pupilR,\pupilAngle\right) 
= \left\{\RingIndex\,\pupilAngle \mid \sqrt{\frac{\RingIndex-1}{\RingNumber}} \le \frac{\pupilR}{\radius} < \sqrt{\frac{\RingIndex}{\RingNumber}}\right\}.
\label{eq:meta:rotationphase}
\end{align}
The y-polarized phase profile $\Fresnelphasey$ is this pattern rotated by $180^\circ$: $\Fresnelphasey(\pupilR,\pupilAngle)=\Fresnelphasex(\pupilR,\pupilAngle-\pi)$. 

\paragraph{Analytical Derivation of Rotating PSF}
Substituting the rotating phase profile (\cref{eq:meta:rotationphase}) of our metalens into the diffraction integral (\cref{eq:meta:psf}) and assuming $\RingNumber\gg1$, we derive an analytic form of the amplitude PSF~\cite{Prasad2013psf}:
\begin{align}
\amplitudePSF_x&(\normimageR,\imageAngle;\defocus)\approx
2\sqrt{\pi}\exp\left[-i\tfrac{\defocus'}{2\RingNumber}\right]
\frac{\sin\left({\defocus'}/{2\RingNumber}\right)}{\defocus}\nonumber\\
&\times\sum_{\RingIndex=1}^{\RingNumber}
  i^\RingIndex
  \exp\left[-i\RingIndex\left(\imageAngle - \tfrac{\defocus'}{\RingNumber}\right)\right]J_{\RingIndex}\left(2\pi\,\sqrt{\RingIndex\RingNumber\normimageR}\right),
\end{align}
where $\normimageR$ and $\imageAngle$ denote the radial and azimuthal coordinates in the normalized image plane, and $J_{\RingIndex}(\cdot)$ is the Bessel function of the first kind of order $\RingIndex$. Here $\defocus'$ is the normalized defocus parameter depending on depth $\depth$:
\begin{align}
\defocus'(\depth)
\;=\;
\frac{\pi\,\radius^{2}}{\lambda}(\frac{1}{\depth}-\frac{1}{\infocus}),
\end{align}
According to these expressions, the PSF rotates by an angle $\Delta\imageAngle(z)$ as the defocus term $\defocus'$ varies with depth $\depth$, given by:
\begin{align}
\Delta\imageAngle (\depth)=\frac{\pi\radius^2}{\RingNumber\lambda}(\frac{1}{\depth}-\frac{1}{\infocus}).
\end{align}
This relationship indicates that a larger aperture radius $\radius$ and a shorter wavelength $\lambda$ increase the rate of PSF rotation. Detailed comparisons of the simulated and experimentally measured rotating PSFs are illustrated in \cref{fig:psfdetail}.

\subsection{Polarization-Multiplexing Depth Encoding}
\label{sec:meta:pair}

\paragraph{Depth-Dependent Image Formation}
We model the image formation process by discretizing the 3D scene $\sceneIrradiance$ into a series of 2D intensity slices at varying depths. As the optical response varies with distance, each slice $\sceneIrradiance(\depth)$ is convolved with its corresponding depth-dependent PSF, $\psfboth(\depth)$. Consequently, the final 2D image $\image_\polarDirection$ is formed by the incoherent superposition of these convolved layers, expressed as $\psfboth(\depth)$:
\begin{equation}
\image_\polarDirection
=
\sum_{\depth}
\sceneIrradiance(\depth) \ast
\psfboth(\depth),
\label{eq:method:psf_convolution_3D}
\end{equation}
where $\ast$ denotes the convolution. The rotating PSF induces slight, depth-dependent position shifts of the objects in the 2D image. Because $\psfX$ and $\psfY$ are 180° apart, the shifts are in opposite directions for the pair of polarized images. This mechanism causes their relative disparity vector to rotate monotonically with depth, providing a geometrically interpretable depth cue. 
\paragraph{Polarization-Multiplexing}
To capture both polarized images in a single shot, we spatially separate them onto the top and bottom halves of the camera sensor by engineering the focusing phase $\phaseLensboth$ to have opposite vertical deflection for the two polarizations:
\begin{align}
\phaseLensboth = -\frac{2\pi}{\wavelength}
\begin{cases}
      \sqrt{\pupilx^2+(\pupily-\sepy)^2+\focus^2}, & \polarDirection=x \\
 \sqrt{\pupilx^2+(\pupily+\sepy)^2+\focus^2}, & \polarDirection=y, \\
\end{cases}
\label{eq:method: focusing}
\end{align}
where $(\pupilx,\pupily)$ are the coordinates on the metalens.

Although our method compares two images, it is fundamentally different from a stereo camera. Our system captures both from a single angle of view with small, several-pixel disparities, keeping it as compact as a monocular camera and avoiding complex stereo matching. Critically, this approach also preserves the underlying image-space structure, unlike computational imaging techniques that introduce blur and distortion. This structural preservation allows our polarization-multiplexed observations to naturally align with the spatial priors of monocular depth foundation models, enabling a seamless transfer of their knowledge to physically-grounded depth estimation.

\section{Metasurface Design and Fabrication}
\label{sec:supp:hw_design_and_fab}
\subsection{Choice of Metasurface Material}
A key enabler of multifunctional metasurfaces is the ability to engineer meta-units with independent control of orthogonal polarization states at subwavelength scales \cite{Mueller2017metapol}. Specifically, by introducing a spatially varying pattern of anisotropic nanostructures (meta-units), one can impart distinct phase shifts on orthogonal polarization components, thus realizing different functions for each polarization channel within a single, ultrathin device \cite{Fan2020arbitrary}. As shown in \Cref{fig:fab}, we employ \matName for its high refractive index and low absorption in the visible regime. These properties simultaneously enable large phase modulation and strong transmission amplitudes for both the x and y polarization channels. We fix a unit-cell (pitch) size that remains subwavelength at the target wavelength, ensuring minimal diffraction orders beyond the zeroth-order transmitted beam.

\subsection{Design of Birefringent Meta-unit Library}
\begin{figure}[]
\centering
    \includegraphics[width=0.5\linewidth]{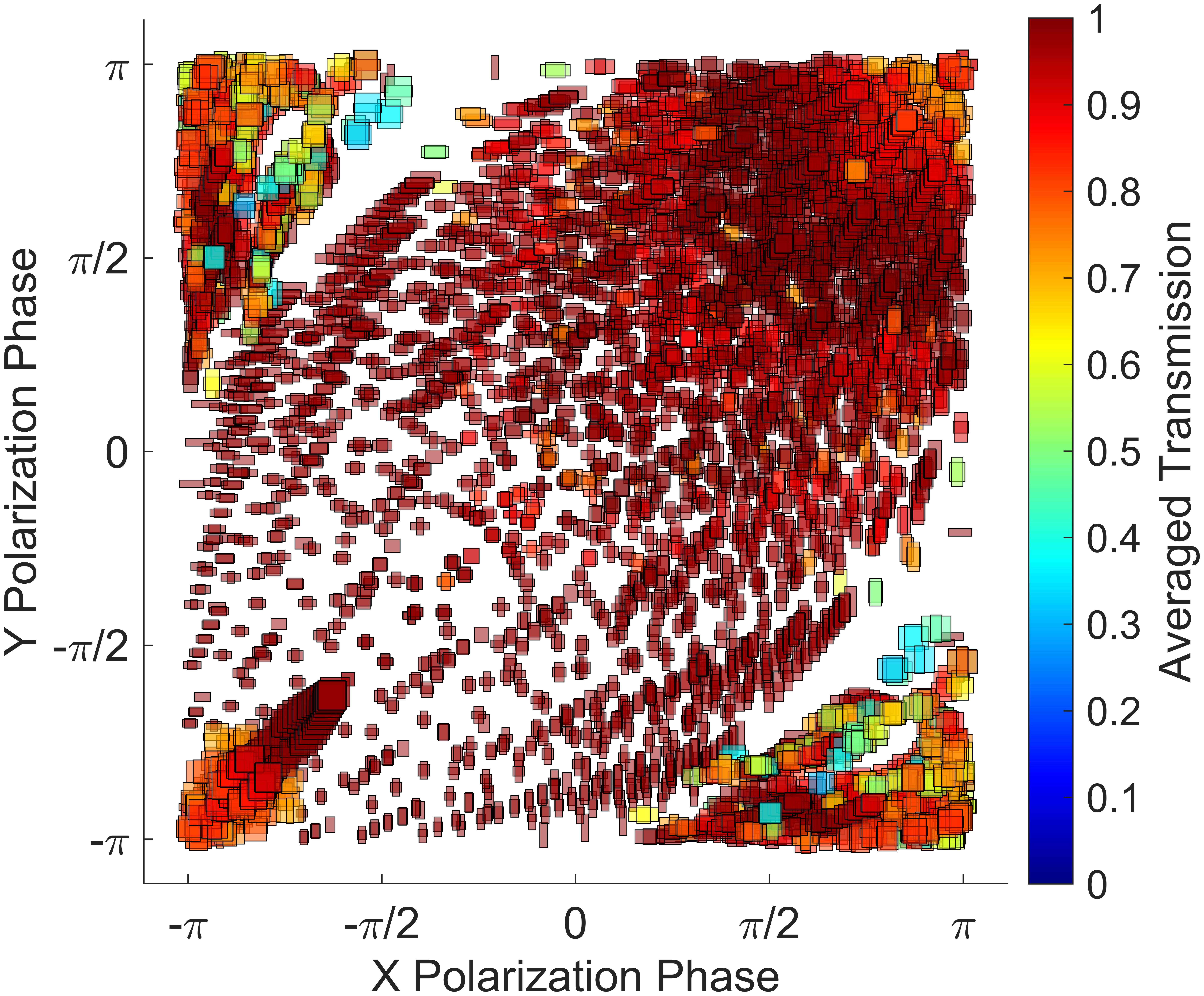}
    
\caption{Each geometry corresponds to a unique type of meta-unit, illustrating the shape of its cross-section. The color represents the transmission efficiency. These meta-units span the entire $2\pi\times2\pi$ phase space while maintaining high transmission.}
\label{fig:library}
\end{figure}
\paragraph{Independency of $x$- and $y$-Polarization Channels}
Polarization multiplexing requires independent phase control for the $x$ and $y$ polarization channels at subwavelength resolution. To achieve this, we seek birefringent ``meta-unit'' structures that can be tuned so that for a specified position $\metaVectorZero$ at the metasurface plane, $\atomPhaseX(\metaVectorZero)$ can take on any desired value over $[-\pi, \pi)$ without constraining the choice of $\atomPhaseY(\metaVectorZero)$. By contrast, metasurfaces lacking sufficient birefringence would impose a correlation between the two polarization channels, thus limiting the efficiency of polarization multiplexing. Hence, the meta-unit library needs to densely sample all possible combinations of $(\atomPhaseX, \atomPhaseY)$ to cover the 2-D phase space $\phaseSpace=\{(\atomPhaseX, \atomPhaseY)|\atomPhaseX,\atomPhaseY\in[-\pi, \pi)\}$ with high transmission in both channels.

\paragraph{Design and Simulation of Meta-unit Library}
The meta-units are designed to be \matName pillars with varying cross-sections and a uniform height. To provide sufficient phase coverage while suppressing the above-zero diffraction orders within our fabrication capability, the pitch and height of our meta-units are chosen to be $\pitch$ = 400 nm and $\height$ = 700 nm, respectively. Within each unit cell, we consider meta-units with rectangular and cross-shaped cross-sections to support different $\atomPhaseX$ and $\atomPhaseY$. The rectangular meta-units are parameterized by their two side lengths $(\slx,\sly)$. The cross meta-units are treated as two overlapping rectangles, resulting in four parameters $(\clx,\cly,\cclx,\ccly)$ that represent the side lengths of each rectangle. These parameters should satisfy the following constraints:
\begin{equation}
\begin{aligned}
\textbf{Square}:&\; (\slx,\sly)\in[\fabCon,\pitch-\fabCon],\\
\textbf{Cross}:&\; (\clx,\cly,\cclx,\ccly)\in[\fabCon,\pitch-\fabCon],\\
               &\; \clx<\cclx,\quad \cly>\ccly.
\end{aligned}
\end{equation}
where $\fabCon$ = 80 nm is the minimum geometry size that can be reliably fabricated within our capability. To construct the whole meta-unit library, we iterate over all the possible geometries generated through the above parameterization and compute the complex transmission coefficients for x and y polarization channels using rigorous coupled-wave analysis (RCWA). The results are provided in \Cref{fig:library}, which clearly shows a comprehensive coverage of the 2-D phase space while maintaining decent transmission for both polarization channels.
\subsection{Metasurface Fabrication Details} 
\begin{figure}[]
\centering
    \includegraphics[width=.8\linewidth]{images/fab_figure.pdf}
\Caption{Illustration of the six-step \matName metasurface fabrication procedure.}{(1) Spin-coat and baking of a 700-nm thick e-beam resist layer. (2) Define the metasurface pattern via e-beam lithography. (3) Develop resist into patterned holes to be filled by \matName. (4) Conformally deposit \matName by ALD. (5) Remove excess \matName layer with reactive ion etching. (6) Remove residual resist to reveal free-standing \matName nanopillars.}
\label{fig:fab}
\end{figure}
As illustrated in \cref{fig:fab}, our $\text{TiO}_2$ metasurfaces are fabricated on 500-\textmu m‑thick, double‑side polished fused silica wafers. A 700-nm ZEP‑520A layer is spin‑coated and baked ($180\ ^{\circ}\text{C}$, 3 min). The thickness of the resist is verified with a stylus profiler (KLA P‑17). After applying an anti-charging layer (DisCharge H2O X2), the nanopillar template is written by 100-KeV electron-beam lithography (EBL; Elionix ELS-G100) with a current of 2 nA and a step size of 4 nm.  The resist is developed in amyl acetate, rinsed in IPA, and nitrogen‑dried, yielding apertures whose depth sets the final $\text{TiO}_2$ pillar height. Amorphous $\text{TiO}_2$ is then conformally deposited at $100\ ^{\circ}\text{C}$ in an ALD reactor (Cambridge NanoTech Savannah 200) until the apertures are fully filled. Excess $\text{TiO}_2$ material on top is removed by inductively coupled plasma (ICP) etching (BCl\textsubscript{3}/Ar, Oxford PlasmaPro 100 Cobra) down to the resist surface.  A final downstream plasma ashing at $600\ \text{W}$ (PVA Tepla IoN 40) removes the resist template, leaving free‑standing $\text{TiO}_2$ nanopillars on the fused‑silica substrate.

\end{document}